\documentclass[aps,twocolumn,prb,superscriptaddress,showpacs]{revtex4}

\usepackage{amsmath,amssymb,bm}
\usepackage{graphicx}
\usepackage{epstopdf}
\usepackage{latexsym}
\usepackage{subfigure}
\usepackage[usenames,dvipsnames]{color}
\usepackage{hyperref}
\usepackage{natbib}

\newcommand{\rvec}{\mathbf{r}}
\newcommand{\qvec}{\mathbf{q}}
\newcommand{\hc}{\mathrm{H.c.}}
\newcommand{\hx}{\hat{x}}
\newcommand{\hy}{\hat{y}}
\newcommand{\abs}[1]{\left\lvert#1\right\rvert}

\def\ET{{$\kappa$-(ET)$_2$Cu$_2$(CN)$_3$}}
\def\dmit{{EtMe$_3$Sb[Pd(dmit)$_2$]$_2$}}

\newcommand{\corrScale}{1.09}

\begin{document}

\title{Bose Metals and Insulators on Multi-Leg Ladders with Ring Exchange}
\author{Ryan V. Mishmash}
\affiliation{Department of Physics, University of California, Santa Barbara, California 93106, USA}
\author{Matthew S. Block}
\affiliation{Department of Physics, University of California, Santa Barbara, California 93106, USA}
\affiliation{Department of Physics and Astronomy, University of Kentucky, Lexington, Kentucky 40506, USA}
\author{Ribhu K. Kaul}
\affiliation{Department of Physics and Astronomy, University of Kentucky, Lexington, Kentucky 40506, USA}
\author{D. N. Sheng}
\affiliation{Department of Physics and Astronomy, California State University, Northridge, California 91330, USA}
\author{\\ Olexei I. Motrunich}
\affiliation{Department of Physics, California Institute of Technology, Pasadena, California 91125, USA}
\author{Matthew P. A. Fisher}
\affiliation{Department of Physics, University of California, Santa Barbara, California 93106, USA}

\date{\today}

\begin{abstract}
We establish compelling evidence for the existence of new quasi-one-dimensional descendants of the $d$-wave Bose liquid (DBL), an exotic two-dimensional quantum phase of uncondensed itinerant bosons characterized by surfaces of gapless excitations in momentum space [O.~I.~Motrunich and M.~P.~A.~Fisher, Phys.~Rev.~B {\bf 75}, 235116 (2007)].  In particular, motivated by a strong-coupling analysis of the gauge theory for the DBL, we study a model of hard-core bosons moving on the $N$-leg square ladder with frustrating four-site ring exchange.  Here, we focus on four- and three-leg systems where we have identified two novel phases:  a compressible gapless Bose metal on the four-leg ladder and an incompressible gapless Mott insulator on the three-leg ladder.  The former is conducting along the ladder and has five gapless modes, one more than the number of legs.  This represents a significant step forward in establishing the potential stability of the DBL in two dimensions.  The latter, on the other hand, is a fundamentally quasi-one-dimensional phase that is insulating along the ladder but has two gapless modes and incommensurate power law transverse density-density correlations.  While we have already presented results on this latter phase elsewhere [M.~S.~Block {\it et~al.}, Phys.~Rev.~Lett.~{\bf 106}, 046402 (2011)], we will expand upon those results in this work.  In both cases, we can understand the nature of the phase using slave-particle-inspired variational wave functions consisting of a product of two distinct Slater determinants, the properties of which compare impressively well to a density matrix renormalization group solution of the model Hamiltonian.  Stability arguments are made in favor of both quantum phases by accessing the universal low-energy physics with a bosonization analysis of the appropriate quasi-1D gauge theory.  We will briefly discuss the potential relevance of these findings to high-temperature superconductors, cold atomic gases, and frustrated quantum magnets.
\end{abstract}


\maketitle

\section{Introduction}

One of the recent challenges in condensed matter physics has been to understand quantum phases characterized by singular surfaces in momentum space.  The canonical example of such a phase is the free Fermi gas (or more generally a Fermi liquid), where the singular surface is simply the Fermi surface.  Ironically, despite our immense theoretical understanding of the free Fermi gas \cite{Ashcroft76_SSP} and the interacting Fermi liquid, \cite{Baym91_FermiLiquids} a controlled and unbiased \emph{numerical} demonstration of such a phase in an interacting microscopic model of itinerant fermions moving in two or more dimensions is still extremely difficult.  The main roadblocks in this numerical pursuit are (1) the infamous sign problem in quantum Monte Carlo simulations \cite{Troyer05_PRL_94_170201} and (2) the anomalously large amount of spatial entanglement present in states with a Fermi surface, \cite{Wolf06_PRL_96_010404, Gioev06_PRL_96_100503, Barthel06_PRA_74_022329, Swingle10_PRL_105_050502} thereby rendering recently developed sign problem-free tensor network state approaches inadequate with current techniques. \cite{Corboz09_PRB_80_165129, Corboz10_PRA_81_010303, Corboz10_PRB_81_165104}  Although the most obvious and familiar, the free Fermi gas and Fermi liquid are not the only examples of phases with singular surfaces in momentum space.

Since the discovery of the cuprate superconductors, interest has emerged in novel two-dimensional (2D) quantum phases that fall outside the paradigm of Fermi liquid theory, but that still have correlations with singularities residing on one-dimensional (1D) surfaces in momentum space.  Perhaps the most prominent example are states with a spinon Fermi surface, so-called ``spin Bose metals'', which have both a long history in the context of the cuprates \cite{Lee06_RevModPhys_78_17} and also a renewed interest in the context of the organic materials \ET\ and \dmit. \cite{Motrunich05_PRB_72_045105, Lee05_PRL_95_036403, Sheng09_PRB_79_205112, Block11_PRL_106_157202}    Here, due to the presence of an emergent gauge field when going beyond mean field, a Fermi-liquid-like quasiparticle description is inapplicable, thus making fully controlled analytics very challenging and further promoting the importance of numerics.  Not surprisingly though, numerics suffers from the same difficulties as before:  (1) the sign problem is likely a necessary condition for realistic parent Hamiltonians, and (2) the beyond boundary law scaling of spatial entanglement is likely the same as that encountered for free fermions. \cite{Grover11_PRL_107_067202}

We focus here on a closely related quantum phase, the ``$d$-wave Bose liquid'' (DBL).  The DBL is an exotic quantum phase of uncondensed itinerant bosons moving on the two-dimensional square lattice first considered in Ref.~\onlinecite{Motrunich07_PRB_75_235116}.  Like states with a spinon Fermi sea, even though the microscopic degrees of freedom are bosonic, the DBL too has a set of gapless excitations residing on 1D surfaces in momentum space, i.e., ``Bose surfaces,'' and so accessing the DBL in 2D using fully controlled techniques, either numerically or analytically, is no more tractable than for the spinon Fermi sea.  However, we make progress studying the DBL in a controlled way by continuing to employ the heretofore fruitful philosophy \cite{Sheng08_PRB_78_054520, Sheng09_PRB_79_205112, Block11_PRL_106_046402, Block11_PRL_106_157202} of building a picture of such a 2D phase through a sequence of controlled quasi-1D ladder studies.  In fact, we believe that the very presence of singular surfaces in momentum space actually renders ladder studies more informative and allows us to circumvent some of the usual numerical and analytical difficulties.  When placed on the $N$-leg ladder, a given 2D phase with a continuous set of gapless modes residing on a 1D surface should enter a phase that is a distinctive multi-mode quasi-1D descendant of the parent 2D phase, with a number of 1D gapless modes that grows linearly with $N$ (used interchangeably with $L_y$ in this paper:  $N=L_y$). \cite{Fisher08_condmat0812.2955}  In such quasi-1D geometries, both numerics and analytics are on much stronger footing:  in principle, potential parent Hamiltonians can be solved numercially with the density matrix renormalization group (DMRG), \cite{White92_PRL_69_2863, White93_PRB_48_10345, Schollwock05_RevModPhys_77_259} which can then be supplemented with variational Monte Carlo (VMC) calculations \cite{Ceperley77_PRB_16_3081} using appropriate projected trial wave functions to help map out the phase diagram.  Despite the fact that the DBL is a strong-coupling phase that cannot be characterized perturbatively in terms of the original bosons, the slave-particle approach on the ladders accesses the phase in a novel way that is tractable via a gauge theory that can be treated using conventional bosonization techniques.

Guided by this line of attack, we aim to accomplish two main goals in this paper:  (1) establish the existence of an exotic metallic DBL phase, i.e., a ``Bose metal,'' on the four-leg ladder, which we will argue is both a close relative of the proposed 2D phase of Ref.~\onlinecite{Motrunich07_PRB_75_235116} and a nontrivial extension of the quasi-1D descendant DBL phase discovered on the two-leg ladder in Ref.~\onlinecite{Sheng08_PRB_78_054520}; and (2) elaborate on our previous work in Ref.~\onlinecite{Block11_PRL_106_046402}, which argued for the existence of a novel, but fundamentally quasi-1D, gapless Mott insulating phase on the three-leg ladder.  We emphasize that while we have already presented highlights of the latter three-leg gapless Mott insulator in a previous work, \cite{Block11_PRL_106_046402} the former four-leg gapless Bose metal constitutes a new and very exciting result.  These two seemingly disparate phases are actually close relatives:  they share a common parton gauge theory description, can be modeled within the same class of projected variational wave functions, and, finally, manifest themselves in the same microscopic model, all of which we will demonstrate in this work.

The model wave function for the DBL is obtained by taking a product of two distinct Slater determinants and evaluating them at identical coordinates (Gutzwiller projection):
\begin{equation}
\Psi_b(\mathbf{r}_1, ...,\mathbf{r}_{N_b}) = \Psi_{d_1}(\mathbf{r}_1, ...,\mathbf{r}_{N_b})\,\Psi_{d_2}(\mathbf{r}_1, ...,\mathbf{r}_{N_b}),
\label{eqn:wf}
\end{equation}
where $N_b$ is the total number of bosons.  In the language of operators, this corresponds to writing the (hard-core) boson operator as a product of two fermionic partons:
\begin{equation}
b^\dagger(\rvec) = d_1^\dagger(\rvec) d_2^\dagger(\rvec),
\label{eqn:partons}
\end{equation}
where the hard-core boson Hilbert space is recovered by requiring that the densities of the two partons, $\rho_{d_\alpha}(\rvec)$, be the same at each lattice site, i.e., $\rho_{d_1}(\rvec)=\rho_{d_2}(\rvec)=\rho(\rvec)$, where $\rho(\rvec)$ is the boson density at site $\rvec$.  Within a gauge theory framework, the mean-field picture of the phase consists of two independent species of fermions hopping on the square lattice with anisotropic hopping; the physical Hilbert space is then obtained by strongly coupling the $d_1$ and $d_2$ fermions with opposite gauge charge to an emergent U(1) gauge field. \cite{Motrunich07_PRB_75_235116, Sheng08_PRB_78_054520}  If one takes the $d_1$ ($d_2$) fermion to hop preferentially in the $\hat{x}$ ($\hat{y}$) direction, then the corresponding wave function [Eq.~(\ref{eqn:wf}) with $\Psi_{d_1}$ ($\Psi_{d_2}$) being a filled Fermi sea compressed in the $\hat{x}$ ($\hat{y}$) direction] has a characteristic $d$-wave sign structure;\cite{Motrunich07_PRB_75_235116} that is, the sign of the wave function goes through a sequence, $+-+-$, upon taking one particle around another and hence the label ``$d$-wave Bose liquid.''  Perhaps more importantly, as alluded to above, this projected wave function has power law singularities of various momentum space correlators, e.g., the boson momentum distribution and density-density structure factor, residing on 1D surfaces in momentum space. \cite{Motrunich07_PRB_75_235116}  These surfaces are perhaps the most distinguishing feature of the DBL and are crucial to our identification of quasi-1D descendants of the phase on the $N$-leg ladder.  A strong-coupling analysis \cite{Motrunich07_PRB_75_235116} of the aforementioned gauge theory motivates a simple microscopic parent Hamiltonian which can potentially harbor the DBL phase.  This Hamiltonian consists of usual nearest-neighbor hard-core boson hopping supplemented with an explicit frustrated four-site ring-exchange interaction (see also Sec.~\ref{sec:model}):
\begin{align}
\label{eqn:model}
H &= H_J+H_K, \\
H_J &= -J\sum_{\rvec;\,\hat{\mu}=\hat{x},\hat{y}}(b^\dagger_\rvec b_{\rvec+\hat{\mu}}+\hc), \\
\label{eqn:modelend}
H_K &= K\sum_{\rvec}(b^\dagger_{\rvec}b_{\rvec+\hat{x}}b^\dagger_{\rvec+\hat{x}+\hat{y}}b_{\rvec+\hat{y}}+\hc),
\end{align}
with $J,K>0$. We focus on this model Hamiltonian, the so-called ``$J$-$K$ model,'' extensively in this paper.  For $K=0$, we expect a generic superfluid with the bosons condensed at $\qvec=0$ and off-diagonal long-range order.  It is in the regime where the ring-exchange term contributes appreciably to the overall Hamiltonian that the $d$-wave Bose liquid is expected to onset.  In the strong-coupling limit of the lattice gauge theory for the DBL, the relative strength of the ring term to the hopping term, $K/J$, increases with increasing hopping anisotropy between the $d_1$ and $d_2$ partons.  Thus, we expect the ring term to potentially stabilize the DBL phase, and, as in Ref.~\onlinecite{Sheng08_PRB_78_054520} on the two-leg ladder, we do in fact remarkably find evidence for such a scenario on three- and four-leg ladder systems.  \\

To conclude this section, we now discuss several lines of motivation behind this work.  Aside from being intrinsically interesting in its own right, we believe the DBL phase is potentially relevant to modern-day experiments and other theoretical pursuits in a number of contexts, including high-temperature superconductors, ultracold atomic gases, and frustrated quantum magnets.  We start by focusing on the first case, where we believe the ideas behind the DBL can be used to describe the charge sector of a particular conducting, non-Fermi liquid phase of itinerant electrons, i.e., a ``strange metal,'' which may possibly be related to the infamous strange metal phase of the cuprates.  Specifically, we have in mind the following scenario.  In the spirit of the slave-boson treatment of the $t$-$J$ model, \cite{Lee06_RevModPhys_78_17} one can (excluding site double occupancy) decompose the electron creation operator as a product of a slave boson (``chargon'') and a fermionic spinon:  $c_\sigma^\dagger(\rvec)=b^\dagger(\rvec)f_\sigma^\dagger(\rvec)$.  This leads to a gauge theory formulation in which the spinons and slave bosons are coupled to an emergent gauge field.  While it is natural in this context for the spinons to form a Fermi sea (see above), the behavior of the slave boson holds the key in determining the properties of the resulting electronic phase:  if the slave bosons condense, a traditional Fermi liquid phase is obtained, whereas if they do not, we may say that the phase is a ``non-Fermi liquid.''  An example of how the latter case can be achieved involves further decomposing the slave boson into two fermionic partons just as we did for the real boson in Eq.~(\ref{eqn:partons}):  $b^\dagger(\rvec) = d_1^\dagger(\rvec) d_2^\dagger(\rvec)$, where again $d_1$ ($d_2$) is taken to fill a Fermi see compressed in the $\hat{x}$ ($\hat{y}$) direction, and now there are two emergent gauge fields needed to enforce the physical electronic Hilbert space.  This setup sets the stage for a theory of a ``$d$-wave metal'' phase.  While we will present work on this phase elsewhere, both in 2D and on the two-leg ladder, we believe scaling up the picture of the charge sector to many legs (as we do in this paper) will be an informative endeavor toward an eventual theoretical understanding of the 2D ``$d$-wave metal.''

A direct experimental realization of the DBL can potentially be achieved in systems of ultracold quantum gases.  Although it has been proposed that the boson ring-exchange model [see Eq.~(\ref{eqn:model})] can be engineered directly in a cold atom system, \cite{Buchler05_PRL_95_040402} perhaps the most feasible scenario involves engineering a pair of mismatched Fermi surfaces in a two-component Fermi gas by introducing a hopping anisotropy between the two species. \cite{Feiguin09_PRL_103_025303}  Such anisotropy can be achieved with a spin-dependent (more precisely ``hyperfine-state-dependent'') optical lattice, a setup that was experimentally first demonstrated several years ago. \cite{Mandel03_Nature_425_937,Mandel03_PRL_91_010407}  In the noninteracting limit, we then have a situation very similar to the mean-field description of the DBL phase as discussed above, i.e., a system of two independent species of fermions characterized by a hopping anisotropy.  However, a fundamental difference is that we are now talking about real fermionic atoms, as opposed to fermionic partons modeling real hard-core bosons [see Eq.~(\ref{eqn:partons})].  What are the potential phases that such a system could enter upon adding local attractive interactions?  This question was first asked by Feiguin and Fisher in Ref.~\onlinecite{Feiguin09_PRL_103_025303}, and although there are several possibilities within a BCS treatment, the most interesting scenario involves formed Cooper pairs entering a ``metallic'' $d$-wave Bose liquid state instead of Bose condensing at a finite set of momenta.  Such a phase is not accessible in a mean-field BCS analysis; however, it can be argued to be a reasonable outcome by observing that when deriving an effective boson Hamiltonian within perturbation theory at strong coupling, a ring term identical to the one we consider in the pure boson context is generated for increased hopping anisotropy between the two fermion species. \cite{Feiguin09_PRL_103_025303}  It is precisely this ring term that drives our boson system into a ``metallic'' DBL phase (see Sec.~\ref{sec:4leg_42} and Ref.~\onlinecite{Sheng08_PRB_78_054520}).  Thus, such a ``Cooper pair Bose metal'' (CPBM) may exist at strong hopping anisotropy and intermediate to strong attractive interactions.  Evidence for such a phase was in fact recently found in a genuine attractive Hubbard model on the two-leg ladder in Ref.~\onlinecite{Feiguin11_PRB_83_115104}, and our results here on three and four legs in the pure boson context may warrant further future studies of the CPBM, both theoretically and experimentally.

A final line of motivation to study the DBL involves thinking of the hard-core boson ring model [see Eq.~(\ref{eqn:model})] as the easy-plane limit of an SU(2) invariant spin-1/2 model with four-spin cyclic ring exchange (see, for example, Ref.~\onlinecite{Chubukov92_PRB_45_7889}).  In such a model, at zero magnetic field (half-filling in the boson language), exact diagonalization (ED) indicates, among other things, the presence of a spin-nematic uniaxial magnet intervening between a four-sublattice biaxial N\'eel state and a fully gapped SU(2) symmetric valence bond solid. \cite{Lauchli05_PRL_95_137206}  The same model has also been studied on the two-leg ladder. \cite{Lauchli03_PRB_67_100409}  Using the same multi-leg ladder approach we employ here, it would be interesting to study this model in the presence of a Zeeman field, which would correspond in the boson language to a finite chemical potential.  Because the 2D DBL is expected to be generically stable only at densities slightly below half-filling (and likely not present at exactly half-filling), \cite{Motrunich07_PRB_75_235116} it is conceivable that such a model in a sector of non-zero net magnetization could enter a spin liquid phase related to our DBL.  Thus, understanding the physics of the DBL and the U(1) symmetric ring-exchange model of Eq.~(\ref{eqn:model}) represents a first step in understanding this putative spin liquid phase.

The rest of the paper is organized as follows. Section~\ref{sec:prelim} describes the basic machinery of our work, including the gauge theory description (Sec.~\ref{sec:gaugetheory}) and the construction of the variational wave functions (Sec.~\ref{sec:wf}); this section also establishes our microscopic model (Sec.~\ref{sec:model}) and provides definitions of the physical measurements we consider (Sec.~\ref{sec:measurements}).  Section~\ref{sec:4leg} addresses in detail the results of our study on the four-leg ladder, including most prominently the detection and characterization of the DBL$[4,2]$ gapless Bose metal phase.  Section~\ref{sec:3leg} gives a similar analysis of our three-leg study, giving special attention to the DBL$[3,0]$ gapless Mott insulator phase.  This phase was the focus of our recent Letter, Ref.~\onlinecite{Block11_PRL_106_046402}.  We first summarize the results of Ref.~\onlinecite{Block11_PRL_106_046402} and then present new, additional evidence and arguments in support of the stability of the gapless Mott insulating phase.  Finally, in Sec.~\ref{sec:conclusions}, we discuss our conclusions and some of our plans for future work and extensions of the ideas presented in this paper.  Appendix~\ref{app:gaugetheory} presents our bosonized solutions to the gauge theories for both the DBL$[4,2]$ and the DBL$[3,0]$ as well as a stability analysis taking into account the effects of short-range interactions.  Appendix~\ref{app:3legIncomm} contains results on the three-leg ladder at incommensurate densities and the successes and failures of the DBL framework in these systems.  Specifically, we find strong evidence for a DBL$[3,1]$ phase for densities $\rho<1/3$, while for $\rho>1/3$ no DBL phase appears to exist; instead we have identified a phase consistent with a three-leg descendent of the ``bond-chiral superfluid'' phase predicted in a recent spin-wave analysis of the 2D $J$-$K$ model. \cite{Schaffer09_PRB_80_014503}  In Appendix~\ref{app:2legHalfFilling} we discuss the situation on the two-leg ladder at half-filling and, in particular, why we fail to find a phase analogous to the DBL$[3,0]$ in that system.

\section{Preliminaries} \label{sec:prelim}

\subsection{Gauge theory description: the DBL on the $N$-leg ladder} \label{sec:gaugetheory}

What follows is a generalization of what has been done in the appendix of Ref.~\onlinecite{Sheng08_PRB_78_054520} for the two-leg ladder.  Here we will summarize the approach and state the key results.  More details can be found in Appendix~\ref{app:gaugetheory}.

\begin{figure}[t]
\centerline{\includegraphics[width=\columnwidth]{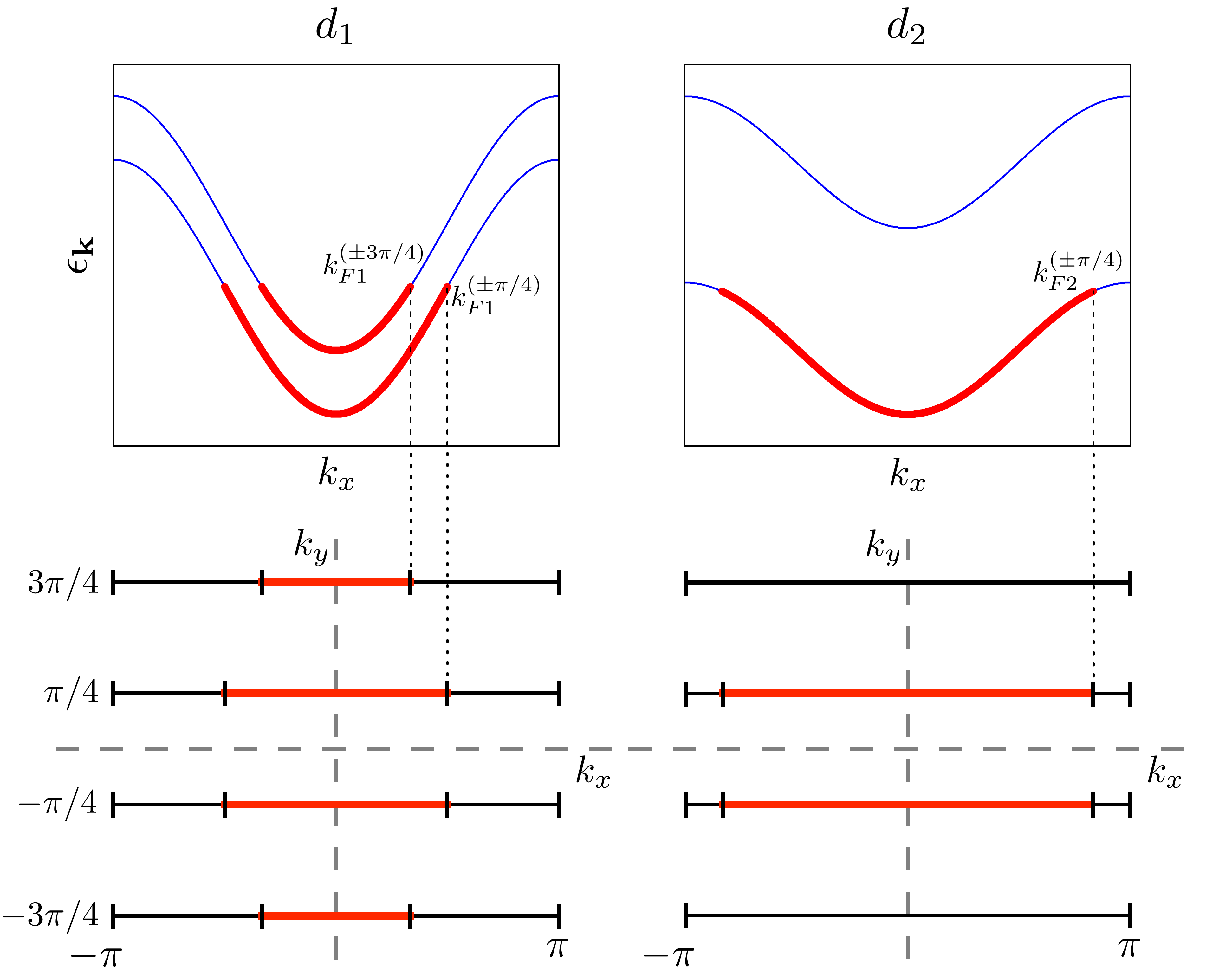}}
\caption{
(color online).  (top panel) Band filling for the DBL$[4,2]$ showing the Fermi wave vectors for the right moving $d_1$ and $d_2$ partons.  (bottom panel)  An overhead view of the occupied momentum states for the DBL$[4,2]$, highlighting the 2D nature of the phase; it is clear that the $d_1$ Fermi sea is compressed along $\hx$, while the $d_2$ Fermi sea is compressed along $\hy$.}
\label{fig:4leg_42bandcurves}
\end{figure}

We can describe the $d$-wave Bose liquid state by first re-expressing the bosonic operators as products of fermionic partons as in Eq.~(\ref{eqn:partons}).  On the $N$-leg ladder, each of the partons has the freedom to fill at most $N$ 1D bands in momentum space corresponding to the $N$ transverse momenta; e.g., choosing periodic boundary conditions in the $\hat{y}$ direction, we have $k_y=2j\pi/N$ for $j=0,\dots,N-1$.  From our mean-field understanding of the state, we expect the partons to occupy contiguous strips centered about $k_x=0$ in each band; such strips can best be described as ``Fermi segments" and the edges of each segment as ``Fermi points," the locations of which are the Fermi wave vectors. These points are the locations where partons can gaplessly be added or removed from the system and lead naturally to the concept of right- and left-movers (see, for example, Fig.~\ref{fig:4leg_42bandcurves}).  Labeling the right-movers' momenta as $k_{F\alpha}^{(k_y)}$, where $\alpha=1,2$ refers to the two flavors of partons and $k_y$ labels the band, we can take the continuum limit in the longitudinal direction ($\hat{x}$) and then linearize the dispersions around each Fermi point, capturing all the relevant low energy physics, and approximately decompose the partons as follows:
\begin{equation}
d_{\alpha}^{\dagger}(x,y)\sim\sum_{k_y,P}\mathrm{exp}\left[iP\left(k_{F\alpha}^{(k_y)}x+k_y(y-1)\right)\right]d_{\alpha P}^{(k_y)\dagger}(x),
\label{eqn:partdecomp}
\end{equation}
where $P=\mathrm{R/L}=+/-$ and denotes the right- and left-movers.  Note that it is not necessary, in general, for either parton to occupy all $N$ bands; therefore, $k_y$ in the sum runs only over the partially filled bands for each flavor $\alpha$.  If the $d_1$ partons partially occupy $n$ bands and the $d_2$ partons partially occupy $m$ bands, we denote the resulting state as DBL$[n,m]$.  Since the two flavors of partons are interchangeable, we can always choose $N\geq n\geq m$.  With this convention, the $d_1$ ($d_2$) partons are most easily associated with those that hop preferentially in the $\hat{x}$ ($\hat{y}$) direction as described in the introduction.  Finally, if we sum up the Fermi wave vectors for each parton, we recover the boson density $\rho$:
\begin{equation}
\label{eqn:fermisum}
\sum_{k_y}k_{F1}^{(k_y)}=\sum_{k_y}k_{F2}^{(k_y)}=N\pi\rho.
\end{equation}

Continuing with our gauge theory description, we assign equal and opposite gauge charges to the $d_1$ and $d_2$ partons and turn on an appropriate U(1) gauge field; in quasi-1D, this behaves like a conventional electric field at long distances attracting the oppositely charged partons toward one another.  In the limit of strong coupling, the partons are bound together on each site such that $d_1^{\dagger}(\rvec)d_1(\rvec)=d_2^{\dagger}(\rvec)d_2(\rvec)=b^{\dagger}(\rvec)b(\rvec)$, thus realizing the physical bosons.  While the mean-field treatment would predict that the DBL$[n,m]$ possesses $n+m$ gapless modes (since the partons are completely free in this case), the effect of the gauge field is to render massive the overall charge mode (see Appendix~\ref{app:gaugetheory}) reducing this number by one.  Hence, a critical feature of the DBL$[n,m]$ is the existence of $n+m-1$ gapless 1D modes.

The gauge theory also offers an explanation for the singular behavior observed at incommensurate wave vectors in momentum-space correlators such as the boson momentum distribution function, $n_b(\qvec)$, and the density-density structure factor, $D_b(\qvec)$, for the DBL (see Sec.~\ref{sec:measurements} for explicit definitions of these quantities).  Consider first $n_b(\qvec)$, the Fourier transform of the boson Green's function, $G_b(\rvec)$.  In the mean-field treatment, $G^{\mathrm{MF}}_b(\rvec)=G^{\mathrm{MF}}_{d_1}(\rvec)G^{\mathrm{MF}}_{d_2}(\rvec)/\rho$; that is, simply the product of the individual fermionic parton Green's functions.  Therefore, $G^{\mathrm{MF}}_b$ will oscillate at various $(Pk_{F1}^{(k_y)}+P' k_{F2}^{(k_y')},Pk_y+P'k_y')$ wave vectors and decay as $1/x^2$; these momenta correspond to the gapless addition/removal of a boson, which involves creating/destroying a $d_1$ and a $d_2$ parton at certain Fermi points.  When gauge field effects are taken into account, we expect these power laws to be altered for the different wave vectors with some oscillations dying off more quickly and some more slowly relative to the mean field.  In Appendix~\ref{app:gaugetheory}, we present a complete low-energy theory for the DBL phases on ladders, which, in principle, allows the exact calculation of all power law exponents.  However, these exponents are also affected by short-range interactions and, due to the large number of parameters in the theory, we do not attempt to do these calculations explicitly.  The gauge theory suggests a mechanism of Amperean enhancement whereby added (or removed) partons with opposite group velocities are favored since their opposite gauge charges generate parallel currents, which will subsequently attract one another thereby satisfying the gauge constraint and realizing the physical bosons.  Assuming that the gauge interactions dominate in our theory, we can rely on the intuition of this Amperean rule despite the effects of the short-range interactions mentioned above.  Thus we expect enhanced singularities in the momentum distribution function at wave vectors $\pm(k_{F1}^{(k_y)}-k_{F2}^{(k_y')},k_y-k_y')$.  As an example, the fermion bilinear $d_{1\mathrm{R}}^{(k_y)\dagger}d_{2\mathrm{L}}^{(k_y')\dagger}$ corresponds to such an enhancement.  These ``Bose points" are the quasi-1D fingerprints of the ``Bose surfaces" of the parent 2D DBL phase; that is, the finite transverse momenta of the ladder slice through the 1D singular curves present in full 2D resulting in these points of singular behavior.  It can be shown within the context of the gauge theory that the other wave vectors, $\pm(k_{F1}^{(k_y)}+k_{F2}^{(k_y')},k_y+k_y')$, are always suppressed relative to the mean field.  In momentum space, $n_b(\qvec)$ shows clearly those wave vectors that are enhanced with sharp peaks or kinks in the curve.

\begin{figure*}[t]
\centerline{\includegraphics[width=\textwidth]{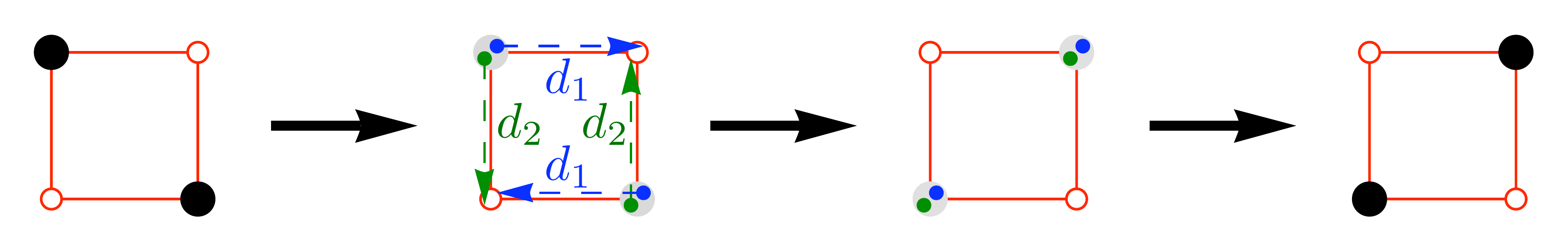}}
\caption{(color online).  Schematic picture of how the anisotropy of the Fermi seas in the parton representation is driven by the ring-exchange term in the model, Eq.~(\ref{eqn:fullmodel}).  If one allows the $d_1$ and $d_2$ partons to virtually hop along their preferred directions, an effective ring exchange for the bosons is generated.  Because we are considering partons with fermionic statistics, the sign of the ring-exchange interaction is rendered positive [cf. Eq.~(\ref{eqn:ringterm})]; in contrast, bosonic partons would generate negative ring exchange.  \cite{Tay11_PRB_83_205107, Tay11_PRB_83_235122}  The resulting sign can intuitively be understood by noting that the ring-exchange process involves one crossing of the partons, as depicted in the second leftmost plaquette shown above, so that the ring-exchange energy is minimized by having a positive (negative) coefficient for fermionic (bosonic) partons.}
\label{fig:ringexch}
\end{figure*}

Turning now to the density-density correlator in real-space, this function can be approximated in the mean-field treatment as the average of the individual parton correlators: $D^{\mathrm{MF}}_b(\rvec)\approx [D^{\mathrm{MF}}_{d_1}(\rvec)+D^{\mathrm{MF}}_{d_2}(\rvec)]/2$, and is therefore expected to oscillate at various ``$2k_F$" wave vectors and decay as $1/x^2$.  These can be thought of as particle-hole excitations between bands of the same flavor of parton.  When the gauge field is turned on, the power law exponents will be modified and the Amperean rule predicts that excitations corresponding to aligned gauge currents will again be enhanced.  Hence, we expect enhanced singularities in the density-density structure factor at wave vectors $\pm(k_{F\alpha}^{(k_y)}+k_{F\alpha}^{(k_y')},k_y+k_y')$.  As an example, the fermion bilinear $d_{1\mathrm{R}}^{(k_y)\dagger}d_{1\mathrm{L}}^{(k_y')}$ corresponds to such an enhancement.  The gauge theory predicts that the non-oscillatory, zero-momentum term in the boson density-density correlator will be unaffected by the gauge interactions, i.e., this term will simply remain $1/x^2$, and the other ``$2k_F$" wave vectors, $\pm(k_{F\alpha}^{(k_y)}-k_{F\alpha}^{(k_y')},k_y-k_y')$, will always be suppressed relative to the mean field.  As in the case of the boson Green's function, considering the momentum space density-density structure factor, $D_b(\qvec)$, allows for a more organized view of those wave vectors that are enhanced by the presence of the gauge field.

The gauge theory predictions involving the number of gapless modes and the locations of singularities allow us to identify the DBL$[n,m]$ if and when it shows up as the ground state of a given model.  We have proposed such a model in Eq.~(\ref{eqn:model}) and will discuss it in greater detail below in Sec.~\ref{sec:model}.

By solving the gauge theory model using bosonization, one can recast the DBL$[n,m]$ as an interacting Luttinger liquid with $n+m-1$ 1D modes with potentially distinct dispersion velocities and corresponding, nontrivial Luttinger parameters characterizing all power law exponents.  This solution, along with a stability analysis, appears in Appendix~\ref{app:gaugetheory}.

\subsection{Wave function:  DBL states on the $N$-leg ladder} \label{sec:wf}

While the DMRG method can readily be employed on any quasi-1D model to extract out measurements of the ground state correlation functions, at best, one would be able to say that these functions decay as power laws and oscillate at incommensurate wave vectors if the ground state were indeed some DBL$[n,m]$.  To make sense of the locations of these wave vectors, we have developed a more direct test for the presence of this phase using variational wave functions and VMC.  These wave functions can be thought of as crude representations of the DBL$[n,m]$ phase as described by the bosonized parton gauge theory in the previous section.  Following the motivation of our mean-field parton description, we start by writing the bosonic wave function as a product of two Slater determinants wherein the orbitals occupied by the partons correspond to the band filling prescription laid out in the previous section (see Fig.~\ref{fig:4leg_42bandcurves}).  But such a wave function without any further constraints would simply correspond to that of two flavors of noninteracting spinless fermions.  We must project this wave function into the space of the physical bosons; i.e., the set of positions of the partons must be equivalent in both determinants [see Eq.~(\ref{eqn:wf})].  This Gutzwiller projection of the wave function plays the role of the strongly-coupled gauge field in the gauge theory.  This sort of variational approach has been rather successful in 1D systems, such as for the original studies of the $t$-$J$ model.\cite{Hellberg91_PRL_67_2080,Hellberg92_PRL_68_3111,Hellberg93_PRB_48_646,Chen93_PRB_47_11548,Chen96_PRB_54_9062,Yang07_CM_19_186218}

The variational parameters in this wave function are simply how one defines the shape of the Fermi seas, built up from 1D segments for $d_1$ and $d_2$ (see Fig.~\ref{fig:4leg_42bandcurves}).  To be truly ``$d$-wave" in nature, there must be some anisotropy in this filling such as in the DBL$[2,1]$, which was identified and characterized in Ref.~\onlinecite{Sheng08_PRB_78_054520}, or the DBL$[4,2]$ and DBL$[3,0]$ states described later in this paper and in Ref.~\onlinecite{Block11_PRL_106_046402}.  But there exists much flexibility in how one does this and many nontrivial fillings are possible in general, each corresponding to different locations of the singular features in the momentum-space correlators.  Indeed, it is the agreement of these locations between the DMRG results and those of the most energetically competitive VMC states that forms the most compelling evidence for the existence of the DBL$[n,m]$ in a given model.

Additional variational freedom can be added by scaling the magnitude of the determinants in the boson wave function; this is one of the fruitful methods borrowed from the 1D $t$-$J$ studies mentioned above.  Explicitly, if we start by writing schematically
\begin{equation}
\Psi_b={\cal P}_G\left[\left(\mathrm{det}D_1\right)\times\left(\mathrm{det}D_2\right)\right],
\label{eqn:wfdets}
\end{equation}
where $D_1$ and $D_2$ are the two matrices populated with $e^{i\qvec_i^{(\alpha)}\cdot\rvec_j}$, the set $\{\qvec_i^{(\alpha)}\}$ is the Fermi sea for $d_{\alpha}$, and ${\cal P}_G$ denotes the Gutzwiller projection, then we can replace each determinant as follows:
\begin{equation}
\mathrm{det}D_{\alpha}=\left\lvert\mathrm{det}D_{\alpha}\right\rvert\frac{\mathrm{det}D_{\alpha}}{\left\lvert\mathrm{det}D_{\alpha}\right\rvert}\rightarrow\left\lvert\mathrm{det}D_{\alpha}\right\rvert^{p_{\alpha}}\frac{\mathrm{det}D_{\alpha}}{\left\lvert\mathrm{det}D_{\alpha}\right\rvert},
\label{eqn:detwithexps}
\end{equation}
such that
\begin{equation}
\Psi_b={\cal P}_G\left[\left(\frac{\mathrm{det}D_1}{\left\lvert\mathrm{det}D_1\right\rvert^{1-p_1}}\right)\times\left(\frac{\mathrm{det}D_2}{\left\lvert\mathrm{det}D_2\right\rvert^{1-p_2}}\right)\right].
\label{eqn:wfexps}
\end{equation}
The exponents $p_1$ and $p_2$ can now be varied; this allows one to tune the $m+n-1$ Luttinger parameters mentioned in the previous section.  
We expect that the ``bare" Gutzwiller wave function, i.e., $p_1=p_2=1$, corresponds to all Luttinger parameters fixed at their trivial values of unity.  \cite{Sheng09_PRB_79_205112}  Varying these exponents can also be seen as augmenting the pure Gutzwiller-projected Slater determinant wave function with a logarithmic Jastrow-type factor.

Under the Gutzwiller projection, different mean-field states can lead to the same physical wave function (gauge redundancy).  For our system and our practical purposes, this can be summed up as follows:  if one shifts the Fermi sea of $d_1$ by momentum $\mathbf{Q}$ and simultaneously shifts the Fermi sea of $d_2$ by momentum $-\mathbf{Q}$, the wave function remains unchanged.  The most significant consequence of this property is that there exists some arbitrariness in the choice of parton boundary conditions.  Since we are going to be considering a finite system with periodic boundary conditions in both directions for the physical bosons, we actually have the option of choosing the parton boundary conditions, in each direction separately, to be both periodic or both antiperiodic.  Either selection is mappable to the other by virtue of this shifting property, and so we will often make this choice based on what is most conceptually clear or aesthetically appealing.

\subsection{Microscopic model:  hard-core bosons with frustrating ring exchange} \label{sec:model}

In the introduction, we stated the model of primary interest in Eqs.~(\ref{eqn:model})-(\ref{eqn:modelend}).  The unfrustrated version of this model ($K\leq0$), with both hard-core and soft-core bosons at and away from half-filling, has a long history in the past decade as a proposed candidate for harboring a deconfined quantum critical point and/or an exotic quantum phase dubbed the ``exciton Bose liquid'' (EBL), \cite{Paramekanti02_PRB_66_054526} a relative of our DBL.  However, for this particular model, both of these scenarios have been largely ruled out in a sequence of quantum Monte Carlo studies, \cite{Sandvik02_PRL_89_247201, Melko04_PRB_69_100408, Sandvik06_AnnPhys_321_1651, Rousseau04_PRL_93_110404, Rousseau05_PRB_72_054524} although recent work has shown that the EBL can be stabilized if one supplements a $K$-only model, Eq.~(\ref{eqn:modelend}), with ring exchange on $1\times 2$ and $2\times 1$ plaquettes. \cite{Tay10_PRL_105_187202, Tay11_PRB_83_205107}

Our focus here is on the explicitly frustrated ($K>0$) case.  This ring-exchange term and its role in manifesting the proposed DBL phase has a simple, intuitive physical explanation, which is visualized in Fig.~\ref{fig:ringexch}.

We can modify Eq.~(\ref{eqn:model}) somewhat to allow for anisotropic hopping of the bosons; that is,
\begin{align}
\label{eqn:fullmodel}
H &= H_{\mathrm{hop}}+H_K, \\
H_{\mathrm{hop}} &= -J\sum_{\rvec}(b^\dagger_\rvec b_{\rvec+\hx}+\hc)-J_{\perp}\sum_{\rvec}(b^\dagger_\rvec b_{\rvec+\hy}+\hc), \\
H_K &= K\sum_{\rvec}(b^\dagger_{\rvec}b_{\rvec+\hat{x}}b^\dagger_{\rvec+\hat{x}+\hat{y}}b_{\rvec+\hat{y}}+\hc). \label{eqn:ringterm}
\end{align}
Now there are two dimensionless parameters, $K/J$ and $J_{\perp}/J$, in addition to the boson density $\rho$.  Thus, for a fixed value of the density, we can explore a two-dimensional phase diagram in search of DBL phases.

Our model lives on an $N$-leg square ladder wherein we can define lattice coordinates: $x=1,\dots,L_x$, where $L_x$ is the length of the chains, and $y=1,\dots,L_y$, where $L_y=N$, the number of legs. The ring term in the Hamiltonian applies to all elementary square plaquettes on the ladder.  We use the convention in all numerical work that each hopping term for a given pair of sites and each ring term for a given plaquette is counted precisely once in the sums.

\subsection{Measurements} \label{sec:measurements}

We examine many quantities in this work and will define them explicitly in this section.  The first and most obvious is the ground state energy.  This can be probed directly with DMRG and ED (for small system sizes).  With VMC, we instead compute the trial energy of the variational wave function using our model [see Eq.~(\ref{eqn:fullmodel})].  Minimizing this quantity with respect to the variational parameters (the band fillings and the exponents on the determinants), allows us to find the most competitive VMC state at each value of $K/J$ and $J_{\perp}/J$ and plot out a VMC phase diagram, which can then be compared against DMRG results (see Secs.~\ref{sec:4legPD} and~\ref{sec:3legPD}).

We examine three main correlators to characterize the ground state for a given set of model parameters.  First is the single-particle Green's function:
\begin{equation}
\label{eqn:greens}
G_b(\rvec,\rvec')\equiv\left\langle b^{\dagger}(\rvec)b(\rvec')\right\rangle,
\end{equation}
and its Fourier transform, the boson momentum distribution function:
\begin{equation}
\label{eqn:bosondist}
n_b(\qvec)\equiv\frac{1}{L_xL_y}\sum_{\rvec,\rvec'}G_b(\rvec,\rvec')e^{i\qvec\cdot(\rvec-\rvec')}=\left\langle b^{\dagger}_{\qvec}b_{\qvec}\right\rangle,
\end{equation}
where $L_y$ is the number of legs of the ladder (interchangeable with $N$ in this paper) and $L_x$ is the number of sites in each chain of the ladder.

Next, we define the boson density-density correlator:
\begin{equation}
\label{eqn:ddcorr}
D_b(\rvec,\rvec')\equiv\left\langle\left[\rho(\rvec)-\rho\right]\left[\rho(\rvec')-\rho\right]\right\rangle,
\end{equation}
where $\rho(\rvec)=b^{\dagger}(\rvec)b(\rvec)$ and $\rho\equiv N_b/(L_xL_y)$ (with no argument) is the average boson density.  Also, its Fourier transform, the density-density structure factor, is defined as
\begin{equation}
\label{eqn:ddstruct}
D_b(\qvec)\equiv\frac{1}{L_xL_y}\sum_{\rvec,\rvec'}D_b(\rvec,\rvec')e^{i\qvec\cdot(\rvec-\rvec')}=\left\langle \delta \rho_{-\qvec}\delta \rho_{\qvec}\right\rangle.
\end{equation}

We also found it useful to consider in some circumstances current-current correlations on our ladder systems.  To this end, we define the current operator as
\begin{equation}
\label{eqn:current}
J_b^{\hat{\mu}}(\rvec)\equiv i\left[b^{\dagger}(\rvec+\hat{\mu})b(\rvec)-b^{\dagger}(\rvec)b(\rvec+\hat{\mu})\right],
\end{equation}
where $\mu=\hx,\hy$, and the current-current correlator as
\begin{equation}
\label{eqn:cccorr}
C_b^{\hat{\mu},\hat{\nu}}(\rvec,\rvec')\equiv\left\langle J_b^{\hat{\mu}}(\rvec)J_b^{\hat{\nu}}(\rvec')\right\rangle.
\end{equation}
The associated structure factor is simply the Fourier transform:
\begin{equation}
\label{eqn:ccstruct}
C_b^{\hat{\mu},\hat{\nu}}(\qvec)\equiv\frac{1}{L_xL_y}\sum_{\rvec,\rvec'}C_b^{\hat{\mu},\hat{\nu}}(\rvec,\rvec')e^{i\qvec\cdot(\rvec-\rvec')}.
\end{equation}
In this work, we have only considered the $\hy$-$\hy$ and $\hx$-$\hx$ correlations, i.e., $C_b^{\hy,\hy}(\qvec)$ and $C_b^{\hx,\hx}(\qvec)$.

As mentioned earlier, having a means by which to determine the number of gapless 1D modes is a critical diagnostic for detecting a DBL$[n,m]$ state.  We can refer to conformal field theory (CFT), the results of which suggest a relationship between the number of gapless modes and the central charge, a quantity that can be extracted from the scaling form of the entanglement entropy.  For a subsystem size $X$ in a system of overall length $L_x$ with periodic boundary conditions, the von Neumann entanglement entropy $S$ is given by the scaling form \cite{Cardy04_JStatMech_P06002}
\begin{equation}
\label{eqn:Sscaling}
S(X,L_x)=\frac{c}{3}\mathrm\log{\left(\frac{L_x}{\pi}\sin{\frac{\pi X}{L_x}}\right)}+A,
\end{equation}
where $A$ is a nonuniversal constant independent of the subsystem length and $c$ is the central charge that we seek.  In our ladder studies, we only consider clean vertical cuts through all $N=L_y$ chains so that the left block contains $NX$ sites ($X$ out of $L_x$ total rungs).  Despite the fact that the bosonization analysis leads to a theory of $n+m-1$ free bosonic modes with generally different velocities such that the full system is not conformally invariant, we still expect the overall measure of gaplessness to be insensitive to this detail:  since we are extracting $c$ from the ground state wave function only, which has no knowledge of mode velocities, the scaling form of $S$ should not depend on the velocities.  This is consistent with the known ``$L\log L$'' scaling for free fermions \cite{Wolf06_PRL_96_010404, Gioev06_PRL_96_100503, Barthel06_PRA_74_022329, Swingle10_PRL_105_050502} and projected Fermi sea states \cite{Grover11_PRL_107_067202} in 2D, in which conformal invariance is not present.  By measuring $S$ using DMRG while varying $X$, we can attempt to fit this form to the data and extract the constant fit parameters $c$ and $A$.  Doing so requires highly converged DMRG data on large system sizes, $L_x$, in order to get reliable estimates of the central charge.  Since this task becomes increasingly more computationally challenging with greater spatial entanglement, our DBL$[n,m]$ phases can be particularly difficult to analyze as $c=n+m-1$ grows large.  For example, while we were successful in measuring a central charge of approximately two for the DBL$[3,0]$ (see Sec.~\ref{sec:3leg_30}), we exhausted the computational resources for the DBL$[4,2]$ (see Sec.~\ref{sec:4leg_42}) where we expect $c=5$.

Finally, we also considered comparisons of the ground state momentum at various points in the phase diagram for the DBL$[3,0]$.  We looked at small system sizes where finite-size effects are expected to be significant and extracted the total momentum from ED calculations, which is easily inferred from ED results since the diagonalization is carried out in blocks diagonal in the momentum quantum number.  We then compared these results to the ground state momenta of the VMC states, which are also trivial to compute since we are using complex-valued orbitals, $e^{i\qvec\cdot\rvec}$: one simply sums up all of the individual momenta of the partons, which fill definite momentum orbitals (see Figs.~\ref{fig:4legDBL_DMRG_VMCcf_1} and \ref{fig:4legDBL_DMRG_VMCcf_2}, bottom panel).

The DMRG calculations presented in this paper were performed with anywhere between $D\simeq1000$-$9000$ states per block and at least 6 finite-size sweeps, where each ``sweep'' traverses the $L_x L_y$ sites of the lattice twice. The accuracy of the results is strongly dependent on the phase being studied, the chosen system size and boundary conditions, as well as the physical quantities being measured.  For example, within the four-leg DBL$[4,2]$ phase discussed in Sec.~\ref{sec:4leg_42}, the momentum space correlators on which we focus, $n_b(\qvec)$ and $D_b(\qvec)$, are converged to a relative error of less than $10^{-2}$, and the ground state energy to a relative error on the order of $10^{-4}$; the density matrix truncation error is on the order of $10^{-5}$.  In all other identified phases, the accuracy of our results is better; e.g., within the four-leg superfluid phase (see Fig.~\ref{fig:4legSF_DMRG}), the truncation error is on the order of $10^{-8}$ keeping only $D=2000$ states, and in the three-leg DBL$[3,0]$ phase of Sec.~\ref{sec:3leg_30} the truncation error is on the order of $10^{-7}$ keeping $D=4000$ states.  These quoted errors are for the case of fully periodic boundary conditions, which we prefer due to the incommensurate nature of the DBL phases.  The entanglement entropy is typically the last quantity to converge, and obtaining highly converged entropy data in the multi-mode critical systems we encounter is an extremely challenging numerical task.  We were able to obtain such converged entropy data (used for determination of the central charge, $c$) within the superfluid, DBL$[3,0]$, and DBL$[3,1]$ phases, but were unfortunately unable to do so within the DBL$[4,2]$ phase.  Finally, we note that in such (potentially) ``very critical'' quasi-1D phases, it is not possible to go to very long systems ($L_x \gtrsim 100$, say, at fixed $L_y=N=3,4$) with the DMRG to definitively rule out eventual small gaps and corresponding long (finite) correlations lengths; this is further compounded by the fact that the finite bond dimension matrix product states produced by DMRG give exponentially decaying correlations by construction.  However, such weak instabilities should not be expected \emph{a priori}, and for the DBL$[4,2]$ and DBL$[3,0]$ phases, we have done the best we can to rule out such scenarios by going to reasonably long systems using both fully periodic and cylindrical boundary conditions.

At this point, it is convenient to mention a subtlety regarding the DMRG and how it handles ground states with non-zero momentum.  To begin with, it is not possible to directly extract the ground state momentum from the DMRG states since the real-space blocking construction necessarily breaks translational invariance.  If this momentum is indeed zero or an integer multiple of $\pi$, there is no ambiguity in our DMRG wave function and the choice of real- versus complex-valued wave functions is irrelevant.  If, however, this is not the case and a given ground state has momentum $\mathbf{Q}$, then it necessarily has a time-reversed partner with momentum $-\mathbf{Q}$.  The DMRG ground state is thus some real-valued combination of these two states.  While the lattice-space measurements discussed above may depend slightly on the details of this combination, the differences become less significant with increased system size.  In all the VMC/DMRG comparisons below, we choose points in the phase diagram where $\mathbf{Q}=0$ so as to completely avoid this ambiguity.

With the preliminary details having now been fully discussed, we are now prepared to give detailed descriptions of the results for the model, Eq.~(\ref{eqn:fullmodel}), on three- and four-leg ladders.

\section{Gapless Bose metal phase on the four-leg ladder} \label{sec:4leg}

\begin{figure}[t]
\centerline{\includegraphics[width=\columnwidth]{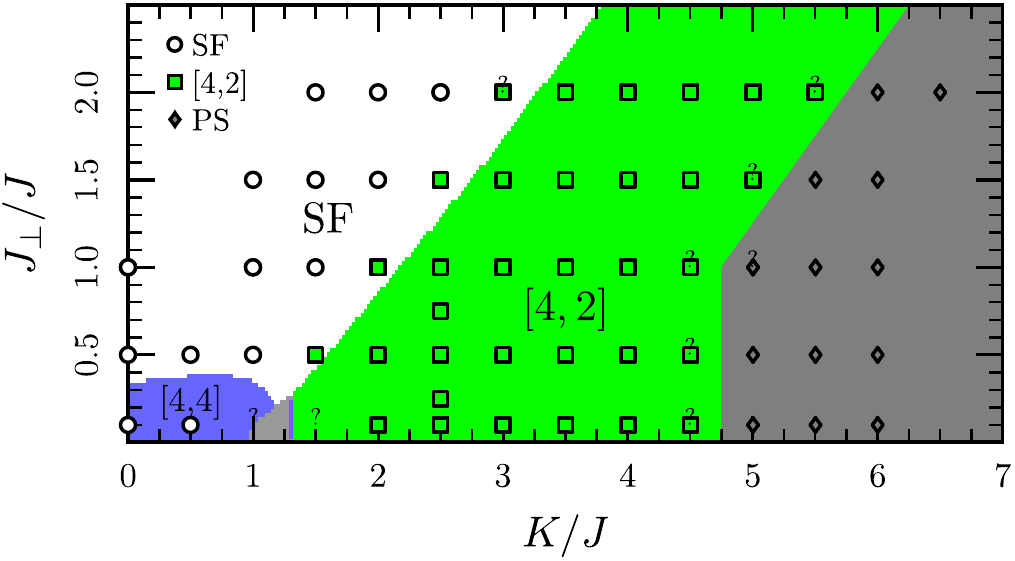}}
\caption{(color online).  Phase diagram of the four-leg system at boson density $\rho=5/12$, using system sizes $12\times4$ with $N_b=20$ bosons, $18\times4$ with $N_b=30$ bosons, and (for some points) $24\times4$ with $N_b=40$ bosons.  The colored regions are delineated using VMC data.  The gray region indicates phase separation and is delineated schematically by considering the boson density in real space with the DMRG approach.  DMRG points for the superfluid, DBL$[4,2]$, and phase separation are indicated by white circles, green squares, and gray diamonds, respectively.  Finally, in the region labeled ``[4,4],'' the VMC finds a DBL state with four equally occupied bands for both the $d_1$ and $d_2$ partons, although this region is likely just a superfluid.}
\label{fig:4legPD}
\end{figure}

Here we present results of our VMC and DMRG study of the model, Eq.~(\ref{eqn:fullmodel}), on the four-leg ladder ($N=L_y=4$) with periodic boundary conditions in both the transverse and longitudinal directions.  Naturally, we first searched for metallic DBL$[n,m]$ phases on the three-leg ladder but were unsuccessful in finding any that were extensible to 2D, most likely due to the non-bipartiteness of the lattice.  We did find some other interesting results, which are discussed in Sec.~\ref{sec:3leg} and Appendix~\ref{app:3legIncomm}.  Here, we focus on $L_x\times4$ system sizes with $L_x=12,18,24$.  We expect qualitatively different DBL$[n,m]$ states in the two naturally defined density regimes: $0<\rho<1/4$ and $1/4<\rho<1/2$.  For small boson densities in the former case, we find that the system readily phase separates as the ring coupling is increased.  For this reason, and that we would like to find metallic DBL$[n,m]$ phases that are extensible to 2D, we considered densities in the latter regime.  While the behavior is expected to be generic for $\rho$ between $1/4$ and $1/2$, for concreteness we will use boson density $\rho=5/12$ for all three system sizes.  The results are summarized in the phase diagram, Fig.~\ref{fig:4legPD}, which we shall discuss first.  The differently colored regions are determined from a VMC study on the $18\times4$ system while the DMRG points are from all three system sizes.  Following that, we shall give an in depth analysis of the DBL$[4,2]$ phase.

\subsection{The four-leg phase diagram at $\rho=5/12$} \label{sec:4legPD}

\begin{figure}[t]
\centerline{\subfigure{\includegraphics[width=\corrScale\columnwidth]{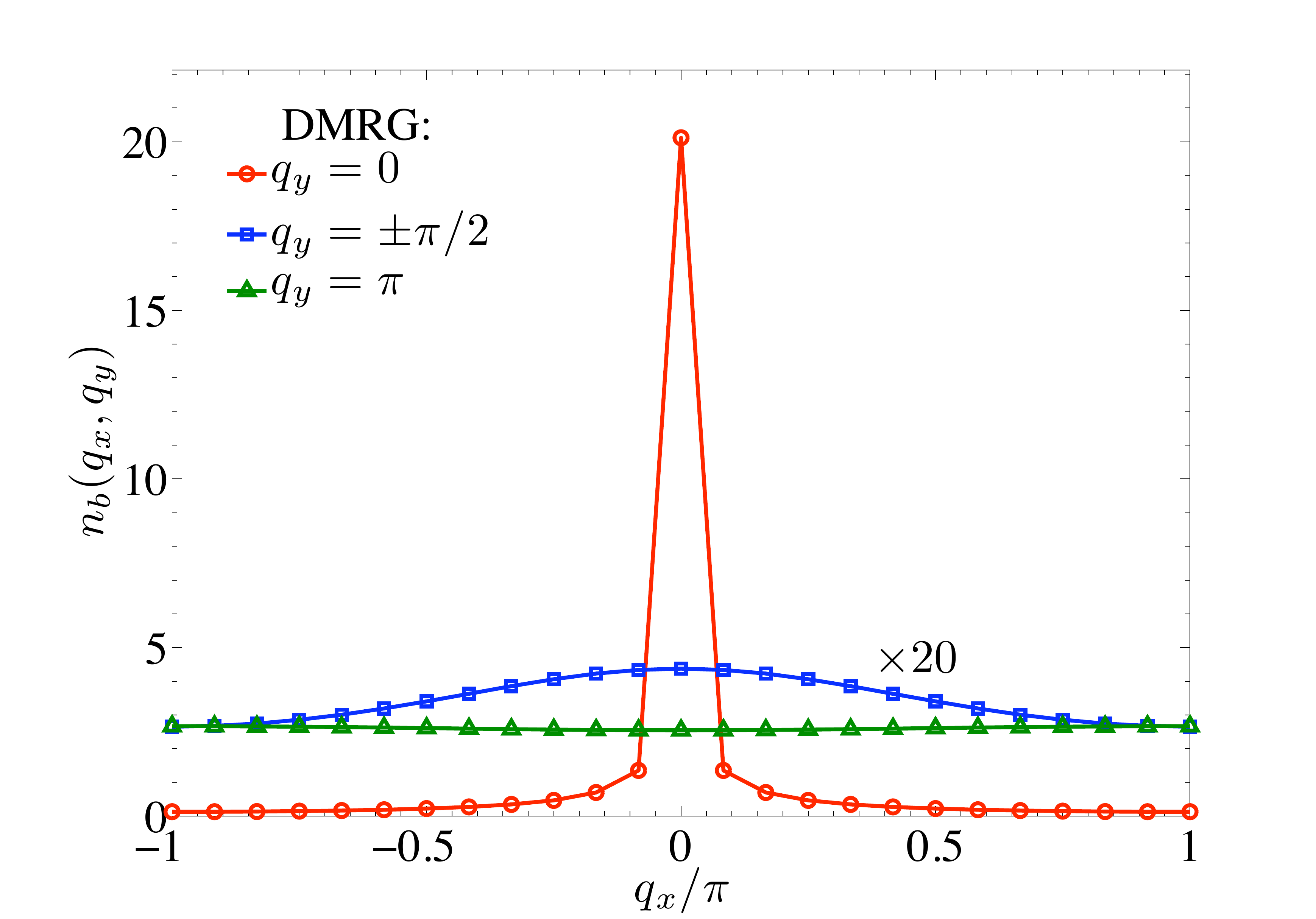}}}
\centerline{\subfigure{\includegraphics[width=\corrScale\columnwidth]{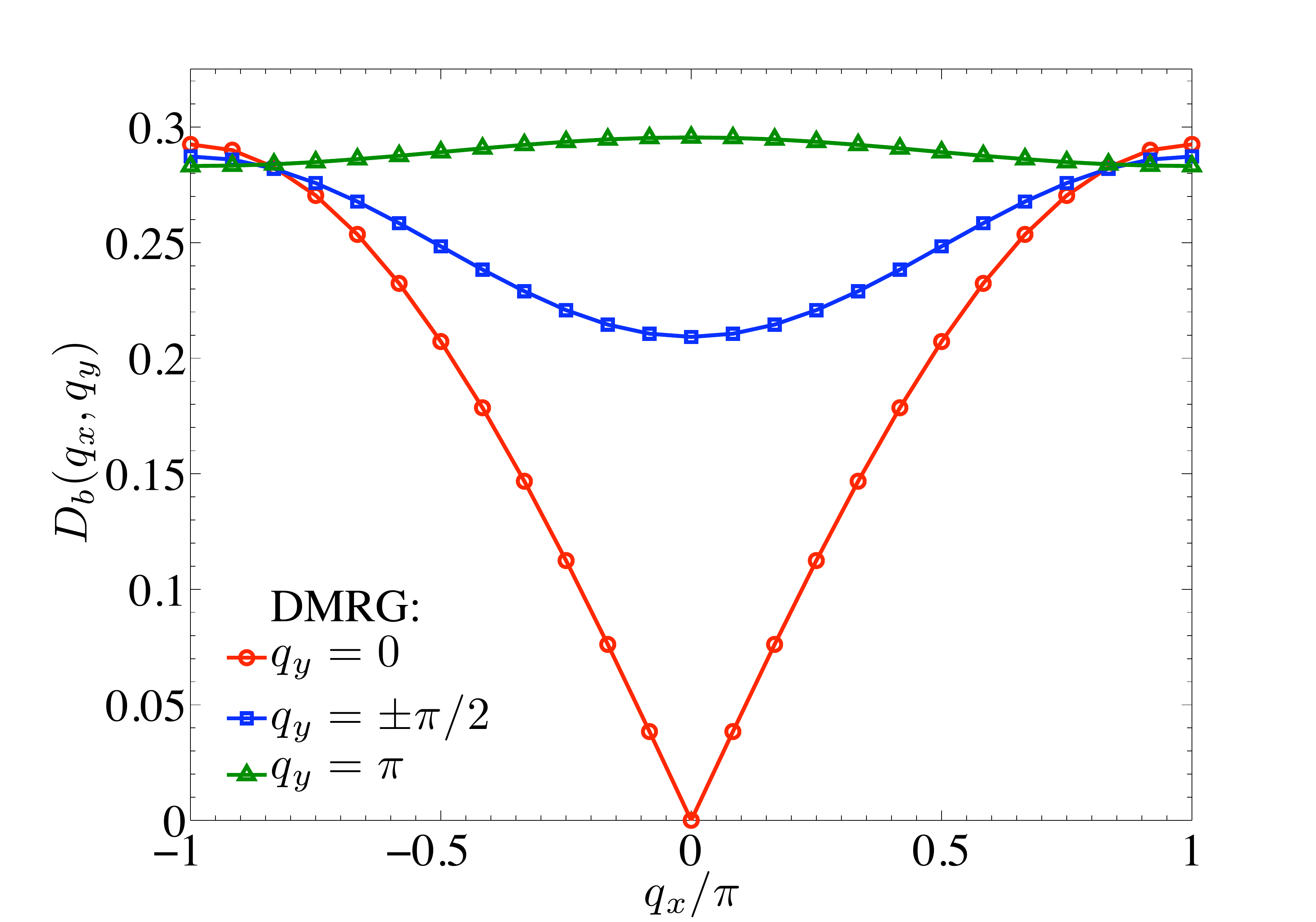}}}
\caption{(color online).  Boson momentum distribution function (top panel) and density-density structure factor (bottom panel) for the superfluid point at $J_{\perp}/J=1$, $K/J=1$ for a $24\times4$ system size with $N_b=40$ ($\rho=5/12$), as obtained by DMRG.  The values of $n_b(\qvec)$ at $q_y=\pm\pi/2,\pi$ have been scaled up by a factor of 20.  The $\qvec=0$ condensation is readily apparent in $n_b(\qvec)$, as is the $|q_x|$ behavior of $D_b(\qvec)$ near $q_x=0$ at $q_y=0$.}
\label{fig:4legSF_DMRG}
\end{figure}

For small $K/J$, the DMRG confirms the existence of a quasi-1D version of the generic superfluid described earlier (points marked with circles in Fig.~\ref{fig:4legPD}) exhibiting quasi-long range order with the bosons ``condensed'' at $\qvec=0$ (see Fig.~\ref{fig:4legSF_DMRG}, top panel) and a central charge of $c\simeq1$ corresponding to the single gapless mode, as determined by measuring the entanglement entropy.  We model this phase in the VMC calculations using a simple Jastrow wave function that simulates the inter-particle repulsion using a constant potential for nearest neighbors and a power law in the separation distance in all other cases.  The Jastrow wave function contains three floating point variational parameters that are optimized by minimizing the trial energy over a finely grained mesh of the phase diagram.  The phase diagram indicates that this modeling is largely successful as it reproduces reasonably well the phase boundary to the DBL$[4,2]$ regime.  The density-density structure factor for a typical point in the superfluid phase is shown in Fig.~\ref{fig:4legSF_DMRG}, bottom panel.  The characteristic $\abs{q_x}$ behavior around $q_x=0$ for $q_y=0$ is clearly present.

Along the cut $J_{\perp}=J$, the DBL$[4,2]$ onsets for roughly $K/J>1.75$ via a first-order phase transition.  This is determined using DMRG (points marked with squares in Fig.~\ref{fig:4legPD}) by looking for the appearance of singular features at incommensurate wave vectors in the boson momentum distribution function and the density-density structure factor (see Figs.~\ref{fig:4legDBL_DMRG_VMCcf_1} and \ref{fig:4legDBL_DMRG_VMCcf_2} for characteristic points).  The precise locations of these features evolve as $K/J$ and $J_{\perp}/J$ are varied within the DBL region of the phase diagram (see Figs.~\ref{fig:4legDBL_DMRG_VMCcf_1} and \ref{fig:4legDBL_DMRG_VMCcf_2}).  This is consistent with the parton picture where larger $K/J$ leads to greater anisotropy in the Fermi seas and hence the number of $d_1$ partons in the $k_y$ band $N_{d_1}^{(k_y)}\rightarrow N_b/4$ for all $k_y$, which drives the evolution of the peak locations.  In the extremal state, $N_{d_1}^{(k_y)}=N_b/4$, which can only be truly realized when $N_b$ is divisible by four, $n_b(\qvec)$ becomes independent of $q_y$ in the variational wave function since there is a conserved number of bosons in each chain (see Fig.~\ref{fig:4legDBL_DMRG_VMCcf_2}).  With VMC, we use the wave functions as described in Sec.~\ref{sec:wf} to model the DBL.  At the onset, we perform an exhaustive search over all possible, unique band filling configurations, the only constraint being that the filled orbitals form a contiguous strip centered at $k_x=0$ within each band.  Note that there are cases where the band occupation numbers are such that no choice of parton longitudinal boundary conditions results in all bands being symmetrically filled.  In such cases, the resulting wave function has a non-zero total momentum in the $\hx$ direction and all possible, unique resolutions of the ``leftover" particles are considered in the search. This first analysis is performed with the ``bare" wave function, i.e., with the exponents on the determinants, $p_1$ and $p_2$, set equal to one.  Using these results, a pool of the most competitive bare VMC states within the parameter regime shown in Fig.~\ref{fig:4legPD} is formed and then a second analysis stage is performed varying these exponents over $0\leq\{p_1,p_2\}\leq1$.  The best DBL state is compared energetically to the best Jastrow state at each point in a finely grained mesh over the phase diagram and this is how the VMC phase boundary is determined.

For still larger values of the ring coupling, roughly $K/J>4.25$, the DMRG data suggest spatial phase separation into regions of zero density and density $1/2$ (points marked with diamonds in Fig.~\ref{fig:4legPD}).  Specifically, our DMRG generally ``gets stuck'' in a nonuniform state with the $(1-2\rho)L_x$ centermost rungs at near zero density and the remaining $2\rho L_x$ outermost rungs at $1/2$ density; there is a corresponding sharp feature in the density-density structure factor, $D_b(\qvec)$, at $\qvec=\pm(2\pi/L_x,0)$.  The actual ground state is presumably an equal superposition of such states with the center of the low-density region at all different $L_x$ rungs of the ladder, to which we could never hope to equilibrate with the DMRG.   This tendency to increase the local density is expected since the ring term induces an effective attraction of the bosons. It is apparently the case that the hopping terms are sufficient to stabilize phases, such as the superfluid and the DBL$[4,2]$, over a large region of the phase diagram for smaller values of $K$.

Throughout our phase diagrams (see Fig.~\ref{fig:4legPD} above and Figs.~\ref{fig:3leg30PD} and \ref{fig:3leg31PD} below), question marks denote a lack of surety in the DMRG data for specifying a phase.  In most cases, however, these points are located near phase boundaries on which we place limited focus in this paper.

\subsection{The DBL$[4,2]$ phase} \label{sec:4leg_42}

\begin{figure}
\centerline{\subfigure{\includegraphics[width=\corrScale\columnwidth]{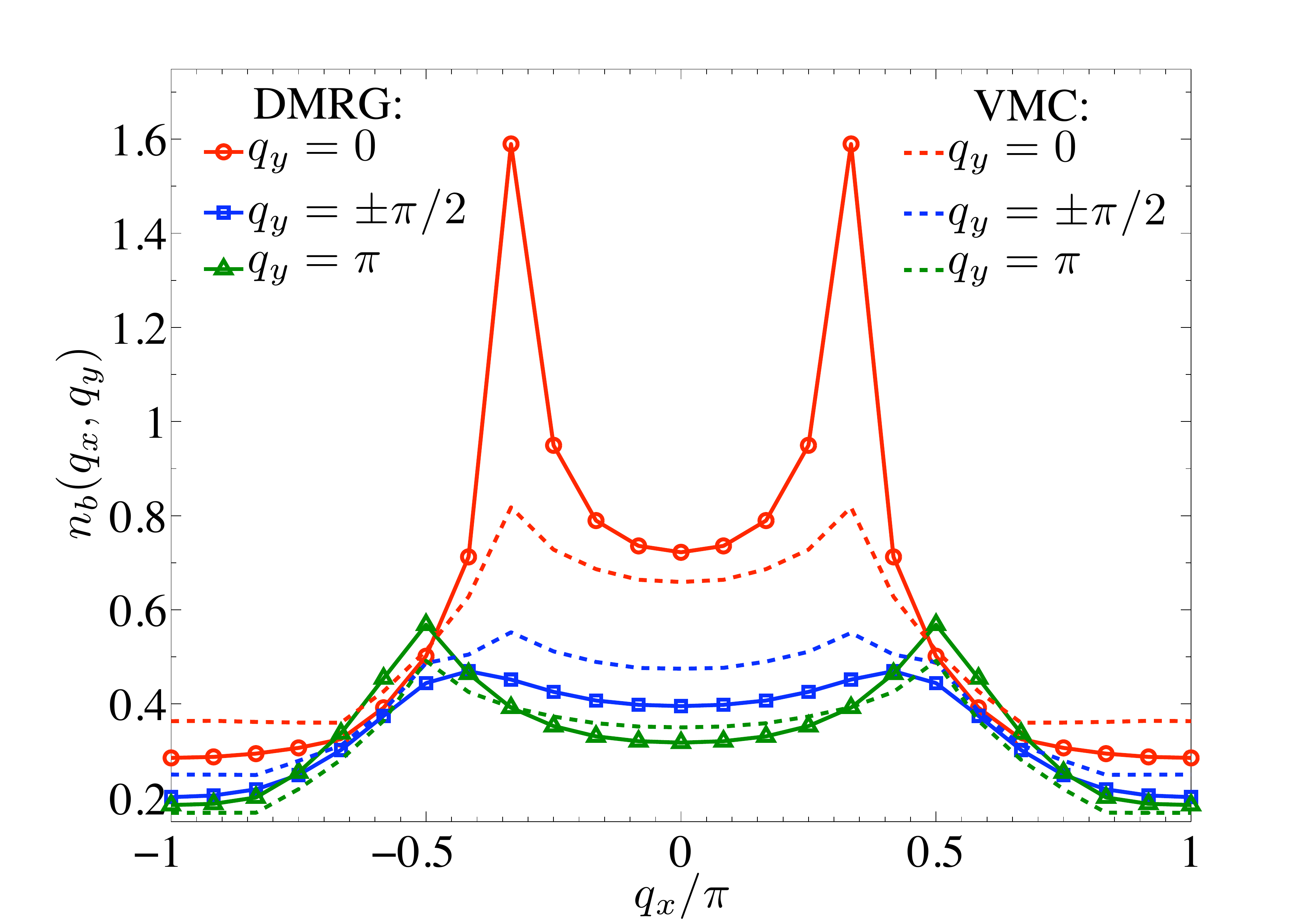}}}
\centerline{\subfigure{\includegraphics[width=\corrScale\columnwidth]{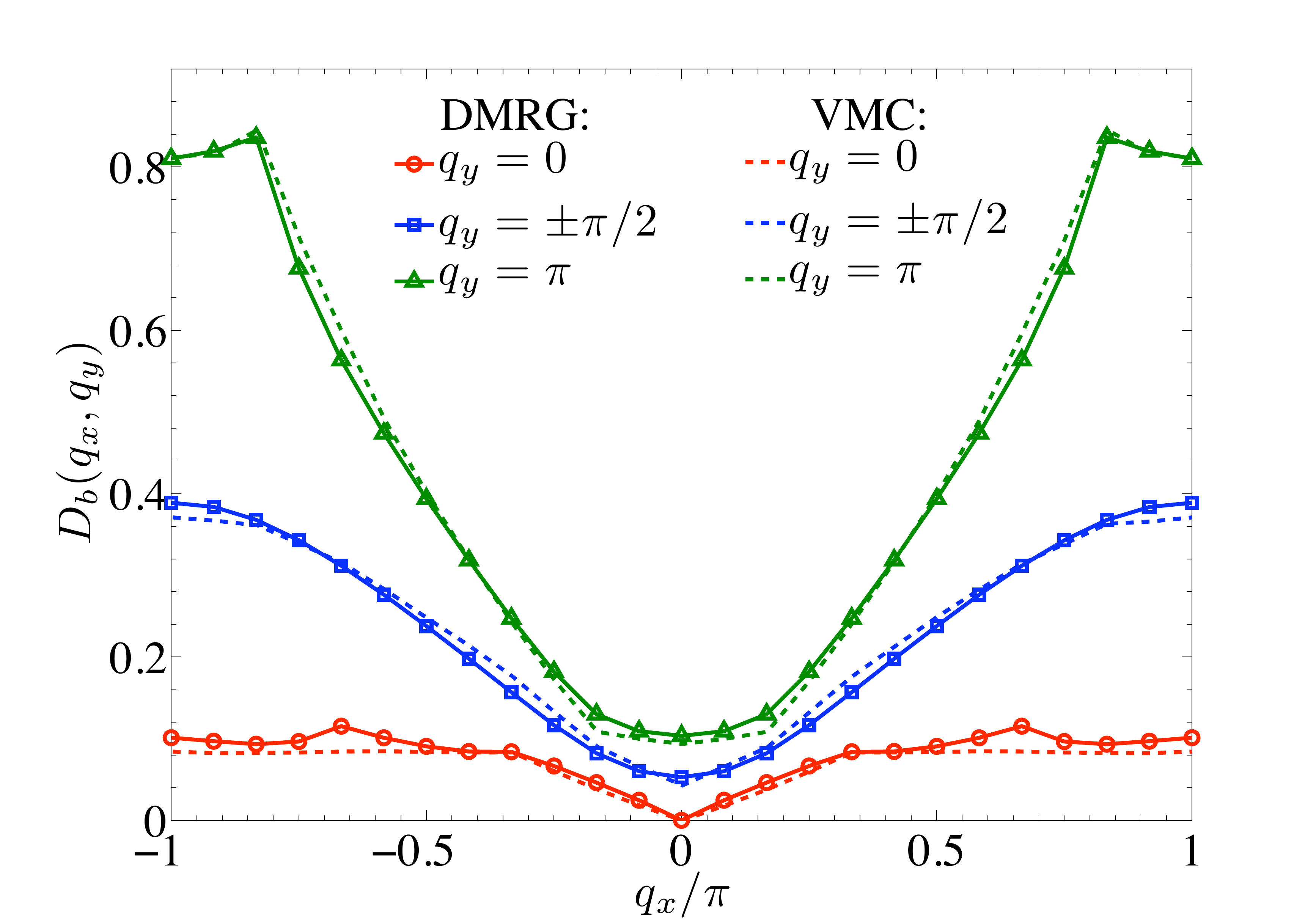}}}
\centerline{\subfigure{\includegraphics[width=\columnwidth]{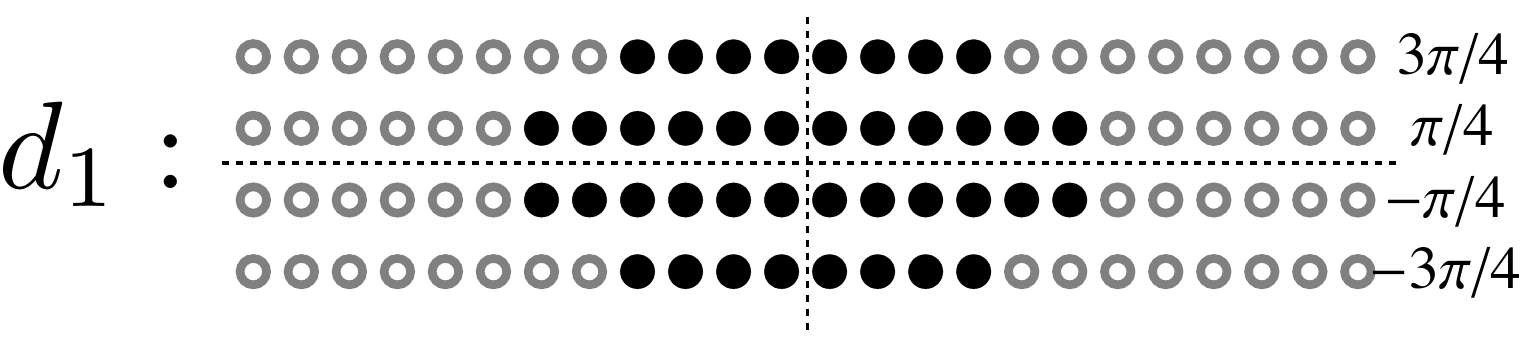}}}
\centerline{\subfigure{\includegraphics[width=\columnwidth]{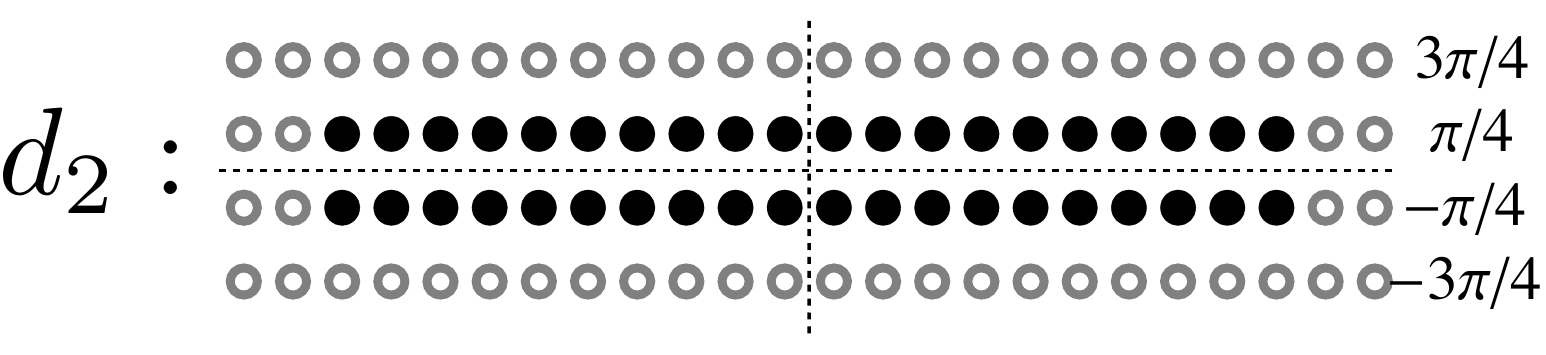}}}
\caption{(color online).  Boson momentum distribution function (top panel) and density-density structure factor (middle panel) for a characteristic DBL$[4,2]$ point: $J_{\perp}/J=1$, $K/J=2.5$.  The system size is $24\times4$ with $N_b=40$ ($\rho=5/12$).  DMRG data are joined with solid curves while the VMC data are joined with dashed curves. (bottom panel) Schematic representation of the energy-optimized VMC state for this point.  The circles represent momentum orbitals in each of the four bands; filled circles denote occupied orbitals.  Each band is centered about $k_x=0$ and we are using antiperiodic boundary conditions in both the $\hx$ and $\hy$ directions for the partons.}
\label{fig:4legDBL_DMRG_VMCcf_1}
\end{figure}

\begin{figure}
\centerline{\subfigure{\includegraphics[width=\corrScale\columnwidth]{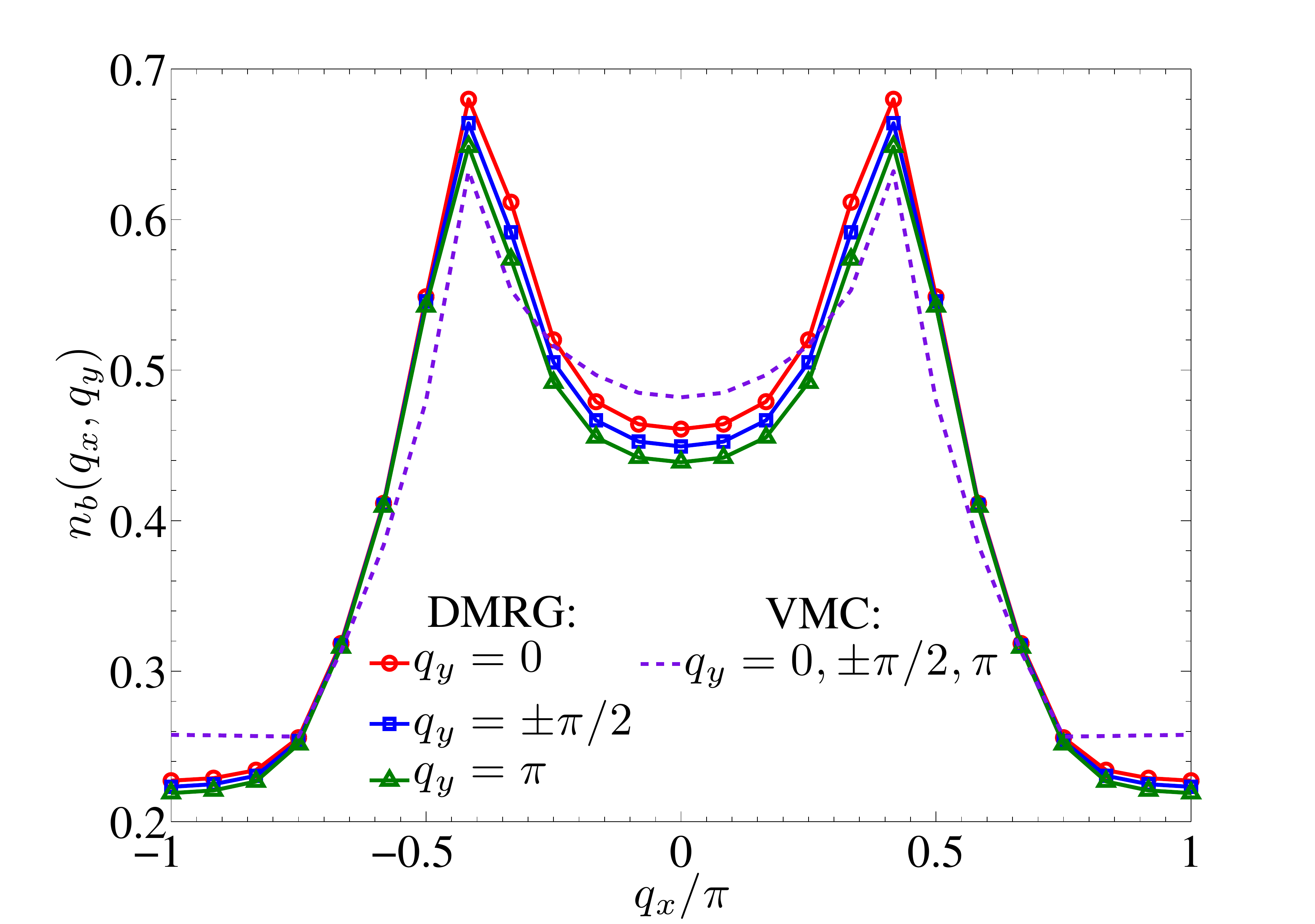}}}
\centerline{\subfigure{\includegraphics[width=\corrScale\columnwidth]{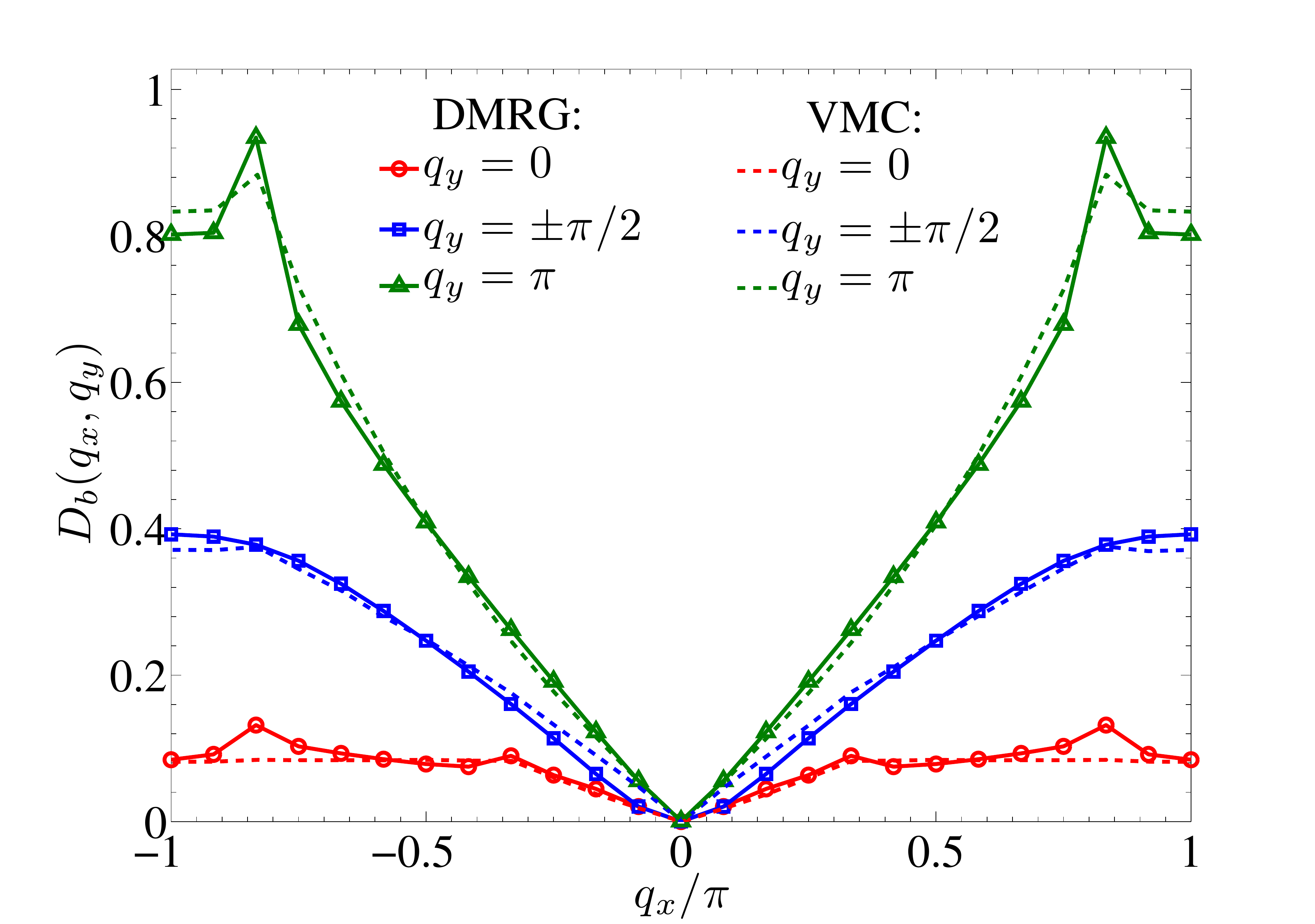}}}
\centerline{\subfigure{\includegraphics[width=\columnwidth]{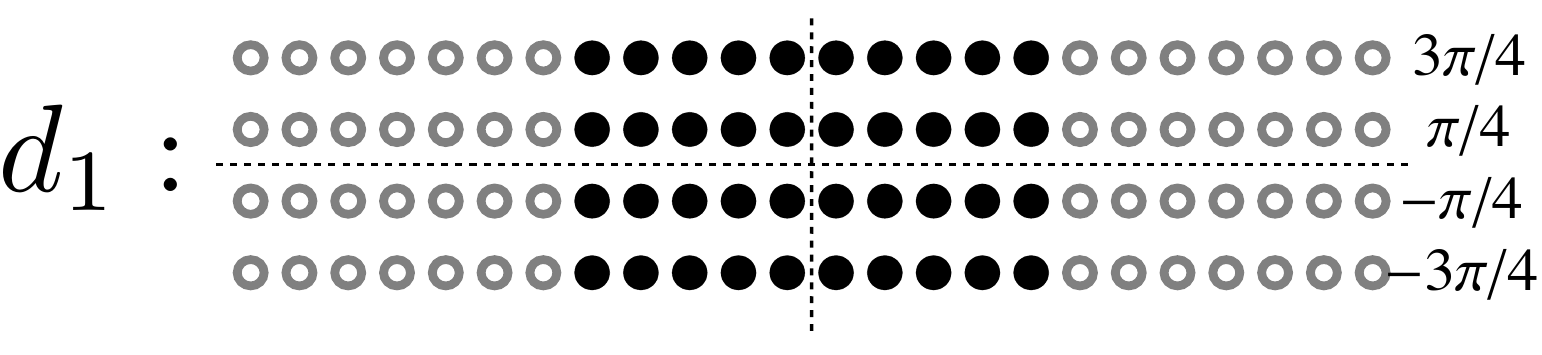}}}
\centerline{\subfigure{\includegraphics[width=\columnwidth]{4leg_42bands_char_d2}}}
\caption{(color online).  The same quantities are plotted as in Fig.~\ref{fig:4legDBL_DMRG_VMCcf_1} for a $24\times4$ system with $N_b=40$ bosons ($\rho=5/12$), except now the measurements are calculated at the point $J_{\perp}/J=0.1$, $K/J=2.5$.  The bottom panel depicts the new energy-optimized VMC state, whose $d_1$ configuration differs from that of Fig.~\ref{fig:4legDBL_DMRG_VMCcf_1}.  Note the feature in the VMC data for $n_b(\qvec)$ at $q_x=\pm3\pi/4$, which is a ``suppressed'' singularity resulting from creating a boson on the same side of the $d_1$ and $d_2$ coincident bands:  $\pm(k_{F1}+k_{F2})=\mp3\pi/4$.  Analogous features are also present in the VMC data of Fig.~\ref{fig:4legDBL_DMRG_VMCcf_1}.  As well as demonstrating the ability to vary the location of the Bose points within the phase diagram, these results in conjunction with those of Fig.~\ref{fig:4legDBL_DMRG_VMCcf_1} also highlight the general nonuniversality of the amplitudes of the power law singularities.}
\label{fig:4legDBL_DMRG_VMCcf_2}
\end{figure}

We now turn our attention to the DBL$[4,2]$ phase itself (see Fig.~\ref{fig:4leg_42bandcurves}), which is a conducting, metal-like phase with one more gapless 1D mode than the number of legs on the ladder.  This phase shows no signs of ordering and breaks no symmetries; in particular, it respects the inversion symmetry of the lattice.  We choose two characteristic points deep within this region of the phase diagram for further analysis: $J_{\perp}/J=1,0.1$ at fixed $K/J=2.5$.  We scale up the most competitive VMC states at these points from the $12\times4$ system by multiplying all band occupation numbers by two in the hopes of accurately capturing the behavior of the phase on a $24\times4$ system with $N_b=40$.  The same points on this larger system size are studied with DMRG and the comparison of these results is shown in Figs.~\ref{fig:4legDBL_DMRG_VMCcf_1} and \ref{fig:4legDBL_DMRG_VMCcf_2}.  We perform this scaling procedure to avoid doing a full VMC energetics study on the $24\times4$ system where the pool of potential trial wave functions is sufficiently large as to render the study intractable.  Naturally, the VMC method is capable of calculating correlators for a single state on system sizes much greater than this, while the DMRG approach is reaching its computational limit.

We now describe in detail the state found at $J_{\perp}/J=1$ and $K/J=2.5$ (see Fig.~\ref{fig:4legDBL_DMRG_VMCcf_1}) as a representative of the DBL$[4,2]$ phase and give an explanation for the locations of the singularities in the structure factors consistent with the gauge theory predictions.  First, we will take antiperiodic boundary conditions for both partons in both the $\hx$ and $\hy$ directions so that the bands will be filled symmetrically; we remind that this is consistent with the physical bosons obeying periodic boundary conditions in both directions.  The band filling situation is displayed visually in Fig.~\ref{fig:4legDBL_DMRG_VMCcf_1}, bottom panel; clearly, this state has zero total momentum.  For the $d_1$ partons, the $k_y=\pm\pi/4$ bands each have 12 orbitals filled, while the $k_y=\pm3\pi/4$ bands each have 8 orbitals filled.  This leads to Fermi wave vectors of $k_{F1}^{(\pm\pi/4)}=12\pi/24=\pi/2$ and $k_{F1}^{(\pm3\pi/4)}=8\pi/24=\pi/3$.  (We point out that there is a technicality here:  in order to be consistent with the continuum limit, it is best to define the Fermi wave vectors as halfway between the last filled orbital and the first unoccupied orbital.)  For the $d_2$ partons, only the $k_y=\pm\pi/4$ bands are occupied and have 20 particles each; hence the relevant Fermi wave vectors are $k_{F2}^{(\pm\pi/4)}=20\pi/24=5\pi/6$.  One can verify explicitly that the sum rule of Eq.~(\ref{eqn:fermisum}) is satisfied for both parton flavors.  The optimal variational exponents on the determinants for the corresponding point in the $12\times4$ system are $p_1=p_2=0.8$, which are close to the bare values of unity; these exponents have been carried over to the scaled up $24\times4$ state.

Looking first to the boson momentum distribution function (Fig.~\ref{fig:4legDBL_DMRG_VMCcf_1}, top panel), we see that two of the predicted enhanced momenta show up:  $\pm(k_{F2}^{(\pm\pi/4)}-k_{F1}^{(\pm\pi/4)},0)=\pm(\pi/3,0)$, $\pm(k_{F2}^{(\pm\pi/4)}-k_{F1}^{(\mp3\pi/4)},\pi)=\pm(\pi/2,\pi)$.  The other two that are predicted to be enhanced are present in the VMC data, although weak, and appear to be smoothed out in the DMRG data:  $\pm(k_{F2}^{(\pm\pi/4)}-k_{F1}^{(\mp\pi/4)},\pm\pi/2)=\pm(\pi/3,\pm\pi/2)$, $\pm(k_{F2}^{(\pm\pi/4)}-k_{F1}^{(\pm3\pi/4)},\mp\pi/2)=\pm(\pi/2,\mp\pi/2)$.  Note that due to the inversion symmetry of the lattice, the $k_y=\pm\pi/2$ bands are always degenerate in all cases.  While the amplitudes of the peaks in the VMC data are not quite right, the agreement of the singular locations is striking.  The fact that the coincident singularities at $q_y=0,\pi$ are quite strong in the DMRG, even more so than in the VMC, is itself encouraging for stability of the DBL phase.

Considering next the density-density structure factor (Fig.~\ref{fig:4legDBL_DMRG_VMCcf_1}, middle panel), the qualitative and quantitative agreement between the VMC and DMRG data is excellent.  In general, we do not expect the peak amplitudes of the VMC results to match those present in the DMRG data; these depend on the details of the gauge theory and the many input parameters about which the trial wave functions have no knowledge.  This is the situation with the boson momentum distribution function data, whereas the $D_b(\qvec)$ curves are actually nearly coincident.  However, the agreement of the locations of the peaks is what matters most in giving credence to the parton description of the phase.  The DMRG results pick up half of the predicted enhanced wave vectors, although some of these features are smoothed out in the VMC results:  $\pm(k_{F1}^{(\pm3\pi/4)}+k_{F1}^{(\mp3\pi/4)},0)=\pm(2\pi/3,0)$, $\pm(k_{F2}^{(\pm\pi/4)}+k_{F2}^{(\mp\pi/4)},0)=\mp(\pi/3,0)$, $\pm(k_{F1}^{(\pm\pi/4)}+k_{F1}^{(\pm\pi/4)},\pm\pi/2)=(\pi,\pm\pi/2)$, $\pm(k_{F1}^{(\pm\pi/4)}+k_{F1}^{(\pm3\pi/4)},\pi)=\pm(5\pi/6,\pi)$.  The first case, $\pm(k_{F1}^{(\pm3\pi/4)}+k_{F1}^{(\mp3\pi/4)},0)=\pm(2\pi/3,0)$, actually shows up clearly in the DMRG but not in the VMC, where it is smoothed by inclusion of the variational exponents [see Eq.~(\ref{eqn:wfexps})].  The VMC alone sees one of the other four wave vectors at which the gauge theory predicts enhanced singularities, but this does not show up in the DMRG data:  $\pm(k_{F1}^{(\pm\pi/4)}+k_{F1}^{(\mp3\pi/4)},\mp\pi/2)=\pm(5\pi/6,\mp\pi/2)$.  The remaining three wave vectors are not seen in either data set:  $\pm(k_{F1}^{(\pm\pi/4)}+k_{F1}^{(\mp\pi/4)},0)=(\pi,0)$, $\pm(k_{F1}^{(\pm3\pi/4)}+k_{F1}^{(\pm3\pi/4)},\mp\pi/2)=\pm(2\pi/3,\mp\pi/2)$, $\pm(k_{F2}^{(\pm\pi/4)}+k_{F2}^{(\pm\pi/4)},\pm\pi/2)=\mp(\pi/3,\pm\pi/2)$.

In Fig.~\ref{fig:4legDBL_DMRG_VMCcf_2}, we show boson momentum distribution and density-density structure factor measurements as obtained by DMRG and VMC at one more point in the phase diagram:  $J_\perp/J=0.1$ and $K/J=2.5$.  Again, we take a $24\times4$ system with $N_b=40$ bosons.  Now, the $d_1$ configuration consists of four equally filled bands, i.e., each band $k_y=\pm\pi/4,\pm3\pi/4$ contains $N_b/4=10$ partons, while the $d_2$ configuration is the same as that in Fig.~\ref{fig:4legDBL_DMRG_VMCcf_1}.  That is, our Fermi wave vectors are now given by $k_{F1}=k_{F1}^{(\pm\pi/4)}=k_{F1}^{(\pm3\pi/4)}=10\pi/24=5\pi/12$ and $k_{F2}=k_{F2}^{(\pm\pi/4)}=20\pi/24=5\pi/6$.  This variational wave function has a conserved number of bosons in each chain and hence $n_b(\qvec)$ is independent of $q_y$ and $D_b(q_x=0,q_y)=0$ for all $q_y$; these properties hold only approximately in the DMRG due to small fluctuations of the single-chain boson number for $J_\perp\neq0$.  As before, the locations of the main singular features are consistent with what we expect from the gauge theory.  Unlike in Fig.~\ref{fig:4legDBL_DMRG_VMCcf_1}, we now see clear features in the DMRG data for $n_b(\qvec)$ at $q_y=\pm\pi/2$, which is to be expected since $n_b(\qvec)$ is approximately independent of $q_y$; however, the features in $D_b(\qvec)$ at $q_y=\pm\pi/2$ are still smoothed in the DMRG data while being present, but very weak, in the VMC data.  Note again the feature in $D_b(\qvec)$ at $\pm(2k_{F1},0)=\pm(5\pi/6,0)$, which is predicted to be enhanced coming from the gauge theory and which actually shows up in the DMRG without being present in the VMC.  Overall, the combined results of Figs.~\ref{fig:4legDBL_DMRG_VMCcf_1} and \ref{fig:4legDBL_DMRG_VMCcf_2} clearly illustrate the tunability of the locations of the Bose points within the DBL$[4,2]$ phase.  In going from Fig.~\ref{fig:4legDBL_DMRG_VMCcf_1} to Fig.~\ref{fig:4legDBL_DMRG_VMCcf_2}, we have chosen to decrease $J_\perp$ instead of increasing $K$ to induce evolution of the Bose points because (upon increasing $K$ at fixed $J_\perp/J=1$) the system phase separates before entering the $d_1$ configuration realized in Fig.~\ref{fig:4legDBL_DMRG_VMCcf_2}.

Considering Figs.~\ref{fig:4legDBL_DMRG_VMCcf_1} and \ref{fig:4legDBL_DMRG_VMCcf_2} as a whole, the fact that not all predicted features are visible is not surprising nor does it detract from our argument in support of the realized phase being a DBL$[4,2]$.  In reality, one must consider all allowed interactions in the full gauge theory, and due to the complexity of this multi-mode Luttinger liquid, it is impossible to know how the short-ranged interactions will affect the anomalous power law exponents.  The gauge theory, insofar as we can interpret it, only predicts potentially enhanced wave vectors, as well as those that are definitely suppressed, i.e., those that do not satisfy the Amperean rule.  Also, it is important to note that the \emph{amplitudes} of the power law singularities are nonuniversal quantities which we could never even hope to predict within our gauge theory framework.  It is entirely possible, perhaps even likely given the overall agreement between the VMC and DMRG, that such matrix element effects could be responsible for the absence of some features in the DMRG momentum space correlators depicted in Figs.~\ref{fig:4legDBL_DMRG_VMCcf_1} and \ref{fig:4legDBL_DMRG_VMCcf_2}; note that even the amplitudes of the corresponding VMC singularities are relatively small in these cases.  It has been suggested that similar effects may be playing a role in smoothening some singular features in the four-leg spin Bose metal phase (so-called SBM-3) of Ref.~\onlinecite{Block11_PRL_106_157202}---note that DBL$[4,2]$ and SBM-3 are closely analogous phases realized in the respective ring models considered here and in Ref.~\onlinecite{Block11_PRL_106_157202}.

We conclude this section by discussing scaling of the entanglement entropy within the DBL$[4,2]$ phase and our efforts to try to extract an effective central charge $c=5$.  Our gauge theory analysis of DBL$[4,2]$, as presented in Appendix~\ref{app:gt42}, predicts $c=4+2-1=5$ 1D gapless modes, and in principle, we should be able to extract this by examining the scaling of the entanglement entropy according to Eq.~(\ref{eqn:Sscaling}).  However, in practice, this is an exceedingly difficult task due to the large amount of spatial entanglement in the ground state as implied by such a phase.  Specifically, to fully converge the von Neumann entanglement entropy the required $D$, which is the number of reduced density matrix basis states retained in the DMRG, becomes prohibitively large given current computational resources.

We have only been able to attain full convergence on small system sizes; however, the results on these small systems are indeed suggestive of $c\simeq5$.  For example, fitting entropy data of an $8\times4$ system with $N_b=12$ bosons at the DBL$[4,2]$ point $J_\perp/J=1$, $K/J=2.5$ gives $c=4.85\pm0.05$, where the error quoted is the error in the fit only (the entropy data itself is converged to about $10^{-4}$ keeping $D=8000$ states).  Unfortunately, we have been unable to fully converge the entropy data for $L_x>8$ systems and have thus not been able to get reliable estimates of $c$ for reasonably long systems.  For $12\times 4$ systems, the entropy [say at the middle of the system, $S(X=L_x/2,L_x)$] is still generally slowly increasing even when keeping up to $D=9300$ states, although fits to the still unconverged data give $c\gtrsim4$.  The very slow increase in $S$ with $D$ is to be expected since $S$ is known to grow only logarithmically with $D$ for 1D critical systems. \cite{Pollman09_PRL_102_255701}  At some points in the phase diagram on the $12\times4$ system, the convergence of $S$ is better due to a smaller value of the constant $A$ in Eq.~(\ref{eqn:Sscaling}).  In these cases, $c$ is even closer to 5; e.g., at the point $J_\perp/J=1.5$, $K/J=2.5$ with $N_b=20$ bosons on a $12\times4$ system, we get $c>4.6$ with the entropy still very slowly increasing.  We caution that in our experience with other realized DBL phases, e.g., DBL$[2,1]$ on two legs (see Ref.~\onlinecite{Sheng08_PRB_78_054520}), DBL$[3,1]$ on three legs (see Appendix~\ref{app:3leg_31}), and DBL$[3,0]$ on three legs (see Ref.~\onlinecite{Block11_PRL_106_046402} and Sec.~\ref{sec:3leg}), small systems tend to overestimate $c$ from its expected value, usually on the order of about $10\%$.  However, DBL$[4,2]$ is a qualitatively distinct phase from those mentioned above, so such trends may no longer apply here.  Finally, we have tried using open boundary conditions (OBC) in the $\hx$ direction (cylindrical boundary conditions overall) in which case it is much less costly to obtain converged entropy data on larger systems; e.g., we were able to get the entropy converged to $1\%$ on systems up to $48\times4$ by keeping $D=6000$ states.  However, this data is plagued by very strong oscillations originating from the finite wave vectors clearly present in the DBL$[4,2]$ phase, and these oscillations ultimately make an accurate determination of $c$ impossible in this case.  All in all, even though we have been unable to definitively confirm $c=5$ gapless modes with the DMRG measurements, our calculations in no way rule out the possibility of this result.

\subsection{Discussion} \label{sec:4legdisc}

It is clear that the agreement of the locations of singularities in the structure factors is excellent, although there remain quantitative differences.  It is important to bear in mind that the Gutzwiller wave functions provide only a caricature of the DBL$[4,2]$ phase (e.g., they should not be expected to accurately capture long-distance properties of the underlying gauge theory \cite{Tay11_PRB_83_235122}) and it is not the goal of this study to find the exact ground state of our ring model using a variational wave function; after all, we already have the ground state using DMRG.  We again emphasize that the lack of some of the expected features in the DMRG measurements is not at all surprising.  The full DBL$[4,2]$ theory, with both gauge and short-range interactions, is an extremely complicated multi-mode Luttinger liquid with five potentially nontrivial Luttinger parameters.  Thus, the manner in which the power law exponents is altered by the allowed short-range interactions is difficult to assess, as is the ultimate effect of nonuniversal power law amplitudes.  Overall, we find the data presented here to be compelling evidence for the validity of our parton description of the remarkable phase found in the four-leg frustrated $J$-$K$ model and its correct identification as a quasi-1D descendent of the $d$-wave Bose liquid.  Furthermore, in Appendix~\ref{app:gt42}, we consider the effects of short-range interactions on the stability of the DBL$[4,2]$ within the scope of the bosonized gauge theory and conclude, from an analytical perspective, that DBL$[4,2]$ is likely a stable quantum phase.

Further numerical confirmation could have in principle been obtained by examining the entanglement entropy in the DMRG and extracting the expected central charge, $c=5=4+2-1$.  However, this study turned out to be inconclusive, mainly due to an inability to converge the entropy on large periodic systems, but also because open (cylindrical) systems are plagued by oscillations originating from the finite Bose wave vectors present in our DBL theory.  Going forward, we believe that there is much room for improvement, both analytically and numerically, using entanglement scaling properties to characterize these systems.  On the numerical front, it would be particularly interesting to use recently developed Monte Carlo techniques \cite{Hastings10_PRL_104_157201, Grover11_PRL_107_067202} to compute the Renyi entropy (as opposed to the von Neumann entropy) directly in our projected variational wave functions; aside from being interesting in its own right, this would also give us a guide as to how much entanglement to expect in the DMRG calculations.  For the Renyi entropy, we may expect subleading oscillatory corrections to scaling \cite{Calabrese10_PRL_104_095701} for the DBL phases, again due to the presence of finite Bose wave vectors, and this could make the data more difficult to analyze.  Finally, it may prove easier to access long-distance properties, e.g., the central charge, of nontrivial quasi-1D gapless phases such as our DBL$[4,2]$ and the proposed SBM-3 phase of Ref.~\onlinecite{Block11_PRL_106_157202} by using recently developed ``entanglement renormalization'' techniques, \cite{Vidal07_PRL_99_220405, Vidal08_PRL_101_110501} an endeavor we leave for future work.

We now compare and contrast the nature of the DBL$[4,2]$ to that of the DBL$[2,1]$ on the two-leg ladder characterized in Ref.~\onlinecite{Sheng08_PRB_78_054520}.  The former can be seen as simply a scaled up version of the latter; the DBL$[4,2]$ has the same band picture as the DBL$[2,1]$ but with pairs of coincident bands replacing the single bands (compare Fig.~\ref{fig:4leg_42bandcurves} of this paper to Fig.~2 of Ref.~\onlinecite{Sheng08_PRB_78_054520}).  In this sense, the DBL$[4,2]$ is precisely the state we would expect in moving toward 2D.  The most significant achievement of the work presented in this paper over what has been done previously is the realization of a DBL$[n,m]$ phase where $n+m-1>N$, that is, where the number of gapless modes exceeds the number of legs on the ladder.  This situation has an interesting consequence:  since $n\leq N$, $m>1$, necessarily, and therefore the determinant for the $d_2$ partons cannot be interpreted as a simple Jordan-Wigner string multiplying the $d_1$ determinant.  For $m=1$, the $d_2$ determinant enforces a condition of no more than one particle per rung; in this case, the physics of this determinant is essentially 1D, allowing for the Jordan-Wigner interpretation.  But when $m>1$, the $d_2$ determinant has a more subtle effect on the sign structure of the boson wave function.  In Ref.~\onlinecite{Sheng08_PRB_78_054520}, the authors searched for a stable DBL$[2,2]$ on the two-leg ladder, which also would have satisfied this condition.  The bosonized gauge theory suggests a strong instability toward an $s$-wave paired phase, and the DMRG confirmed that this was indeed the case.  Thus, no DBL$[2,2]$ was ever stabilized.  Finally, we point out that the prevalence of phase separation in the phase diagram (see Fig.~\ref{fig:4legSF_DMRG}) does not seem to have grown in going from $N=2$ to 4 legs, which is encouraging for the eventual stability of the DBL in 2D.

Prior to studying $N=4$, we thoroughly explored the three-leg ladder in the hopes of finding a stable DBL$[3,2]$ phase, which would have also satisfied $n+m-1>N$, for densities greater than the commensurate value $\rho>1/3$.  With our chosen model, Eq.~(\ref{eqn:fullmodel}), we were not successful and instead found what appears to be a three-leg descendant of an unusual superfluid predicted in a recent spin-wave treatment of the $J$-$K$ model on the 2D square lattice. \cite{Schaffer09_PRB_80_014503}  Although this phase can be qualitatively understood with an exceedingly simple classical analysis of Eq.~(\ref{eqn:fullmodel}), it is itself rather exotic as it is a gapless phase that breaks both time reversal and translational symmetry (see Appendix~\ref{app:3leg_noBMs} for details).  We did, however, find a stable DBL$[3,1]$ at density $\rho=1/4<1/3$, which is a natural extension of the DBL$[2,1]$ on two legs and which we describe in greater detail in Appendix~\ref{app:3leg_31}.  The DBL$[3,1]$ has exactly as many gapless modes as the number of legs, same as the DBL$[2,1]$.  The study of the three-leg ladder also led to our discovery of a novel, truly quasi-1D, insulating phase with two gapless modes and a $d$-wave sign structure in the wave function; that is, a DBL$[3,0]$, which exists only for the special, commensurate case $\rho=1/3$.  We now discuss this phase further in the following section.

\section{Gapless Bose insulator phase on the three-leg ladder} \label{sec:3leg}

\begin{figure}[b]
\centerline{\includegraphics[width=\columnwidth]{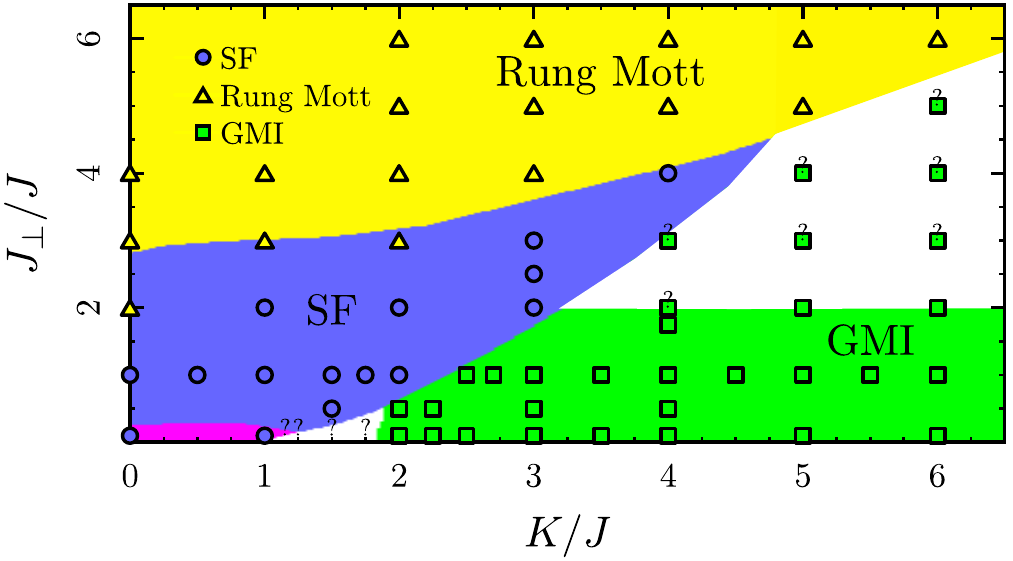}}
\caption{(color online).  Phase diagram of the three-leg system at boson density $\rho=1/3$, using a system of size $24\times3$ with $N_b=24$ bosons.  The colored regions are delineated using VMC data; the white regions indicate where our understanding of the phase diagram is limited.  DMRG points for the superfluid, rung Mott insulator, and DBL$[3,0]$ are indicated by blue circles, green squares, and yellow triangles, respectively.  This figure has been reproduced from Ref.~\onlinecite{Block11_PRL_106_046402}.}
\label{fig:3leg30PD}
\end{figure}

In this section, we present DMRG, VMC, and ED results for our study of the same model [see Eq.~(\ref{eqn:fullmodel})] on the three-leg ladder ($N=L_y=3$) with periodic boundary conditions in both the longitudinal and transverse directions.  Primarily, we focus on the DBL$[3,0]$ phase, which is stable at boson density $\rho=1/3$.  This phase was characterized and discussed in detail in our recent Letter, Ref.~\onlinecite{Block11_PRL_106_046402}, in which it was referred to as a gapless Mott insulator (GMI).  For completeness, we now summarize those results as well as present additional and compelling data and arguments in support of this exotic state as a stable, quasi-1D phase realized by our ring model.  First, we will tour the phase diagram for a model finite-size system, and then we will focus the discussion on the DBL$[3,0]$.  Please note that Figs.~\ref{fig:3leg30PD} and \ref{fig:3leg30_DMRG_VMCcf} were presented in our previous work, \cite{Block11_PRL_106_046402} but are reproduced here to make the present paper self-contained.

\subsection{The three-leg phase diagram at $\rho=1/3$} \label{sec:3legPD}

The three-leg phase diagram at 1/3 filling is presented in Fig.~\ref{fig:3leg30PD}, where we show VMC and DMRG results for a $24\times3$ system size with $N_b=24$.  There are three major phases plus some other regions where our understanding is currently limited.  First, a large portion of the diagram, for small to moderate values of $K/J$ and $J_{\perp}/J$, is the generic superfluid phase, which has the same properties as in the four-leg case above.  The DMRG points that we have identified as SF are marked with circles.  Again, the DMRG confirms a central charge of $c\simeq1$ consistent with the predicted single gapless mode.  This phase is modeled in the VMC using the same type of Jastrow wave functions as in the four-leg case.

\begin{figure}[b]
\centerline{\includegraphics[width=\corrScale\columnwidth]{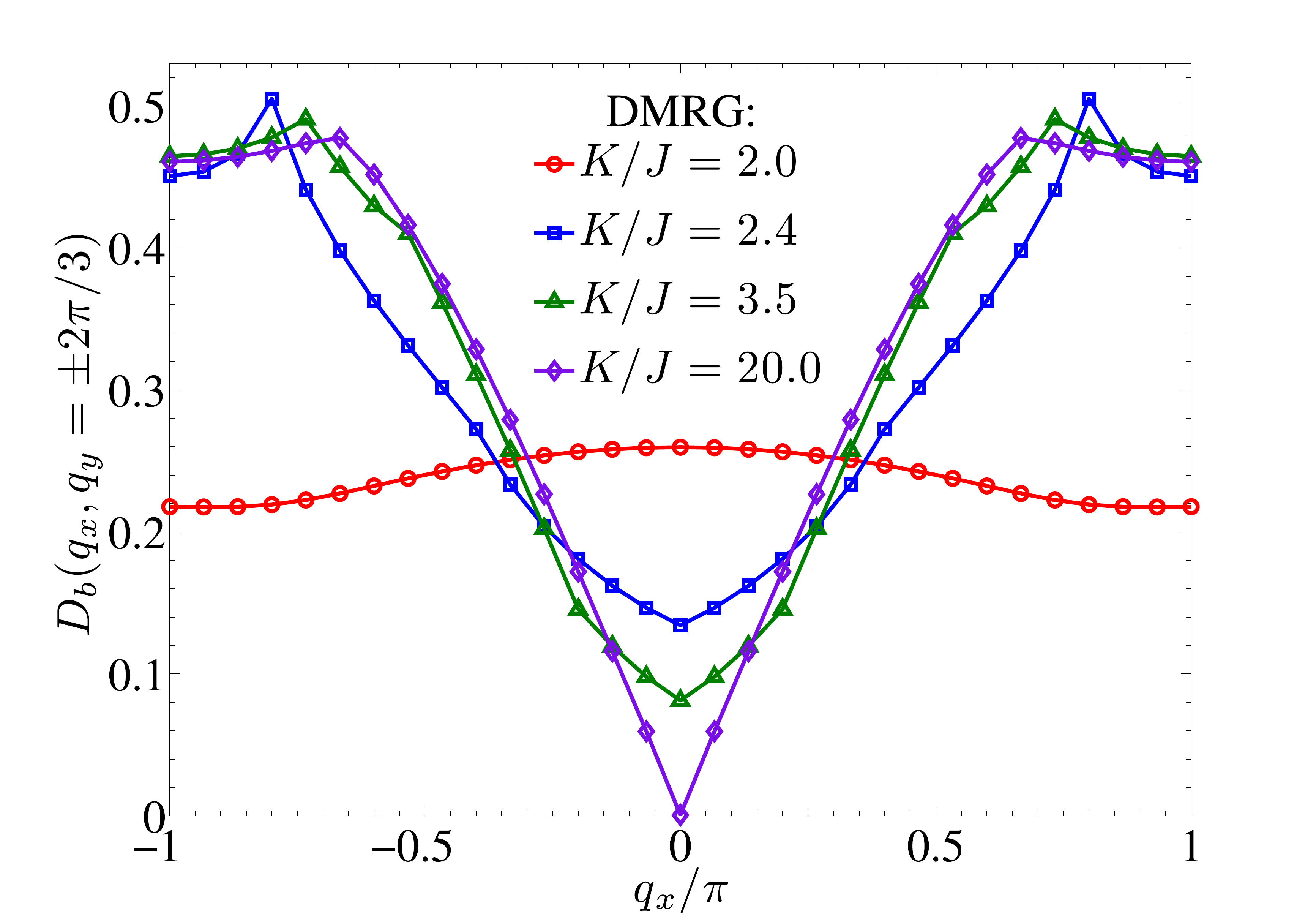}}
\caption{(color online).  Density-density structure factor evaluated at $q_y=\pm 2\pi/3$, i.e., $D_b(q_x, q_y=\pm 2\pi/3)$, for various $K/J$ and fixed $J_\perp/J=1$ on a $30\times 3$ system with $N_b=30$ ($\rho=1/3$).  The qualitative change in this function is striking as one crosses the first-order phase transition from the superfluid to the DBL$[3,0]$ at around $K/J\approx 2.2$.  In particular, $D_b(q_x, q_y=\pm 2\pi/3)$ is a featureless function in the superfluid phase, but has singular structure in the DBL$[3,0]$ at wave vectors originating from the ``$2k_F$'' wave vectors within the $d_1$ parton band filling configuration.  Furthermore, the evolution of the singular features within the DBL$[3,0]$ phase is fully consistent with the three $d_1$ parton bands becoming more equally occupied as $K$ is increased.  For very large $K$, e.g., $K/J=20$, the system is in a phase with an approximately conserved number of bosons in each rung and chain, which is well represented by three equally filled $d_1$ bands.  These calculations were done with DMRG.}
\label{fig:3leg30peakevolution}
\end{figure}

For larger values of $K/J$ with $J_{\perp}/J<2$, we see evidence for the gapless Bose insulating phase: the DBL$[3,0]$ (marked with squares in Fig.~\ref{fig:3leg30PD}; see Figs.~\ref{fig:3leg30peakevolution} and \ref{fig:3leg30_DMRG_VMCcf} for characteristic points).  This phase is a particularly interesting incarnation of the DBL in that one parton ($d_1$) is gapless while the other parton ($d_2$) is gapped, the latter being due to a fully filled $k_y=0$ band at $\rho=1/3$.  Due to the presence of this fully filled $d_2$ band and the absence of any partially filled bands (a situation that can only occur at commensurate densities; see Sec.~\ref{sec:3leg_30}, in particular the bottom panel of Fig.~\ref{fig:3leg30_DMRG_VMCcf}, for more details), there exists a gap to adding a boson and hence we expect the real-space single-boson Green's function to decay exponentially, a prediction that is confirmed by the featureless boson momentum distribution function (top panel of Fig.~\ref{fig:3leg30_DMRG_VMCcf}).  The density-density structure factor (middle panel of Fig.~\ref{fig:3leg30_DMRG_VMCcf}), however, shows singular peaks at incommensurate wave vectors; these are, in fact, the signatures for which we look in the DMRG measurements to detect the onset of the DBL$[3,0]$.  As with the DBL$[4,2]$, the specific locations of these features evolve as we vary $K/J$ and $J_{\perp}/J$ within the DBL$[3,0]$ region.  As usual, within our theory, this evolution is a result of increasing ring exchange inducing an increase in hopping anisotropy between the $d_1$ and $d_2$ partons (in DBL$[3,0]$, the $d_2$ configuration is a fixed fully filled $k_y=0$ band, so only the $d_1$ configuration varies), which in turn results in different ``$2k_F$'' wave vectors as detected by a measurement of $D_b(\qvec)$.  We demonstrate this phenomenon clearly in Fig.~\ref{fig:3leg30peakevolution}, where DMRG data for $D_b(q_x, q_y=\pm 2\pi/3)$ is plotted for various values of ring-exchange coupling.

Another test considers the entanglement entropy.  For the DBL$[3,0]$, we expect only two gapless modes and hence it is computationally feasible to extract the central charge from the scaling form of the entanglement entropy for comparison.  Figure~4 of Ref.~\onlinecite{Block11_PRL_106_046402} shows the fit value for the central charge, $c$, for a range of values of $K/J$ for fixed $J_{\perp}/J=1$ on a $24\times 3$ system.  The jump from $c\simeq1$ to $c\simeq2$ at roughly the same value of $K/J$ at which the DBL phase onsets in the phase diagram is striking.  The inset of Fig.~4 in Ref.~\onlinecite{Block11_PRL_106_046402} shows the actual entanglement entropy and fitted curves for both the superfluid and DBL phases as a function of subsystem size $X$.  In this paper, we discuss finite-size effects on entanglement entropy scaling below in Sec.~\ref{sec:3leg_30} (see Fig.~\ref{fig:3legEEscaling}).

For larger values of $J_{\perp}/J$, the system goes into a conventional rung Mott insulator phase, so called due to the decoupling of the rungs; in the caricature of this phase, each rung contains exactly one boson which is in the zero $y$-momentum state.  These points are marked with triangles in Fig.~\ref{fig:3leg30PD}.  Our determinantal wave function actually does an excellent job modeling this phase by filling only a single band for each parton.  This phase has a gap to all excitations.  We note that very recent work on the two-leg XY model [Eq.~(\ref{eqn:fullmodel}) with $K=0$] has surprisingly revealed that at half-filling the superfluid phase is weakly unstable to such a rung Mott insulator for any finite $J_\perp$. \cite{Carrasquilla11_PRB_83_245101, Crepin11_PRB_84_054517}

There are two regions, one large above the DBL$[3,0]$ and one small at intermediate $K/J$ and very small $J_{\perp}/J$, filled with white in the phase diagram.  The larger region has DMRG points marked as squares with question marks.  The VMC data suggest that the DBL$[3,0]$ wins energetically in this region, but the DMRG data show that the identifying features fail to persist; i.e., the singular peaks in the density-density structure factor smoothen and the central charge falls below $c=2$.  An analysis of the spectral gap at various values of $L_x$ using ED and DMRG reveals that the system is either gapless or has a very small gap in this parameter regime, as we demonstrate explicitly in the Supplemental Material of Ref.~\onlinecite{Block11_PRL_106_046402}.  Also, we have not been able to identify any obvious ordering in the ground state within this region.

Finally, there is no phase separation for this system.  The particle density of $\rho=1/3$ is highly stable due to the fact that it is commensurate with the number of legs.

\subsection{The DBL$[3,0]$ phase} \label{sec:3leg_30}

\begin{figure}
\centerline{\subfigure{\includegraphics[width=\corrScale\columnwidth]{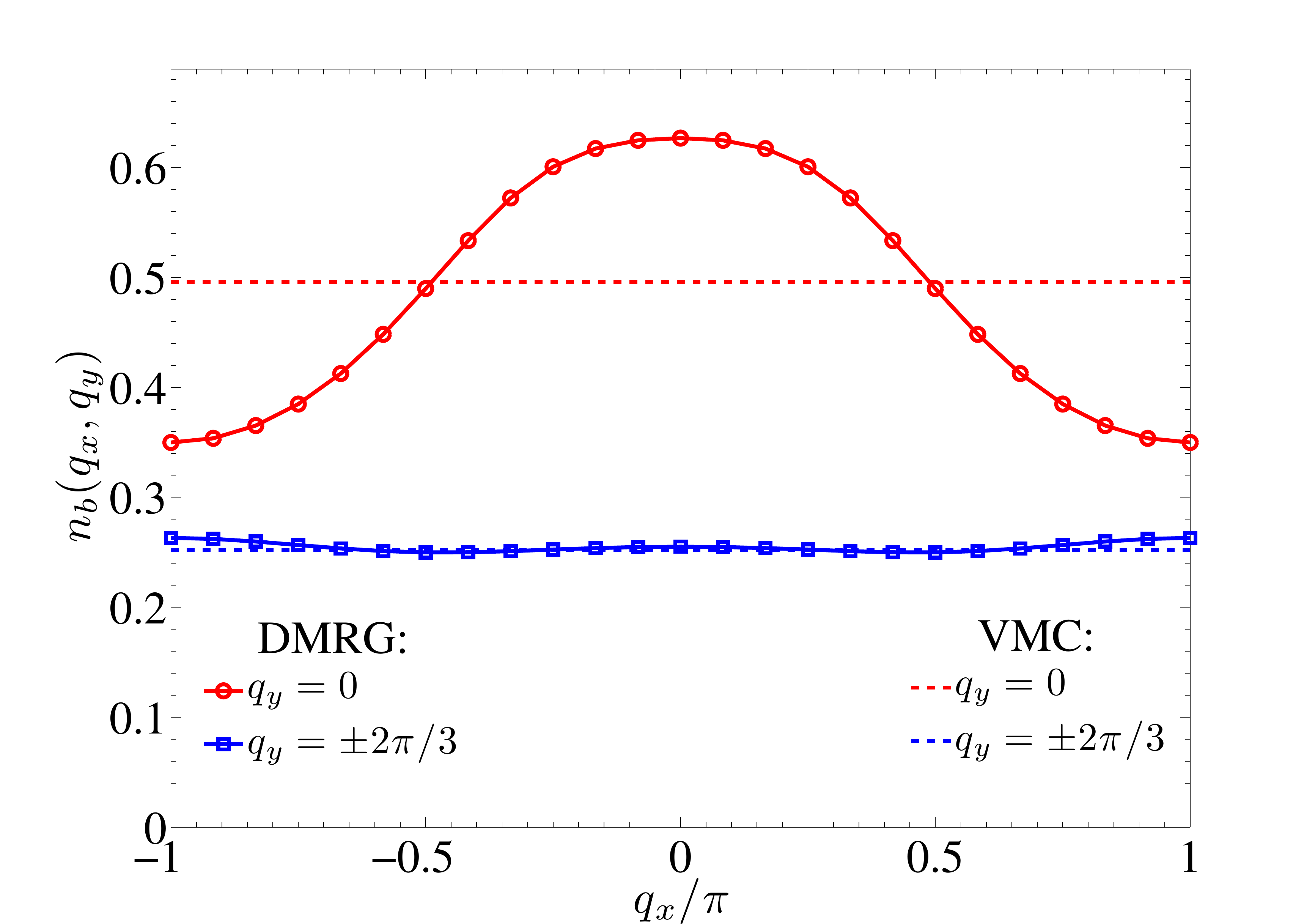}}}
\centerline{\subfigure{\includegraphics[width=\corrScale\columnwidth]{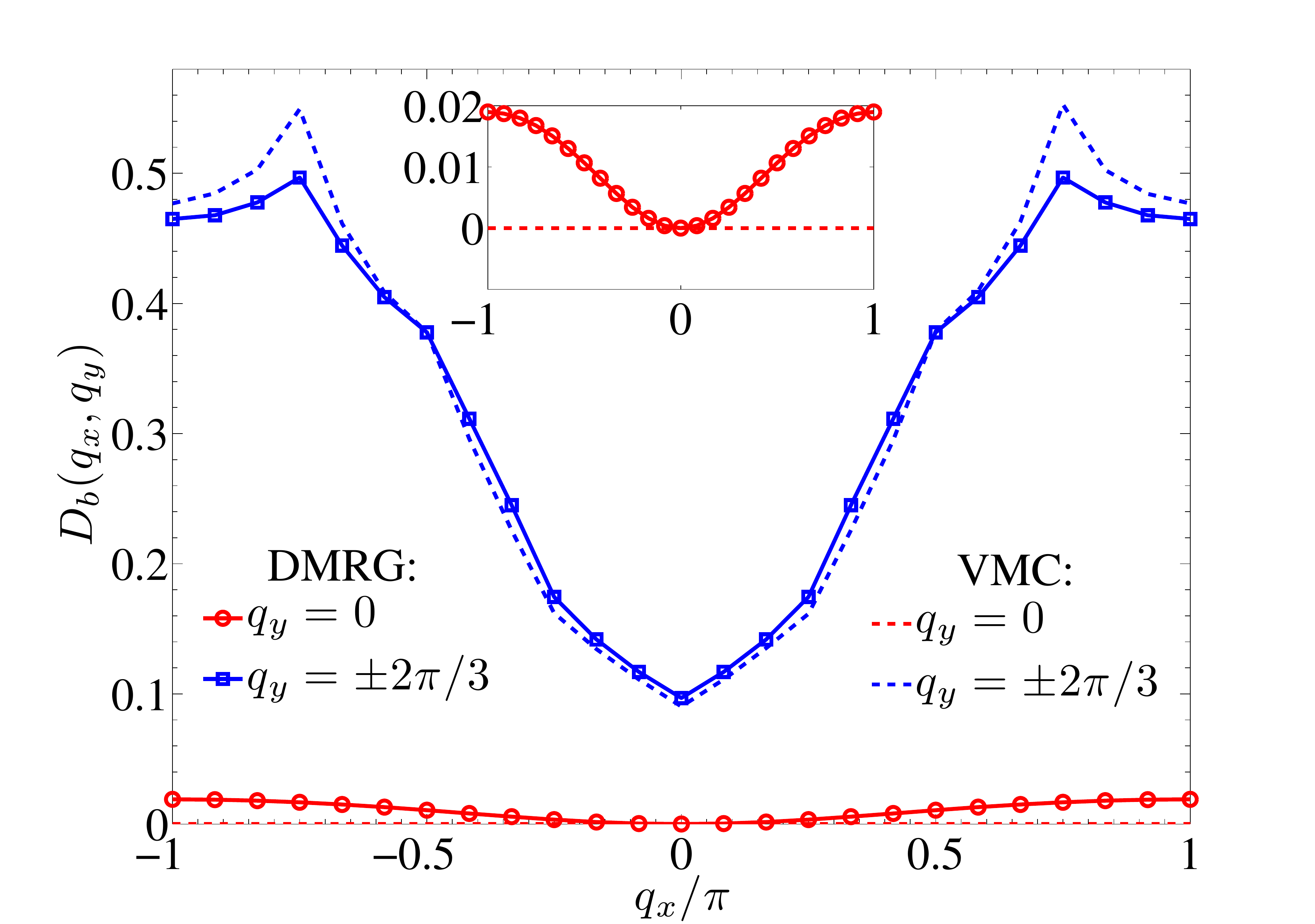}}}
\centerline{\subfigure{\includegraphics[width=\columnwidth]{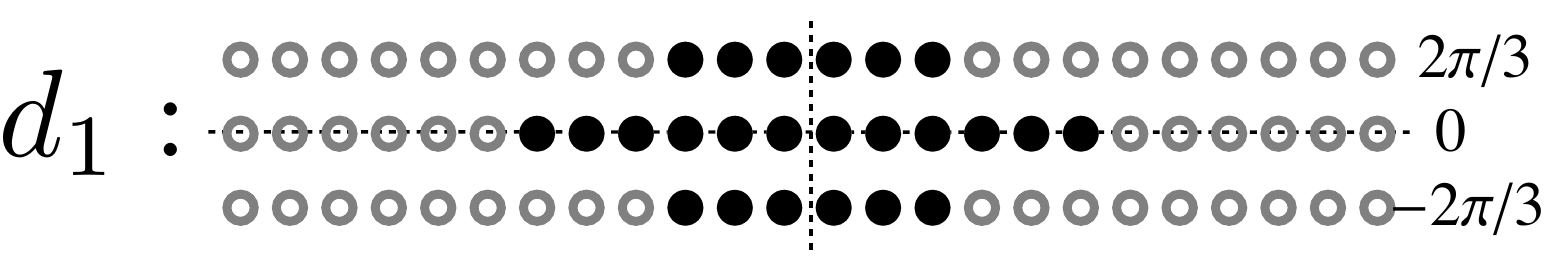}}}
\centerline{\subfigure{\includegraphics[width=\columnwidth]{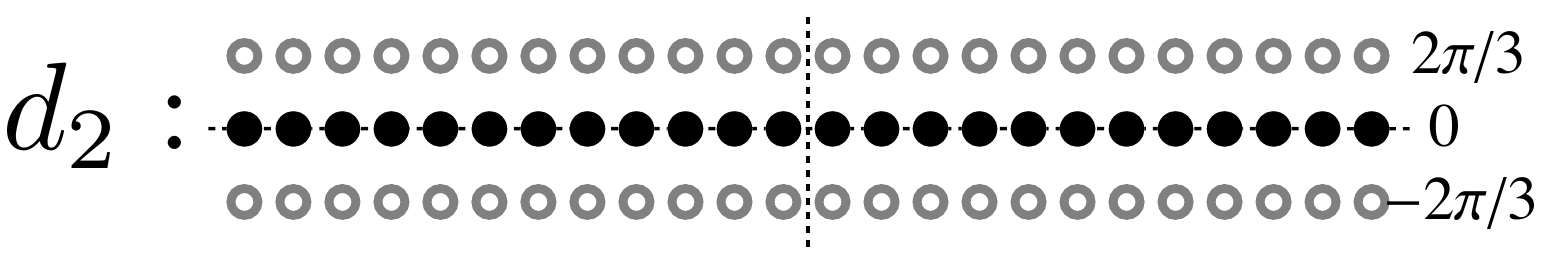}}}
\caption{(color online).  Boson momentum distribution function (top panel) and density-density structure factor (middle panel) for a characteristic DBL$[3,0]$ point: $J_{\perp}/J=1$, $K/J=2.7$.  The system size is $24\times3$ with $N_b=24$ ($\rho=1/3$).  DMRG data are joined with solid curves while the VMC data are joined with dashed curves. (bottom panel) Schematic representation of the energy-optimized VMC state for this point, shown in an analogous way to the DBL$[4,2]$ state in the bottom panels of Figs.~\ref{fig:4legDBL_DMRG_VMCcf_1} and \ref{fig:4legDBL_DMRG_VMCcf_2}.  Here, the $d_1$ configuration resembles a traditional gapless parton Fermi sea, while the $d_2$ configuration is gapped due to the commensurate $\rho=1/3$ density.  This figure has been reproduced from Ref.~\onlinecite{Block11_PRL_106_046402}.}
\label{fig:3leg30_DMRG_VMCcf}
\end{figure}

We focus now in greater detail on the DBL$[3,0]$ phase, which is an incompressible, gapless Mott insulator (GMI) that shows strong density-density correlations at finite transverse momenta and at incommensurate longitudinal wave vectors.  We choose a characteristic point, $K/J=2.7$ and $J_{\perp}/J=1$, within the GMI region of the phase diagram for careful comparison of DMRG and VMC results on a $24\times 3$ system.  For the VMC wave function, we again use antiperiodic boundary conditions on the partons in the longitudinal ($\hx$) direction, but this time take periodic boundary conditions in the transverse ($\hy$) direction such that $k_y=0,\pm2\pi/3$. The $d_1$ and $d_2$ occupations for the VMC state at this point are shown visually in the bottom panel of Fig.~\ref{fig:3leg30_DMRG_VMCcf}; the $k_y=0$ band for the $d_2$ partons is fully filled with a non-zero gap to the other two bands, while the $d_1$ partons partially occupy all three bands.  The $k_y=0$ band has 12 particles and the $k_y=\pm2\pi/3$ bands (which are always degenerate due to the inversion symmetry of the lattice) have 6 particles each.  Therefore, there are no Fermi wave vectors for $d_2$ (i.e., the band is dispersionless) and we have $k_{F1}^{(0)}=12\pi/24=\pi/2$ and $k_{F1}^{(\pm2\pi/3)}=6\pi/24=\pi/4$.

As mentioned earlier, the boson momentum distribution function (top panel of Fig.~\ref{fig:3leg30_DMRG_VMCcf}) is featureless, the VMC results not showing any structure at all.  This is because the condition of one particle per rung is exact in the determinantal wave function whereas, in the DMRG measurements, the number of particles per rung does fluctuate slightly.  The density-density structure factor (middle panel of Fig.~\ref{fig:3leg30_DMRG_VMCcf}), on the other hand, shows singular features at $\pm(k_{F1}^{(0)}+k_{F1}^{(\pm2\pi/3)},\pm2\pi/3)=\pm(3\pi/4,\pm2\pi/3)$ and $\pm(2k_{F1}^{(\pm2\pi/3)},\mp2\pi/3)=\pm(\pi/2,\mp2\pi/3)$.  The other two potentially enhanced wave vectors with $k_y=0$, $\pm(2k_{F1}^{(0)},0)=(\pi,0)$ and $\pm(k_{F1}^{(\pm2\pi/3)}+k_{F1}^{(\mp2\pi/3)},0)=\pm(\pi/2,0)$, are suppressed by the $d_2$ partons.  In the VMC data, this is clear due to the fact that the single boson per rung condition yields $D_b(q_x,q_y=0)=0$ exactly.  The DMRG results show small fluctuations away from zero in this quantity with $q_x^2$ dependence near $q_x=0$.  This is evidence of the phase's incompressibility and rules out the possibility of it being some sort of superfluid, either conventional or paired.  Also, a finite-size study of one- and two-boson gaps \cite{Block11_PRL_106_046402} strongly suggests that the realized phase is indeed insulating, as does a direct measure of the compressibility \cite{Song10_PRB_82_012405} ($\kappa\rightarrow0$) obtained by a scaling analysis of bipartite number fluctuations (data not shown).

\begin{figure}[t]
\vspace{0.1in}
\centerline{\includegraphics[width=1.05\columnwidth]{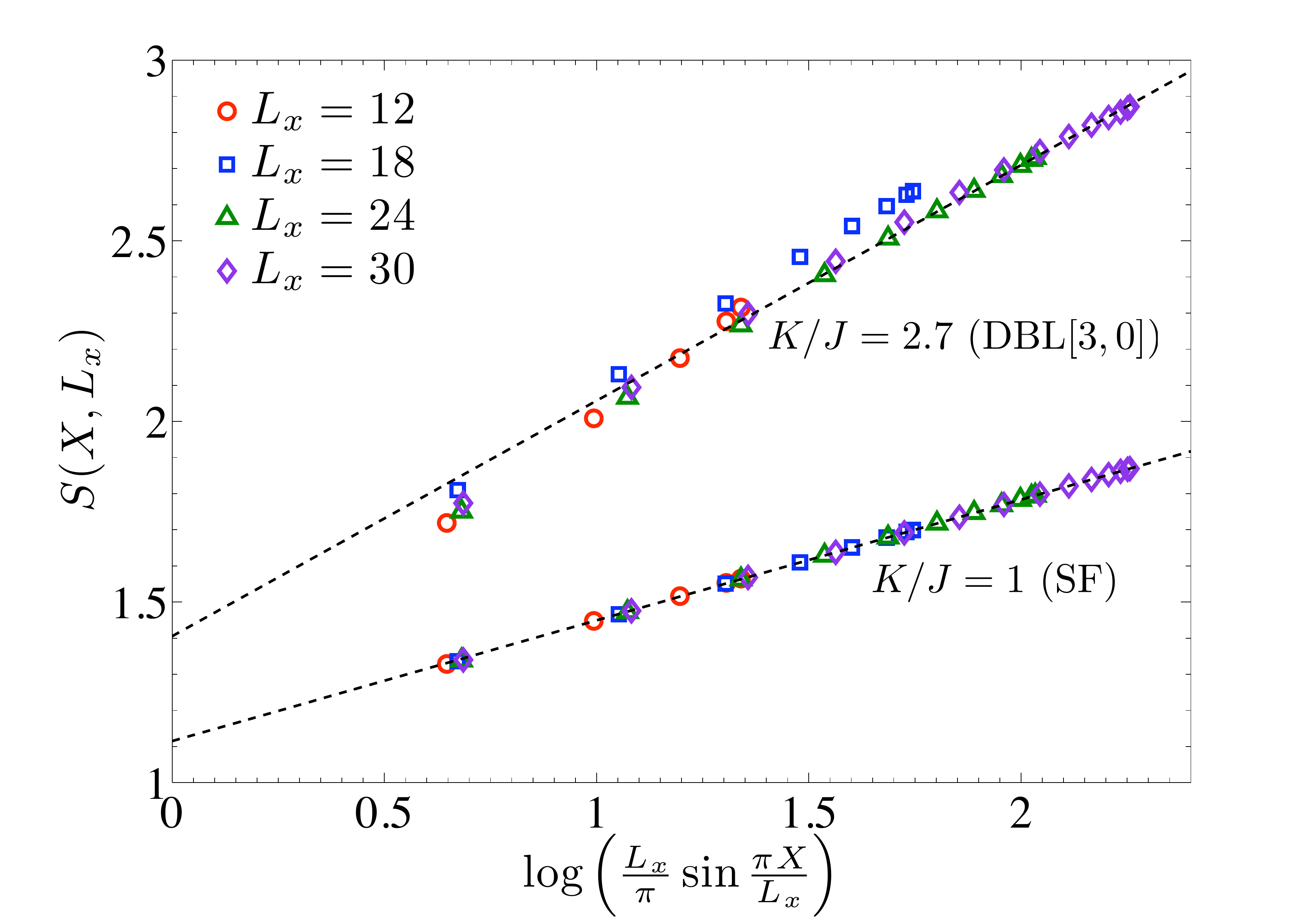}}
\caption{(color online).  Entanglement entropy scaling for different system sizes on the three-leg system at $\rho=1/3$ in the superfluid ($K/J=1$) and DBL$[3,0]$ ($K/J=2.7$) phases, as obtained by DMRG with fully periodic boundary conditions; $J_\perp/J=1$ in both cases.  The dashed lines are fits to the scaling form, Eq.~(\ref{eqn:Sscaling}).  For the superfluid point, all the data collapse extremely well onto a curve with $c=1.00$ and $A=1.12$, which is the lower dashed line.  The dashed line for the DBL$[3,0]$ data is that used in Fig.~4 of Ref.~\onlinecite{Block11_PRL_106_046402} in which $24\times 3$ data was used and the four smallest/largest $X$ values were discarded in the fit; the obtained fit parameters are $c=1.96$ and $A=1.41$.  The collapse of the $L_x=24$ and $L_x=30$ data is reasonable, although there is a very weak downward shift of the slope in the latter case for the largest $X$.  Shell-filling effects or other corrections to the scaling form (which we have not considered) are likely the cause of the less impressive data collapse in the DBL$[3,0]$ as compared to that observed in the superfluid.}
\label{fig:3legEEscaling}
\end{figure}

\begin{figure}[t]
\centerline{\includegraphics[width=1.05\columnwidth]{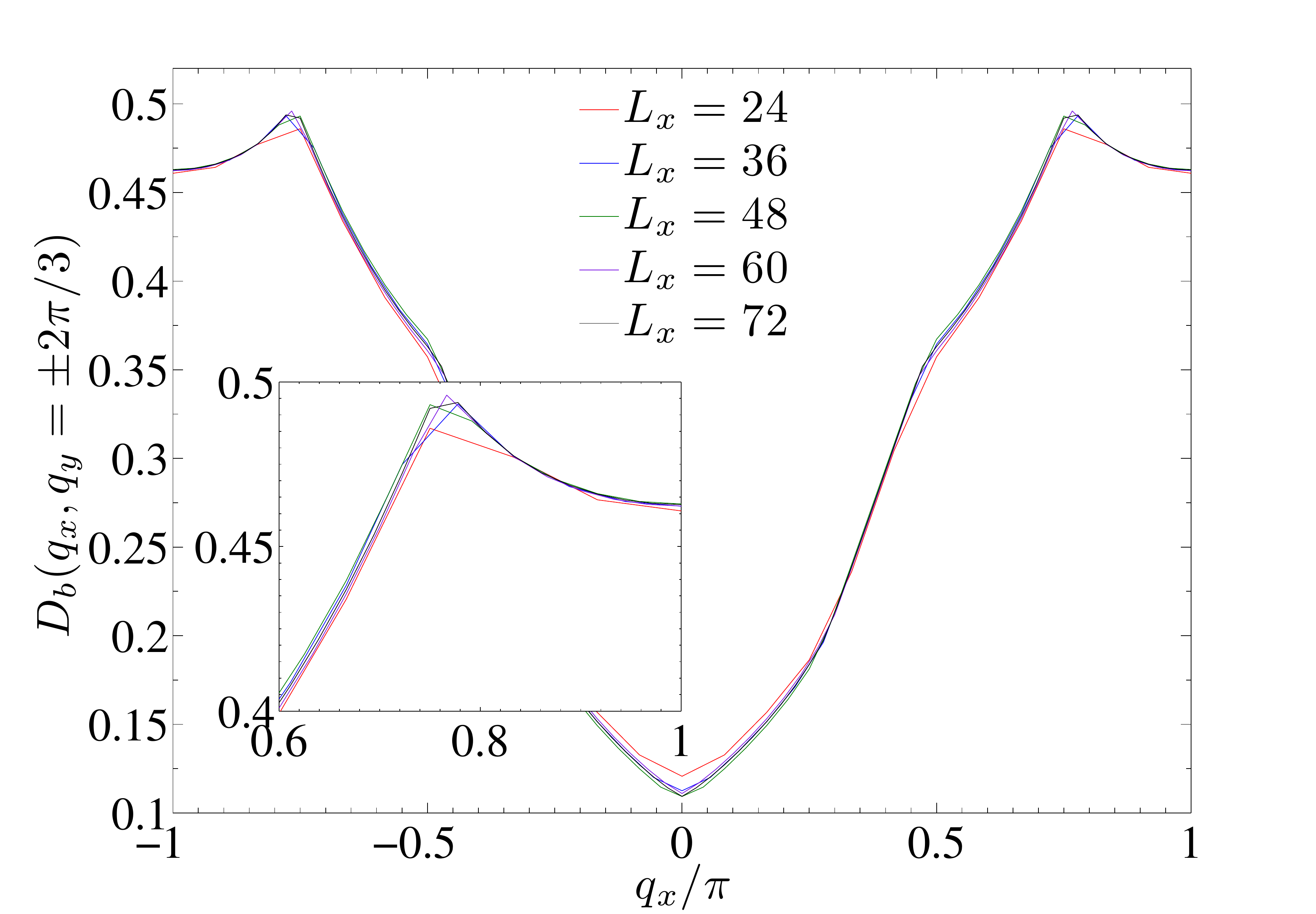}}
\caption{(color online).  Density-density structure factor at $q_y=\pm 2\pi/3$ in the DBL$[3,0]$ phase ($J_\perp/J=1, K/J=2.7$) for various system sizes, $L_x\times 3$, using DMRG with cylindrical boundary conditions.  The main power law singularity (see inset) is robust up to systems of size $72\times 3$.  This is consistent with the long-distance properties of the gapless DBL$[3,0]$ phase, although we have not pursued a full characterization of the long-distance power law behavior.  There is some small irregular behavior of the main singularity on the scale of $2\pi/L_x$, but this is to be expected with cylindrical boundary conditions and shell-filling effects present in our DBL$[3,0]$ theory.}
\label{fig:3leg30_BS}
\end{figure}

\begin{figure}[t]
\centerline{\includegraphics[width=\columnwidth]{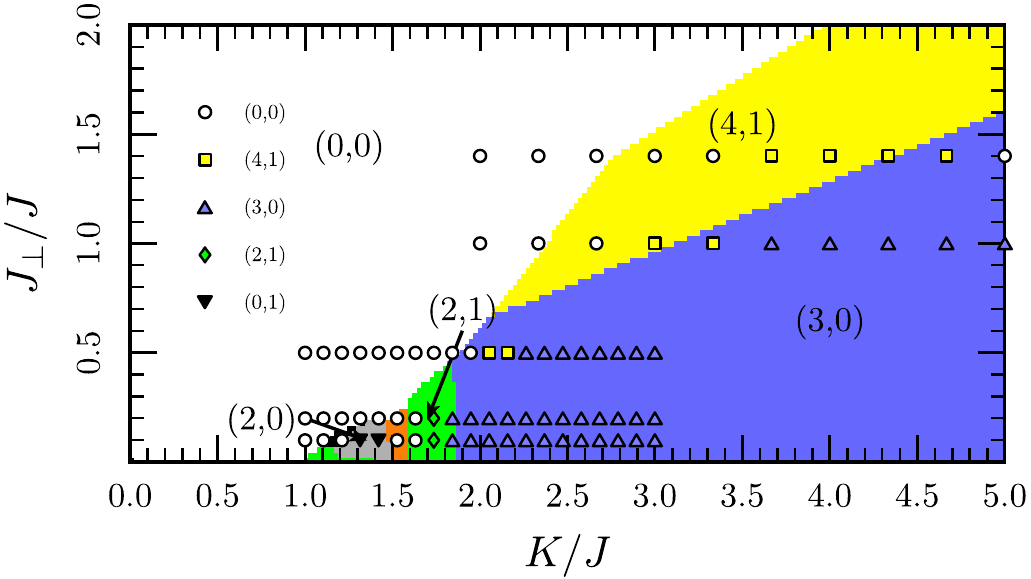}}
\caption{(color online).  Ground state momentum diagram for a $10\times3$ system with $N_b=10$ ($\rho=1/3$).  The colored regions denote the ground state momentum of the variational wave function that optimizes the trial energy at each point in parameter space.  The symbols indicate ED data for the true ground state momentum of the system.  The notation for the quantities in parentheses is $(Q_x,Q_y)$, where $Q_x$ and $Q_y$ are measured in units of $2\pi/L_x$ and $2\pi/L_y$, respectively.  For example, $(4,1)\rightarrow(4\times2\pi/10,1\times2\pi/3)=(4\pi/5,2\pi/3)$.  The agreement between the ED and VMC predictions of ground state momenta at this system size is rather remarkable.}
\label{fig:30_gsm}
\end{figure}

\begin{figure}[t]
\centerline{\includegraphics[width=\columnwidth]{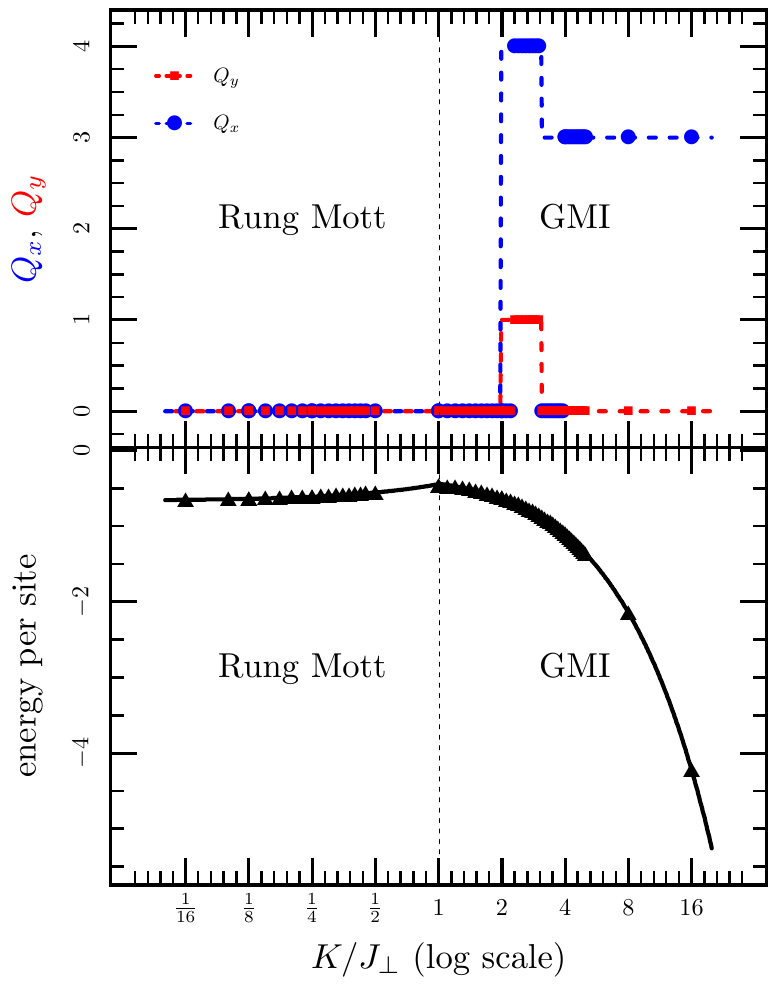}}
\caption{(color online).  Three quantities are plotted on these axes for a $10\times3$ system with $N_b=10$ ($\rho=1/3$) with $J=0$ where $K/J_{\perp}$ is the only dimensionless parameter.  We plot the independent variable on a log scale to better interpolate between the two limits.  Small $K/J_{\perp}$ corresponds to the rung Mott phase where the rungs of the ladder effectively decouple and there exists a gap to all excitations; large $K/J_\perp$ should be the gapless Mott insulator phase, or DBL$[3,0]$.  The solid and dashed curves are the VMC results, while the points are ED results; the vertical dashed line is the rung Mott-GMI phase boundary as determined by VMC.  The blue curve with blue circles is the $x$-component of the ground state momentum measured in units of $2\pi/10$; the red curve with red squares is the $y$-component of the ground state momentum measured in units of $2\pi/3$; finally, the black curve with black triangles is the ground state energy per site.  Remarkably, the variational wave functions capture the energetics exceedingly well for the rung Mott and the DBL$[3,0]$ phases, much more so than for the DBL$[4,2]$ and other conducting DBL$[n,m]$ states.}
\label{fig:J0_gsm}
\end{figure}

In this paper, we go beyond the results presented in Ref.~\onlinecite{Block11_PRL_106_046402} in three ways.  First, we investigate finite-size effects in the DBL$[3,0]$ phase with respect to scaling of the entanglement entropy and robustness of the singular features in the density-density structure factor; these results are highlighted in Figs.~\ref{fig:3legEEscaling} and \ref{fig:3leg30_BS}.  Second, we make detailed comparisons of ED and VMC results, focusing on the ground state energy and momenta as obtained from both methods; these results are highlighted in Figs.~\ref{fig:30_gsm} and \ref{fig:J0_gsm}.  Finally, in the limit $J=0$, we map our three-leg $J$-$J_\perp$-$K$ model at $\rho=1/3$ filling to an equivalent spin-1 model and suggest a potential connection between our $c=2$ gapless DBL$[3,0]$ phase and the $c=2$ gapless phase known to exist in the bilinear-biquadratic spin-1 chain. \cite{Lauchli06_PRB_74_144426}

In Fig.~\ref{fig:3legEEscaling}, we plot for various $L_x$ the bipartite entanglement entropy, $S$, versus $\log(d)$, where $d\equiv(L_x/\pi)\sin(\pi X/L_x)$ is the conformal length and $X$ is the subsystem length.  According to Eq.~(\ref{eqn:Sscaling}), at a given point in the phase diagram we should observe data collapse onto a linear function with universal slope given by $c/3$.  At the point $K/J=J_\perp/J=1$ deep within the superfluid phase, the collapse is excellent for all system sizes $L_x=12,18,24,30$ onto a curve $S(X,L_x)=(1.00/3)\log(d)+1.12$, which is strongly suggestive of $c=1$ gapless mode as we should expect in the superfluid phase.  On the other hand, at the point $K/J=2.7$ and $J_\perp/J=1$, i.e., the representative DBL$[3,0]$ point presented above and in Ref.~\onlinecite{Block11_PRL_106_046402}, the data collapse is less impressive, but is still very suggestive of the $c=2$ gapless modes predicted by our gauge theory description of DBL$[3,0]$ (see Appendix~\ref{app:gt30}).  In fact, we are still likely observing some finite-size ``shell-filling'' effects up to systems of size $30\times 3$ which makes finite-size scaling of any quantity difficult, including the entanglement entropy.  We note that such shell-filling effects are consistent with our parton theory of the DBL$[3,0]$ phase, and such effects actually support our identification of the phase found in our model as a DBL$[3,0]$.  Similar shell-filling effects were also seen in Ref.~\onlinecite{Sheng09_PRB_79_205112} within the two-leg spin Bose metal phase when investigating entanglement entropy scaling in that system.

Obtaining the highly converged entanglement entropy data necessary for convincing determination of $c$ becomes exceedingly difficult for larger system sizes, especially when using fully periodic boundary conditions (PBC) as we have employed to this point.  We have thus also considered OBC in the $\hx$ direction (cylindrical boundary conditions overall), in which case it is much less computationally costly to obtain very accurate DMRG data of any observable on much larger systems.  For example, at the DBL$[3,0]$ point considered in Fig.~\ref{fig:3legEEscaling} on a $24\times 3$ system, retaining $D=4000$ ($D=1500$) states in the DMRG gives a density matrix truncation error of $10^{-7}$ ($10^{-10}$) with PBC (OBC).  As we also experienced with DBL$[4,2]$ in Sec.~\ref{sec:4leg_42}, when using OBC within DBL$[3,0]$ we observe subleading but apparent oscillatory corrections to the usual CFT scaling form of the entanglement entropy [Eq.~(\ref{eqn:Sscaling})], which are to be expected given the finite Bose wave vectors present in the density-density structure factor.  However, for DBL$[3,0]$ the amplitude of these oscillations is significantly weaker than in the case of DBL$[4,2]$ and practically nonexistent for the largest subsystem sizes $X\simeq L_x/2$.  Then, ignoring the oscillatory piece, and fitting the data to the usual form for OBC, $S(X,L_x)=(c/6)\mathrm\log{\left[(2L_x/\pi)\sin{(\pi X/L_x)}\right]}+A'$,
we find a robust $c\simeq 2$ up to systems of size $72\times 3$, with no evidence of a downward shifting $c$ which could signify the eventual opening of a small gap.

To provide further evidence for stability of the DBL$[3,0]$ phase, we plot in Fig.~\ref{fig:3leg30_BS} the density-density structure factor at $q_y=\pm 2\pi/3$ on several large system sizes using cylindrical boundary conditions.  As explained above and in Ref.~\onlinecite{Block11_PRL_106_046402}, this measurement detects the allowed ``$2 k_F$'' wave vectors within the $d_1$ parton band-filling configuration and should contain power law singularities at various incommensurate wave vectors.  Indeed, up to systems of size $72\times 3$, the dominant singularity (see inset of Fig.~\ref{fig:3leg30_BS}) does not smoothen as $L_x$ is increased, indicating that the real-space measurements look like a power law on these system sizes.  Although there is some irregular behavior of the singularity on the order of $2\pi/L_x$ momenta, this is likely due to shell-filling effects and our use of an open boundary condition in the $\hx$ direction.  We caution that it is still possible that the DBL$[3,0]$ is weakly unstable in our model on very long length scales beyond those considered here, although we have not observed evidence for such a trend.  As with DBL$[4,2]$, the long-distance properties considered above can perhaps be investigated further using entanglement renormalization techniques, \cite{Vidal07_PRL_99_220405, Vidal08_PRL_101_110501} which we will leave for future work.

In Appendix~\ref{app:gt30}, we present a bosonized gauge theory analysis of DBL$[3,0]$ and argue that it is, at the very least, a potentially stable quantum phase.  However, as we explain in Appendix~\ref{app:gt30}, there are instabilities out of DBL$[3,0]$ that lead to a fully gapped theory but that are difficult to characterize because one cannot construct any local observables that correspondingly obtain a finite expectation value.  The rung Mott insulator is one such featureless phase, but it is not the only possibility since the sign of the Cooper channel interaction in question may lead to different states.  It is conceivable that the uncharacterized (white) region above the GMI in the phase diagram (see Fig.~\ref{fig:3leg30PD}) is in a fully gapped phase (topologically) distinct from the conventional rung Mott phase; however, we have been unsuccessful in identifying it as such.

We have also performed a detailed comparison of VMC and ED results for the ground state momentum and energy using small system sizes.  First, we would like to emphasize that ED alone on small clusters can ultimately never have the final say regarding the stability of gapless Bose metal-like states, e.g., the $d$-wave Bose liquid or spin Bose metal, in a given model.  This is clearly due to an insufficient resolution of the Brillioun zone for detecting a critical, singular surface, but also because the presence of incommensurate wave vectors makes finite-size scaling difficult to interpret, especially on the system sizes accessible to ED.  For these small systems, we do expect finite-size effects to be significant in our DBL$[3,0]$ phase, and thus the inability to symmetrically fill bands in the VMC wave functions to be more prevalent.  If these variational wave functions truly do capture the main physics of the DBL$[3,0]$ phase, the resulting finite momentum in the ground state should be detectible using ED methods.  In Fig.~\ref{fig:30_gsm}, we show the ground state momentum comparison for a $10\times3$ system with 10 bosons.  Although the VMC wave functions do not quantitatively reproduce all transitions between different regions of ground state momentum, it is clear that the ED data are reflecting the finite-size effects as manifested in the ground state momenta of the nearby GMI states.  The particular system size that we are presenting actually reflects the best agreement over a range of sizes considered (for system sizes $L_x\times3$, we considered $L_x=4,5,6,8,9,10,12$).  Taking these data sets as a whole, we can only say that the ED measurements are sensitive to the finite ground state momenta and show some agreement with VMC results, while the predictive power of the VMC data for the boundaries between various momentum eigenstates in the finite system size corresponding to different band fillings is not always compelling.  Since we expect our DBL predictions to be more accurate as finite-size effects are diminished, these results are perhaps not surprising.  Finally, we note that a recent, analogous consideration of momentum quantum numbers in ED ground states tantalizingly points to the possibility of a spin Bose metal state existing near the Mott transition in the half-filled triangular lattice Hubbard model, \cite{Lauchli10_PRL_105_267204} which serves as a nice example of using ED to at least suggest the potential stability of a gapless Bose metal-like phase.

We also performed a study of the $J=0$ case, where the condition of one particle per rung is strictly enforced by the model itself, using VMC and ED.  In this extreme limit, there is no superfluid phase and we expect a (nearly) direct transition from the rung Mott insulating phase to the DBL$[3,0]$ phase as we increase the single control parameter $K/J_{\perp}$.  However, as discussed above, DMRG on larger systems indicates that there is likely an intermediate phase between the rung Mott insulator and DBL$[3,0]$, \cite{Block11_PRL_106_046402} a phase which we have been yet unable to characterize fully; this phase also appears at $J=0$ in the DMRG at larger systems.  In Fig.~\ref{fig:J0_gsm}, we show a comparison of the VMC and ED ground state momenta as we vary this parameter for a $10\times3$ system with 10 bosons.  The agreement is excellent over the entire range of values studied suggesting that, in this limit, the VMC wave functions capture the main physics exceptionally well, even though the DMRG and VMC do not agree as favorably in the transition region on larger system sizes.

Finally, in this limit of $J=0$, we have also considered mapping our three-leg model at $\rho=1/3$ to an equivalent model of interacting spin-1 degrees of freedom living on a 1D chain.  Working as always with periodic boundary conditions in the $\hy$ direction, the state of each rung can be labeled by the $y$ momentum of the single boson on that rung:  let $|0\rangle_i$, $|+\rangle_i$, $|-\rangle_i$ denote the $q_y=0,+2\pi/3,-2\pi/3$ states of rung $i$, respectively.  Our three-leg $J_\perp$-$K$ model at $\rho=1/3$ can then, up to a constant, be written in terms of spin-1 operators as follows:
\begin{equation}
H = K\sum_i\left[P_{i,i+1}(\theta)+\lambda Q_{i,i+1}\right] + 3J_\perp\sum_i(S^z_i)^2,
\label{eqn:spin1model}
\end{equation}
where we have defined
\begin{align}
P_{i,j}(\theta) &\equiv \sqrt{2}\left[\cos\theta(\mathbf{S}_i\cdot\mathbf{S}_j) + \sin\theta(\mathbf{S}_i\cdot\mathbf{S}_j)^2\right], \\
Q_{i,j} &\equiv -\frac{1}{3}\sum_{\alpha=1,2,3}|\phi_\alpha\rangle_{ij}~_{ij}\langle\phi_\alpha|,
\end{align}
with
\begin{align}
|\phi_1\rangle_{ij} \equiv |0\rangle_i |0\rangle_j + |+\rangle_i |-\rangle_j + |-\rangle_i |+\rangle_j\,, \\
|\phi_2\rangle_{ij} \equiv |+\rangle_i |+\rangle_j + |0\rangle_i |-\rangle_j + |-\rangle_i |0\rangle_j\,, \\
|\phi_3\rangle_{ij} \equiv |-\rangle_i |-\rangle_j + |0\rangle_i |+\rangle_j + |+\rangle_i |0\rangle_j\,,
\end{align}
and introduced the parameters $\lambda$ and $\theta$ to connect to previous studies of similar models; our $J_\perp$-$K$ model corresponds to $\theta=\pi/4$ and $\lambda=1$.  The operators $S^{x,y,z}_i$ are the usual spin-1 operators for the pseudospin on rung $i$.

Up to a constant, $P_{i,j}(\theta=\pi/4)$ is the SU(3) symmetric spin-exchange operator for spin-1.  Hence, Eq.~(\ref{eqn:spin1model}) with $\theta=\pi/4$, $\lambda=0$, and $J_\perp=0$ is equivalent to the SU(3) symmetric Heisenberg model, i.e., the spin-1 Lai-Sutherland model, \cite{Lai74_JMathPhys_15_1675, Sutherland75_PRB_12_3795} which is known to be critical with $c=2$ gapless modes. \cite{Sutherland75_PRB_12_3795, Fath91_PRB_44_11836, Lauchli06_PRB_74_144426, Corboz07_PRB_76_220404, Rachel08_AnnPhys_17_922} There is also known to be an extended critical phase in this model, still at $\lambda=0$ and $J_\perp=0$, but away from the SU(3) point in the parameter range $\theta\in[\pi/4,\pi/2)$. \cite{Fath91_PRB_44_11836, Itoi97_PRB_55_8295, Lauchli06_PRB_74_144426, Corboz07_PRB_76_220404}  Interestingly, this critical phase has soft modes \cite{Fath91_PRB_44_11836} at $q_x=0,\pm2\pi/3$ with dominant spin quadrupolar correlations at wave vector $2\pi/3$, \cite{Lauchli06_PRB_74_144426} which is precisely the ``$2k_F$'' wave vector in the $J=J_\perp=0$ limit of our DBL$[3,0]$ theory (i.e., three equally filled $d_1$ bands) at which we should expect singularities in, e.g., $D_b(q_x,q_y=\pm2\pi/3)$.  It thus seems plausible that the well-known extended critical phase in the bilinear-biquadratic spin-1 chain \cite{Lauchli06_PRB_74_144426} is connected to the decoupled chains limit of DBL$[3,0]$, although we can't be sure without a direct study of Eq.~(\ref{eqn:spin1model}) at variable $\lambda$.  Finally, the SU(3) symmetric Lai-Sutherland model ($\lambda=0$, $\theta=\pi/4$) has also been studied in the presence of single-site anisotropy, \cite{Schmitt96_JPhysA_29_3951} which is precisely our $J_\perp$ term.  As in the pure SU(3) symmetric case, this model is integrable via the Bethe ansatz, using which it was shown that the $c=2$ gapless SU(3) phase exists at finite $J_\perp$ before a transition to a large $J_\perp$ gapped phase, \cite{Schmitt96_JPhysA_29_3951} which corresponds in the boson language to our rung Mott insulator.  This is consistent with our finding that DBL$[3,0]$ is stable for finite values of $J_\perp$.

All that being said, our actual $J_\perp$-$K$ model maps to Eq.~(\ref{eqn:spin1model}) at finite $\lambda=1$ and fixed $\theta=\pi/4$.  While the operator $Q_{i,j}$ is rather unnatural for a spin system and we have not pursued a direct study of Eq.~(\ref{eqn:spin1model}) with variable $\lambda$, it seems reasonable that the known $c=2$ gapless phase \cite{Lauchli06_PRB_74_144426} present in the extended parameter space discussed above could be stable at $\lambda=1$ and thus adiabatically connected to our DBL$[3,0]$.  However, we note that, unlike the gapless phase in the bilinear-biquadratic spin-1 model, DBL$[3,0]$ at finite $J_\perp$ generally exhibits correlations with incommensurate wave vectors.

Rather remarkably, we have been able to make a (potential) connection to an exotic spin-1 system \cite{Lauchli06_PRB_74_144426} coming from a considerably distant starting point, i.e., itinerant bosons hopping with ring exchange on a three-leg ladder. The above considerations elucidate that it is ultimately the $y$ coordinate of the bosons responsible for the gaplessness of DBL$[3,0]$.  In fact, the situation of entering DBL$[3,0]$ out of the superfluid by increasing $K$ is somewhat reminiscent of interactions inducing a metal-insulator transition in a half-filled Hubbard chain, wherein the charge degrees of freedom get immobilized while the spin degrees of freedom remain gapless.  In the case of the superfluid-DBL$[3,0]$ transition, the $x$ coordinate of the bosons becomes localized, while, very loosely speaking, the $y$ coordinate remains free---a more rigorous interpretation of the phase in terms of bosons only is unavailable, hence our fermionic parton description.  Ultimately, we find it quite profound that the number of gapless modes actually increases by one upon entering the insulator in the latter case, whereas it always decreases in the well-known former case.

\subsection{Discussion} \label{3legdisc}

Again, the qualitative agreement of the VMC and DMRG results for the DBL$[3,0]$ is excellent.  The quantitative agreement, in fact, is better than in the case of the DBL$[4,2]$.  Furthermore, our confirmation of the central charge of $c\simeq2$ for the DBL$[3,0]$ and the compelling results of the VMC/ED comparison for small system sizes serve to reinforce our proposed realization of this phase in our model.  We also carefully considered finite-size effects using DMRG, and, although we can never rule out eventual small gaps appearing on very long length scales, \cite{Crepin11_PRB_84_054517} we believe the data presented strongly supports identification of the remarkable phase found in our ring model as an exotic gapless Mott insulator.

Subsequent to our discovery and characterization of the DBL$[3,0]$ phase, we revisited the two-leg ladder and the possibility of realizing the analogous phase at half-filling ($\rho=1/2$), i.e., a DBL$[2,0]$ phase.  Unfortunately, only the most mundane incarnation of this phase is realized for $J_{\perp}=0$ and sufficiently large $K/J$.  For any finite $J_{\perp}$, the DBL$[2,0]$ phase is unstable toward either a $(\pi,\pi)$ charge density-wave (CDW) (for moderate $K/J$ and small $J_{\perp}/J$) or a state with staggered currents on the rungs (for larger $K/J$).  We discuss the situation on two legs in greater detail in Appendix~\ref{app:2legHalfFilling}.

We also considered the possibility of finding a DBL$[4,0]$ phase on the four-leg ladder at either $\rho=1/4$ or $\rho=1/2$.  At filling $\rho=1/4$, DMRG indicates that the system phase separates into $\rho=0$ and $\rho=1/2$ filled regions immediately outside of the superfluid, so no GMI-type phase is observed.  On the other hand, at $\rho=1/2$, we observe $(\pi,\pi)$ CDW ordering even in the DBL$[4,0]$ VMC wave functions, thus giving little hope for realizing such a phase in the DMRG.  However, the DMRG  itself does not show robust long-range CDW order even at very large $K$ and reasonably large systems (e.g., at $K/J=64$ up to systems of size $48\times 4$).  Instead of a clear Bragg peak, density-density correlations [see Eq.~(\ref{eqn:ddcorr})] appear to decay as an oscillatory power law with wave vector $(\pi,\pi)$.  This same behavior is also observed in our four-leg half-filled model with unfrustrated ring exchange ($K<0$) on the same system sizes (e.g., at $K/J=-64$ up to systems of size $48\times 4$).  Recall that for the $K$-only model ($J=0$) the sign of the ring-exchange term can be changed by dividing the square lattice into four sublattices and performing the canonical transformation $b\rightarrow-b$ on one sublattice, hence the sign of $K$ is irrelevant in this limit.  We believe the lack of long-range CDW order in our half-filled four-leg system, for both signs of $K$, is possibly related to the unusual size dependence of the $(\pi,\pi)$ CDW order parameter observed in the original studies of the 2D unfrustrated $J$-$K$ model with quantum Monte Carlo. \cite{Sandvik02_PRL_89_247201, Sandvik06_AnnPhys_321_1651}  For example, at $K/J=-64$ in 2D, the CDW order parameter shows distinct non-monontonic behavior with $1/L$, and extrapolations performed using only system sizes up to $8\times8$ would lead to the ultimately incorrect conclusion that the system lacks long-range $(\pi,\pi)$ CDW order. \cite{Sandvik02_PRL_89_247201, Sandvik06_AnnPhys_321_1651}  We conjecture that a similar effect is occurring in our four-leg system due to the small transverse size of the ladder.

This serves as a reminder that the four-leg system is in reality still not 2D, and that we should always be cautious about extrapolating results from $N=2,3,4,\dots$ legs to full 2D.  Exactly what these results at $\rho=1/2$ on four legs, which fail to connect particularly well to the known 2D unfrustrated results, mean for the eventual stability of metallic DBL phases in full 2D at $\rho<1/2$ (e.g., 2D extensions of DBL$[4,2]$ from Sec.~\ref{sec:4leg}) is presently still unclear and deserving of future investigation, although the final answer may not emerge until there exists a scalable numerical method suited for $L\log L$ scaling of entanglement entropy in 2D.  However, the DBL$[4,2]$ at density $\rho=5/12<1/2$ on four legs seems very robust in our model, and the behavior of the model at $\rho=1/2$ for $K \gg J$ is a special case not necessarily related to the stability of the DBL, which is expected to exist only for $\rho<1/2$ and intermediate $K/J$.

\section{Conclusions and Future Directions} \label{sec:conclusions}

In Ref.~\onlinecite{Sheng08_PRB_78_054520}, the authors presented compelling evidence for the existence of a quasi-1D descendent of the exotic, strongly correlated, two-dimensional $d$-wave Bose metal phase.  The so-called DBL$[2,1]$ exhibited the distinctive singular fingerprints of the parent 2D phase and possessed the two gapless 1D modes (same as the number of legs) predicted by the bosonized gauge theory.  In this paper, we have taken a significant step toward the 2D limit by finding and characterizing a stable DBL$[4,2]$ phase for our ring-exchange model, which has five gapless modes, one more than the number of legs.  In particular, both of the slave fermions, $d_1$ and $d_2$, occupy more than a single momentum band and thus manifest the $d$-wave sign structure of the wave function in a highly nontrivial manner, demonstrating extensibility to two dimensions.  It may yet be possible to consider a system of six legs ($N=6$), using the results obtained thus far for $N=2,4$ to guide the search for perhaps a DBL$[6,2]$ or even DBL$[6,4]$ phase.  The great challenge here is that the spatial entanglement becomes distressingly large as the gaplessness increases and DMRG methods will have great difficulty converging meaningful results on systems of reasonable length.  Also, as we discuss further in Appendix~\ref{app:3legIncomm}, it may be possible to stabilize DBL phases on odd leg ladders with $\rho>1/N$ by modifying the transverse boundary conditions; this would potentially allow us to realize DBL phases on three- and five-leg ladders that are also extensible to 2D.  [In the case of the three-leg ladder with periodic transverse boundary conditions, our numerics indicates that such Bose liquid phases are not stabilized, and instead an exotic (``bond chiral'') superfluid phase \cite{Schaffer09_PRB_80_014503} is likely realized by our ring model (see Appendix~\ref{app:3leg_noBMs} for details).]

Our tried and true method of approaching 2D via a series of quasi-1D ladders has remarkably yielded great success in demonstrating the DBL as a potentially stable, gapless, and strongly interacting state of hard-core bosons for which no perturbative understanding currently exists.  The quasi-1D nature of the systems we have studied has allowed us to access a very challenging realm of condensed matter physics both with exact numerical techniques (DMRG/ED) and with a robust gauge theory that, in principle, allows for the exact calculation of all anomalous power law exponents.  It is worth noting that it is precisely the existence of singular surfaces in momentum space and the lack of a quasiparticle description in the 2D phase that simultaneously render the numerical and analytical analysis extremely formidable while paving the way for our quasi-1D methods to be especially effective in gaining a physical intuition for these strongly correlated phases.  It is only by virtue of the ``Bose surfaces" that we can reliably and meaningfully detect the corresponding ``Bose points" on the ladders.

However, we still caution that the ladder approach is not perfect, and studies at $N=2,3,4,\dots$ legs hardly represent crossing over to full 2D.  Indeed our results on ladders of this size are somewhat irregular.  As discussed further in Appendix~\ref{app:3leg_noBMs}, we do see a strong even-odd effect (with the number of legs $N$) in the stability of compressible Bose-metal phases; the transverse boundary condition is a substantial modification to the model on these few-leg ladders, and we conjecture that altering it could change the story considerably.  We note that in some cases it is possible to truly solve frustrated 2D models with ingenious and systematic use of the DMRG, \cite{White11_Science_332_1173,Stoudenmire11_condmat1105.1374} but unfortunately, those techniques are not practical for use with our Bose-metal-type phases in our ring model.  Nonetheless, the local physics is expected to become more 2D-like as $N$ increases, and thus we believe our four-leg results in Sec.~\ref{sec:4leg} are at the least very suggestive of eventual stability of the DBL in 2D.

In the course of our search for gapless $d$-wave Bose metals, we came across the highly novel gapless Mott insulating phase at commensurate density of $\rho=1/3$ on the three-leg ladder.  The so-called DBL$[3,0]$ gave rise to oscillations at incommensurate wave vectors in its density-density correlator, possessed two gapless modes, and exhibited the characteristic $d$-wave correlations common to all of the DBL$[n,m]$ phases.  Despite being truly quasi-1D in nature, this strange insulator is very interesting in its own right as it can be described by a parton gauge theory in which one parton is gapless while the other is gapped (see Appendix \ref{app:gt30}).  We expect that it may be possible to find a quasi-2D analog of this phase by studying an $N$-\emph{layer} square lattice system at density $\rho=j/N$ for $j=1,2,\dots$ where we believe that strongly coupled ring exchanges between the layers may give rise to the desired behavior.  Additionally, while our study of the four-leg ladder did not discover a stable DBL$[4,0]$ phase, it may be the additional frustration of an odd number of legs that allowed the DBL$[3,0]$ to exist stably.  Therefore, it may be possible to search for an analogous DBL$[5,0]$ phase on the five-leg ladder at densities $\rho=1/5,2/5$.  The smaller density will likely phase separate right out of the superfluid phase as was the case for $\rho=1/4$ on the four-leg ladder, but the larger one may not.

Finally, we should here again mention the physical problem that inspired the original search for the DBL in the first place: itinerant electrons with real-space $d$-wave correlations; that is, a $d$-wave metal phase.  If one imagines fractionalizing the electron creation operator into a charge-carrying hard-core bosonic piece and a spin-carrying fermionic piece (i.e., $c_\sigma^\dagger=b^\dagger f^\dagger_\sigma$) and then frustrating the system sufficiently (with an appropriate ring term) such that the bosons fail to condense and instead realize a DBL phase (i.e., $b^\dagger=d_1^\dagger d_2^\dagger$), the resulting electronic metal should be accessible by all of the methods employed herein.  Namely, we can place such a system on a ladder and study it numerically using DMRG and analytically using a similar bosonized gauge theory approach as that presented in this work.  Such an investigation is currently underway with promising results thus far.

\acknowledgments

This work was supported by the NSF under grants DMR-1101912 (R.V.M., M.S.B., and M.P.A.F.), DMR-1056536 (R.K.K.), DMR-0906816, DMR-0611562 (D.N.S.), and DMR-0907145 (O.I.M.); and Microsoft Station Q (R.V.M. and R.K.K.).

\appendix

\section{Gauge Theory Description and Solution by Bosonization} \label{app:gaugetheory}

Here we describe the analytical approach for the $N$-leg ladder where $N\geq3$, which is generic; the formulations for $N=1,2$ are special cases.  To accurately capture the behavior of the physical bosons within the slave particle picture [see Eq.~(\ref{eqn:partons})] requires a compact U(1) lattice gauge theory.  First, we consider the gauge sector: we denote the integer-valued electric field on links between sites $x$ and $x+1$ on the same chain of the ladder by $E_{\ell,x}$, where $\ell=1,\dots,N$ labels the chains, and on links between chains $\ell$ and $\ell+1$ at the rung located at $x$ by $E_{\ell,y}$.  We then introduce a $2\pi$-periodic vector potential on each link of the lattice, which is denoted the same way as the electric field: $a_{\ell,x}$ for links within chain $\ell$ and $a_{\ell,y}$ for links between chains $\ell$ and $\ell+1$.  The magnetic field, which lives on the elementary plaquettes is then given by the lattice curl of the vector potential:
\begin{equation}
\label{eqn:Bfield}
(\nabla\times\mathbf{a}_{\ell})=a_{\ell,x}(x)+a_{\ell,y}(x+1)-a_{\ell+1,x}(x)-a_{\ell,y}(x),
\end{equation}
where $\mathbf{a}_{\ell}=(a_{\ell,x},a_{\ell,y})$.  The Hamiltonian density for the gauge sector is thus
\begin{equation}
\label{eqn:Hgauge}
h_{\mathrm{gauge}}=\sum_{\ell=1}^N\left[h\frac{\mathbf{E}_{\ell}^2}{2}-K\cos(\nabla\times\mathbf{a}_{\ell})\right],
\end{equation}
where $\mathbf{E}_{\ell}=(E_{\ell,x},E_{\ell,y})$ and where we identify $\ell=N+1=1$ due to the transverse periodic boundary conditions.  We also include here arbitrary coupling parameters for the electric and magnetic fields, $h$ and $K$.  Note that the contribution due to the magnetic field is encoded in a compact form due to the $2\pi$-periodicity of the vector potential.

The purpose of the gauge field in our slave particle theory is to bind the $d_1$ and $d_2$ partons together at each site to realize the hard-core bosons.  To this end, we assign equal and opposite gauge charges to the partons; specifically, charge $e_1=+1$ to the $d_1$ partons and charge $e_2=-1$ to the $d_2$ partons.  The lattice divergence of the electric field gives the total free charge:
\begin{align}
\label{eqn:chargeconstraint}
(\nabla\cdot\mathbf{E}_{\ell})&=E_{\ell,x}(x)-E_{\ell,x}(x-1)+E_{\ell,y}(x)-E_{\ell-1,y}(x),\\
(\nabla\cdot\mathbf{E}_{\ell})&=d_1^{\dagger}d_1-d_2^{\dagger}d_2.
\end{align}
We therefore wish to enforce the constraint that $(\nabla\cdot\mathbf{E}_{\ell})=0$ at each site.  This is accomplished by adding a constraint term to the action.

We work with a Euclidean path integral and denote the temporal components of the vector potential at each site on chain $\ell$ as $a_{\ell,\tau}$.  The electric field and vector potential are canonically conjugate quantum operators: $[E_{\ell,\mu},a_{\ell,\nu}]=i\delta_{\mu\nu}$ and so the full gauge sector Lagrangian is thus
\begin{align}
\label{eqn:gaugesector}
{\cal L}_{\mathrm{gauge}}=\sum_{\ell=1}^N[i\mathbf{E}_{\ell}\cdot\nabla_{\tau}\mathbf{a}_{\ell}+h\frac{\mathbf{E}_{\ell}^2}{2}-K\cos(\nabla\times\mathbf{a}_{\ell})+\nonumber\\+i(\nabla\cdot\mathbf{E}_{\ell})a_{\ell,\tau}],
\end{align}
where $\nabla_{\tau}a_{\ell,\nu}=a_{\ell,\nu}(x,\tau+1)-a_{\ell,\nu}(x,\tau)$.  The first term is the ``$ip\partial_{\tau}q$" term where $[q,p]=i$.  The second and third terms are the Hamiltonian.  The final term is the constraint that will enforce $(\nabla\cdot\mathbf{E}_{\ell})=0$ upon integrating out $a_{\ell,\tau}$.  By taking advantage of the $2\pi$-periodicity of the resulting action in the spatial components of the vector potential, we can choose to integrate out the electric field over all real numbers despite the fact that it is strictly integer-valued.  The resulting Lagrangian is
\begin{align}
\label{eqn:gaugesectornoE}
{\cal L}'_{\mathrm{gauge}}&=\sum_{\ell=1}^N\left[\frac{1}{2h}(\nabla_{\tau}a_{\ell,x}-\nabla_xa_{\ell,\tau})^2+\right.\nonumber\\&\left.+\frac{1}{2h}(\nabla_{\tau}a_{\ell,y}-a_{\ell+1,\tau}+a_{\ell,\tau})^2-K\cos(\nabla\times\mathbf{a}_{\ell})\right],
\end{align}
where $\nabla_xa_{\ell,\nu}=a_{\ell,\nu}(x+1,\tau)-a_{\ell,\nu}(x,\tau)$.

We now turn our attention to gauge fixing.  In principle, we have the freedom to specify a scalar field $\Lambda_{\ell}(x,\tau)$ at every site in the space-time lattice; however, in anticipation of taking the continuum limit in $x$ and $\tau$, we only have the freedom to choose $N$ functions of $x$ and $\tau$, which are precisely the functions $\Lambda_{\ell}$.  Let us first consider how the $a_{\ell,y}$ fields transform under a gauge transformation,
\begin{equation}
\label{eqn:ygaugetrans}
a_{\ell,y}\rightarrow a_{\ell,y}+\Lambda_{\ell+1}(x,\tau)-\Lambda_{\ell}(x,\tau).
\end{equation}
Interestingly, the combination $a_y\equiv(1/N)[\sum_{\ell=1}^{N}a_{\ell,y}]$ is thus inherently gauge invariant.  If we imagine our ladder as a cylinder due to the strictly finite extent and periodic boundary condition in the transverse direction, this combination of gauge fields corresponds to the magnetic field through the cylinder, which we can choose to set to zero \emph{without expending any gauge freedom}.  We can then choose $N-1$ constraints on the scalar fields $\Lambda_{\ell}$ so as to fix all the other $y$-component gauge fields to zero: $a_{\ell,y}=0$ for all $\ell$.  This leaves us with exactly one constraint left to choose.  The Lagrangian can be simplified at this stage as follows:
\begin{align}
\label{eqn:gaugesectory0}
{\cal L}'_{\mathrm{gauge}}=&\sum_{\ell=1}^N\left[\frac{1}{2h}(\nabla_{\tau}a_{\ell,x}-\nabla_xa_{\ell,\tau})^2+\right.\nonumber\\&\left.+\frac{1}{2h}(a_{\ell+1,\tau}-a_{\ell,\tau})^2-K\cos(a_{\ell+1,x}-a_{\ell,x})\right].
\end{align}

In formulating the fermion sector of our gauge theory, which we will do below, we will consider as independent flavors the partons corresponding to different momentum bands.  These partons couple to the gauge field combination $a_{\mu}\equiv(1/N)[\sum_{\ell=1}^{N}a_{\ell,\mu}]$, where $\mu=x,\tau$, as to the usual gauge field.  The other linearly independent combinations of the gauge fields, $a_{\ell,\mu}$, are all massive in Eq.~(\ref{eqn:gaugesectory0}) and can thus safely be integrated out of the effective Lagrangian, leaving only $a_x$ and $a_{\tau}$.  We can now use our final gauge constraint to fix $a_x=0$.  Thus, the final, effective gauge sector Lagrangian is quite simple:
\begin{align}
\label{eqn:gaugesectoreff}
{\cal L}''_{\mathrm{gauge}}=\frac{N}{2}(\nabla_xa_{\tau})^2,
\end{align}
where we have taken the parameter $h=1$.

We now consider the fermion sector, in which we start with just the two partons that compose the hard-core bosons in the slave-particle picture: $d_{\alpha}(\rvec)$ for $\alpha=1,2$.  Then we decompose each such operator into its discrete Fourier components in the $\hy$ direction:
\begin{equation}
\label{eqn:partonsFTy}
d_{\alpha}^\dagger(x,y)=\frac{1}{\sqrt{N}}\sum_{k_y}e^{ik_y(y-1)}d_{\alpha}^{(k_y)\dagger}(x),
\end{equation}
where $k_y$ runs over the partially occupied bands for each value of $\alpha$.  For the DBL$[n,m]$, $k_y$ varies over $n$ discrete values of the transverse momenta for $\alpha=1$ and over $m$ discrete values for $\alpha=2$.  From this point on, we take the continuum limit in the longitudinal direction and treat $x$ as a continuous variable.  All operators in what follows will be functions of this single variable $x$ and so it will be suppressed henceforth.

Each partially occupied band has orbitals filled on the interval $[-k_{F\alpha}^{(k_y)},k_{F\alpha}^{(k_y)}]$ and this defines the Fermi wave vectors.  We are primarily interested in the low-energy physics of the DBL phases and so our next step is to split the fermion operators into right- and left-movers linearizing about the right and left Fermi points for each parton:
\begin{equation}
\label{eqn:rightleftmovers}
d_{\alpha}^{(k_y)\dagger}(x)\approx d_{\alpha R}^{(k_y)\dagger}e^{ik_{F\alpha}^{(k_y)}x} + d_{\alpha L}^{(-k_y)\dagger}e^{-ik_{F\alpha}^{(-k_y)}x},
\end{equation}
which gives the full right-left decomposition shown in Eq.~(\ref{eqn:partdecomp}).  Throughout this paper, we use the standard convention \cite{Lin97_PRB_56_6569} that the grouping of a pair of right- and left-movers corresponds to a pair of fields related by time reversal:  $(k_x,k_y)\rightarrow(-k_x,-k_y)$.  From now on, $k_y$ on the fields is a label:  right-movers carry the same $y$ momentum as the label, while left-movers carry opposite $y$ momentum.  Each right- and left-moving fermion behaves like a massless Dirac fermion.  With the spatial components of the gauge field ($a_x$ and $a_y$) fixed at zero by our chosen gauge constraints, the free fermionic action is governed by a Lagrangian of the form
\begin{align}
\label{eqn:fermionsector}
{\cal L}_{\mathrm{ferm}}=\sum_{\alpha,k_y,P}&\left[d_{\alpha P}^{(k_y)\dagger}(\partial_{\tau}-ie_{\alpha}a_{\tau})d_{\alpha P}^{(k_y)}\right.\nonumber\\&\left.-iPv_{F\alpha}^{(k_y)}d_{\alpha P}^{(k_y)\dagger}\partial_xd_{\alpha P}^{(k_y)}\right],
\end{align}
where $v_{F\alpha}^{(k_y)}$ are the dispersion velocities corresponding to each Fermi point.  In a theory free of any four-fermion interactions, these velocities are the only input parameters.  We could, in principle, use variational wave functions, compared against DMRG results, to determine on finite-size systems the approximate filling of the bands for particular points in our $J$-$J_{\perp}$-$K$ model [see Eq.~(\ref{eqn:fullmodel})].  Assuming free fermion band dispersions, we could then estimate the dispersion velocities as simply the slopes of the band curves at the Fermi points.

We now consider potential four-fermion interactions, starting with density-density interactions:
\begin{equation}
\label{eqn:densityint}
{\cal L}_{\rho\mathrm{-int}}=\sum_{b,b',P}[B_{b,b'}\rho_{bP}\rho_{b'P}+C_{b,b'}\rho_{bP}\rho_{b'-P}],
\end{equation}
where $b,b'$ are composite indices for $\alpha$ and $k_y$ and run over all possible values of these.  The density operators are simply $\rho_{bP}=d_{bP}^{\dagger}d_{bP}$.  The couplings $B_{b,b'},C_{b,b'}$ can be thought of as $(m+n)\times(m+n)$ symmetric matrices for a DBL$[m,n]$.  Such terms are strictly marginal with respect to the DBL fixed point, but will shift and renormalize the initial dispersion velocities.

Other, potentially more worrisome, four-fermion interactions do exist and will be considered within the context of particular DBL$[n,m]$ phases below.

The gauge theory as presented thus far is entirely treatable by conventional bosonization techniques.  This controlled approach allows one, in principle, to compute all power law exponents thus capturing all the long-range, low-energy physics of the putative DBL$[n,m]$ phases.  We begin with the bosonization formula
\begin{equation}
d_{\alpha P}^{(k_y)}=\eta_{\alpha}^{(k_y)}\mathrm{exp}\left[i\left(\phi_{\alpha}^{(k_y)}+P\theta_{\alpha}^{(k_y)}\right)\right],
\label{eqn:bosonization}
\end{equation}
where $\phi_{\alpha}^{(k_y)}$ and $\theta_{\alpha}^{(k_y)}$ are the conjugate phase and phonon fields for each band satisfying
\begin{equation}
\label{eqn:bosecommutators}
\left[\phi_{\alpha}^{(k_y)}(x),\theta_{\alpha'}^{(k_y')}(x')\right]=i\pi\delta_{\alpha\alpha'}\delta_{k_yk_y'}\Theta(x-x'),
\end{equation}
while $\eta_{\alpha}^{(k_y)}$ are Klein factors, which preserve the fermion anticommutation relations, i.e., they commute with the bosonic fields and anticommute among themselves.  The fermion sector Lagrangian, cast in bosonic fields, becomes
\begin{align}
\label{eqn:fermionsectorbosonized}
{\cal L}_{\mathrm{ferm}}=\sum_{\alpha,k_y}\frac{i}{\pi}\left(\partial_x\theta_{\alpha}^{(k_y)}\right)\left(\partial_{\tau}\phi_{\alpha}^{(k_y)}-e_{\alpha}a_{\tau}\right)+{\cal H}_{\mathrm{kin}}\,,
\end{align}
where
\begin{equation}
\label{eqn:Hkin}
{\cal H}_{\mathrm{kin}}=\sum_{\alpha,k_y}\frac{v_{F\alpha}^{(k_y)}}{2\pi}\left[\left(\partial_x\phi_{\alpha}^{(k_y)}\right)^2+\left(\partial_x\theta_{\alpha}^{(k_y)}\right)^2\right].
\end{equation}
The first term in the Lagrangian can be broken into two, one of which does not involve the gauge field.  This term is the ``$ip\partial_{\tau}q$" term where we identify $q\leftrightarrow\phi$ and $\tilde{\rho}=(\partial_x\theta)/\pi\leftrightarrow p$; indeed $[\phi,\tilde{\rho}]=i$.  The other piece of the first term, which does involve the gauge field, is a descendent of the constraint that will end up fixing the $d_1$ and $d_2$ occupation numbers to one another at every position.  The rest of the Lagrangian is the Hamiltonian.

Taking now the full action given by ${\cal L}={\cal L}''_{\mathrm{gauge}}+{\cal L}_{\mathrm{ferm}}$, we can integrate out the gauge field $a_{\tau}$, which presumes the strong-coupling limit, and arrive at
\begin{equation}
\label{eqn:Lmassive}
{\cal L}'=\sum_{\alpha,k_y}\frac{i}{\pi}\left(\partial_x\theta_{\alpha}^{(k_y)}\right)\left(\partial_{\tau}\phi_{\alpha}^{(k_y)}\right)+M\theta_c^2+{\cal H}_{\mathrm{kin}},
\end{equation}
where
\begin{equation}
\label{eqn:thetac}
\theta_c\equiv\frac{1}{\sqrt{n+m}}\sum_{\alpha,k_y}e_{\alpha}\theta_{\alpha}^{(k_y)}
\end{equation}
with corresponding mass $M=M(N,n+m,h)$, and $n$ and $m$ refer to the labels of the DBL$[n,m]$.  The normalization factor is chosen in anticipation of performing a unitary transformation on the bosonic fields.  This overall charge mode, $\theta_c$, has been rendered massive by the constraint placed by the gauge field.  We can thus integrate out the $(\phi_c,\theta_c)$ mode effectively pinning $\theta_c$ to zero throughout the theory.  This condition implies that
\begin{equation}
\label{eqn:pinning}
\rho_{d_1}=\frac{1}{\pi}\sum_{k_y}\partial_x\theta_1^{(k_y)}=\frac{1}{\pi}\sum_{k_y}\partial_x\theta_2^{(k_y)}=\rho_{d_2},
\end{equation}
where the $k_y$ index in each sum varies over only the partially occupied bands for each flavor of parton.  Therefore, there will be no free gauge charge in the system (i.e., confinement), which is equivalent to saying that either a position will be unoccupied or it will have a hard-core boson.

We now have a theory of $m+n-1$ gapless 1D modes, and, after diagonalizing to find the normal modes, the effective Lagrangian can be written as follows:
\begin{equation}
\label{eqn:luttingerL}
{\cal L}_{\mathrm{eff}}=\sum_{\nu=1}^{n+m-1}\frac{i}{\pi}(\partial_x\theta_{\nu})(\partial_{\tau}\phi_{\nu})+{\cal H}_{\mathrm{eff}}\,,
\end{equation}
with
\begin{equation}
\label{eqn:luttingerH}
{\cal H}_{\mathrm{eff}}=\sum_{\nu=1}^{n+m-1}\frac{v_{\nu}}{2\pi}\left[g_{\nu}(\partial_x\phi_{\nu})^2+\frac{1}{g_{\nu}}(\partial_x\theta_{\nu})^2\right].
\end{equation}
The quantities $g_{\nu}$ are the Luttinger parameters and $v_{\nu}$ the dispersion velocities for each normal mode.  With these ingredients, plus the transformation matrices that diagonalize the Lagrangian leading to Eq.~(\ref{eqn:luttingerL}), one can compute any quantity of physical interest (ignoring four-fermion interactions).

Finally, it is worth noting that the density-density interactions are trivially bosonized using
\begin{equation}
\label{eqn:densitybosonized}
\rho_{bP}=\frac{1}{2\pi}(\partial_x\theta_b+P\partial_x\phi_b).
\end{equation}

\subsection{DBL[4,2]} \label{app:gt42}

We work on the four-leg ladder ($N=4$).  Motivated by the numerical VMC and DMRG results, we conclude that for the DBL$[4,2]$, the $d_1$ partons fill all four bands in such a way that pairs of neighboring bands have the same occupation; that is, there are two pairs of coincident bands.  The $d_2$ partons fill only two bands and these two are also coincident.  Thus, the most sensible choice of transverse boundary conditions for the partons is antiperiodic (this still results in the bosons obeying periodic boundary conditions) wherein the $\pm\pi/4$ bands are coincident as are the $\pm3\pi/4$ bands, consistent with symmetry of the lattice.  Hence, there are only three unique Fermi wave vectors, which we designate as follows: $k_{F11}\equiv k_{F1}^{(\pm\pi/4)}$, $k_{F13}\equiv k_{F1}^{(\pm3\pi/4)}$, and $k_{F2}\equiv k_{F2}^{(\pm\pi/4)}$ as well as the corresponding Fermi velocities $v_{F11}\equiv v_{F1}^{(\pm\pi/4)}$, $v_{F13}\equiv v_{F1}^{(\pm3\pi/4)}$, and $v_{F2}\equiv v_{F2}^{(\pm\pi/4)}$.  For boson density $\rho$, the Fermi wave vectors satisfy $k_{F11}+k_{F13}=2\pi\rho$ and $k_{F2}=2\pi\rho$.  In the mean-field treatment of the slave particle theory, $d_1$ and $d_2$ simply fill up energy bands about $k_x=0$; for free fermions with nearest-neighbor hopping, these band curves are given by cosine functions and hence we can sensibly assume that the band velocities can be calculated simply by evaluating the derivatives of these curves at the Fermi points.  There are only two distinct ratios of velocities in general:
\begin{align}
\label{eqn:42fermivels}
\frac{v_{F11}}{v_{F2}}&=\frac{\sin(k_{F11})}{\sin(k_{F2})}=\frac{\sin(k_{F11})}{\sin(2\pi\rho)},\\
\frac{v_{F13}}{v_{F2}}&=\frac{\sin(k_{F13})}{\sin(k_{F2})}=\frac{\sin(2\pi\rho-k_{F11})}{\sin(2\pi\rho)},
\end{align}
and hence, once one specifies the density $\rho$, there is only one input to the theory, $k_{F11}$ (ignoring all four-fermion interactions).  Since we were very successful in finding the DBL$[4,2]$ at $\rho=5/12$, we shall use this density here.  A thorough analysis of the scaling dimensions of various boson operators, density excitation operators, and four-fermion interactions can now be performed with respect to the DBL fixed point while varying the input parameter $k_{F11}$ from $\pi\rho$, which occurs when the four bands are equally filled (this limit corresponds to the most anisotropic Fermi seas for the partons and is expected for large $K$ and/or small $J_{\perp}$ in our model), to $2\pi\rho$, which occurs when the $\pm3\pi/4$ bands are no longer occupied for the $d_1$ partons (in this limit, both flavors of parton fill the same Fermi sea and the variational wave function is simply a determinant squared; the resulting phase has no $d$-wave character at all and is likely unstable to a superfluid \cite{Motrunich07_PRB_75_235116}).

This analysis confirms that the boson operators expected to have enhanced power law behavior (relative to the mean field) via the Amperean rule do indeed have scaling dimensions less than one for the range of $k_{F11}$ given above, which, for the single-boson correlator, results in exponents less than the mean-field value of two.  These operators, listed by the $y$ momentum they carry, are as follows:
\begin{align}
\label{eqn:bosonops}
q_y=0&:d_{1R}^{(\pm\pi/4)\dagger}d_{2L}^{(\pm\pi/4)\dagger}\,;\\
q_y=\pm\frac{\pi}{2}&:d_{1R}^{(\pm\pi/4)\dagger}d_{2L}^{(\mp\pi/4)\dagger}\,;\\
q_y=\pm\frac{\pi}{2}&:d_{1R}^{(\pm3\pi/4)\dagger}d_{2L}^{(\pm\pi/4)\dagger}\,;\\
q_y=\pi&:d_{1R}^{(\pm3\pi/4)\dagger}d_{2L}^{(\mp\pi/4)\dagger}.
\end{align}
We do not list here operators that can be obtained by exchanging $R\leftrightarrow L$.

Next, considering the density excitation operators, the bosonization analysis confirms that the scaling dimensions of the operators expected to have enhanced power laws are indeed less than one.  These operators, again listed by their $y$ momentum, are as follows:
\begin{align}
\label{eqn:densityops}
q_y=0&:d_{1R}^{(\pm\pi/4)\dagger}d_{1L}^{(\mp\pi/4)}\,;\\
q_y=0&:d_{1R}^{(\pm3\pi/4)\dagger}d_{1L}^{(\mp3\pi/4)}\,;\\
q_y=0&:d_{2R}^{(\pm\pi/4)\dagger}d_{2L}^{(\mp\pi/4)}\,;\\
q_y=\pm\frac{\pi}{2}&:d_{1R}^{(\pm\pi/4)\dagger}d_{1L}^{(\pm\pi/4)}\,;\\
q_y=\pm\frac{\pi}{2}&:d_{1R}^{(\pm3\pi/4)\dagger}d_{1L}^{(\mp\pi/4)}\,;\\
q_y=\pm\frac{\pi}{2}&:d_{1R}^{(\mp3\pi/4)\dagger}d_{1L}^{(\mp3\pi/4)}\,;\\
q_y=\pm\frac{\pi}{2}&:d_{2R}^{(\pm\pi/4)\dagger}d_{2L}^{(\pm\pi/4)}\,;\\
q_y=\pi&:d_{1R}^{(\pm3\pi/4)\dagger}d_{1L}^{(\pm\pi/4)}.
\end{align}
Again, we do not list the $q_x$ reversed partners of each of these terms.

We have also considered all possible four-fermion interactions allowed by the symmetries of the system (i.e., parton number conservation and crystal momentum conservation).  There are the density-density interactions of Eq.~(\ref{eqn:densityint}), which are strictly marginal for any value of $k_{F11}$ since their bosonized form [see Eq.~(\ref{eqn:densitybosonized})] is linear in the bosonic fields; therefore, density-density products only contribute to the quadratic terms in the theory.  There exist other terms, however, which yield exponentials of the bosonic fields and therefore interact beyond quadratic order.  The first class of such terms have the following form:
\begin{equation}
\label{eqn:4ferm1}
h_{\alpha\beta\gamma\delta}=d_{1R}^{(\alpha\pi/4)\dagger}d_{1L}^{(\beta\pi/4)\dagger}d_{1R}^{(\gamma3\pi/4)}d_{1L}^{(\delta3\pi/4)}+\hc\,,
\end{equation}
where $\alpha,\beta,\gamma,\delta=\pm1$ and where we must have $\alpha-\beta-3\gamma+3\delta=8n$ for $n$ an integer in order to conserve momentum.  There are thus three terms, $h_{++--}$, $h_{++++}$, and $h_{+--+}$, along with their $q_y$ reversed partners.  Ignoring all four-fermion interactions in the Lagrangian, we have performed an analysis of the scaling dimension of these terms and found that they are strictly irrelevant for the entire range of $k_{F11}$ except for at $\pi\rho$ (four equally filled bands; large $K$ and/or small $J_{\perp}$) where they are marginal.  Finally, one can form valid four-fermion terms from combinations of parton transfers on the same side of coincident bands:
\begin{equation}
\label{eqn:4ferm2}
h_{(\alpha,k_y,P),(\alpha',k_y',P')}=B_{\alpha,k_y,P}B_{\alpha',k_y',P'}+\hc\,,
\end{equation}
where
\begin{equation}
\label{eqn:4ferm2B}
B_{\alpha,k_y,P}=d_{\alpha P}^{(k_y)\dagger}d_{\alpha P}^{(-k_y)}.
\end{equation}
To conserve momentum, we must have $Pk_y+P'k_y'=\pi n$ with $n$ an integer.  Without doing any work, we would expect that these terms are not relevant since the $B_{\alpha,k_y,P}$ bilinears are precisely the suppressed density excitation operators that do not follow the Amperean rule.  We expect their scaling dimensions to thus be greater than or equal to one and therefore that the four-fermion products of such bilinears will have scaling dimensions greater than or equal to two.  Bosonization analysis indeed reveals that these terms are strictly marginal over the range of $k_{F11}$ studied.

While there are certainly several different phases that can be realized by the full phase space of quadratic bosonized models, we have showed convincingly that with only gauge projection and no residual forward scattering interactions, our parton model lies entirely within the DBL$[4,2]$ phase.  We have also analyzed the scaling dimension of the cosine terms of Eqs.~(\ref{eqn:4ferm1}) and (\ref{eqn:4ferm2}) while including various density-density interactions [see Eq.~(\ref{eqn:densityint})] in the Lagrangian.  For the interactions considered, e.g., attraction between bosons, we again found the cosine terms to be at worst marginal; however, we caution that this study was not exhaustive, and it is presumably possible to add interactions that will render the cosines relevant, especially in the decoupled-chains limit where Eq.~(\ref{eqn:4ferm1}) is already marginal in the pure gauge theory.  All in all, our analysis still strongly suggests that DBL$[4,2]$ is a stable quantum phase over a large range of reasonable parameters.  These results, combined with the encouraging agreement of the DMRG and VMC results presented in Sec.~\ref{sec:4leg_42}, suggest that the DBL$[4,2]$ is a stable phase, the qualitative features of which are faithfully manifested by our proposed model, Eq.~(\ref{eqn:fullmodel}), over a large region of the $K/J-J_{\perp}/J$ parameter space.

\subsection{DBL[3,0]} \label{app:gt30}

In this section we work on the three-leg ladder ($N=3$) at commensurate density $\rho=1/3$.  The qualitative agreement of the VMC and DMRG data suggests a DBL$[3,0]$ phase where the $d_1$ partons occupy all three bands, two of which are coincident, while the $d_2$ partons occupy fully a single band.  This structure suggests that periodic transverse boundary conditions for the partons are appropriate such that the $k_y=\pm2\pi/3$ bands are degenerate and it is the $k_y=0$ band that the $d_2$ partons fully occupy.  The lack of any partially filled $d_2$ bands implies that the addition of a $d_2$ parton is gapped, and the same must then be true for the boson in the strong-coupling limit where the partons are bound together.  This fact reveals itself in the featureless boson momentum distribution function obtained via DMRG (see Fig.~\ref{fig:3leg30_DMRG_VMCcf}, top panel).  Additionally, the single fully filled band enforces a condition of exactly one particle per rung on the ladder.  This constraint is exact in the VMC wave functions and even in the DMRG, the rung occupation number in the putative DBL$[3,0]$ region rarely differs from one.

The bosonization analysis of this phase is similar in many regards to that of the DBL$[4,2]$ above, but here there are no fields corresponding to the $d_2$ partons since there are no $d_2$ Fermi points.  This requires a careful encoding of the one-particle-per-rung constraint.  We denote the Fermi wave vectors in the three $d_1$ bands
as $k_{F0}\equiv k_{F1}^{(0)}$ and $k_{F}\equiv k_{F1}^{(\pm2\pi/3)}$ and the corresponding Fermi velocities as $v_{0}\equiv v_{F1}^{(0)}$ and $v_{\sigma}\equiv v_{F1}^{(\pm2\pi/3)}$.  At $\rho=1/3$, we have the condition, $k_{F0} + 2 k_{F} = \pi$.  If we again assume that the Fermi velocities are given simply by the slope of nearest-neighbor hopping band curves at the Fermi points, then there is only one distinct ratio of velocities given by
\begin{equation}
\label{eqn:30fermivels}
\frac{v_\sigma}{v_{0}}=\frac{\sin[(\pi-k_{F0})/2]}{\sin(k_{F0})},
\end{equation}
and so, ignoring residual, short-range interactions beyond the gauge field, the theory has only one input parameter, $k_{F0}$, which can be varied from $\pi/3$ where the three $d_1$ bands are equally occupied (i.e., the large $K$ and/or small $J_{\perp}$ limit) to $\pi$ where the state becomes a rung Mott phase with the $k_y=0$ band filled fully for both partons.  This setup allows us to compute various scaling dimensions numerically as we did for the DBL$[4,2]$ above.  However, the two-mode harmonic liquid theory for DBL$[3,0]$ is sufficiently simple that it can be treated analytically, as we will now explain.

We do not expect there to be any long-range boson correlations since the $d_2$ dispersion is gapped; furthermore, the fully filled $k_y=0$ band for the $d_2$ partons suppresses any density-density correlations at $k_y=0$.  We do, however, expect there to be singular density-density correlations at $k_y=\pm2\pi/3$.  Also, there are potentially relevant four-fermion interactions that must be considered in discussing the stability of the DBL$[3,0]$.

The gauge field construction for this phase follows the same approach as for the DBL$[4,2]$ with the gauge constraint requiring
that $d^\dagger_1 d_1 = d^\dagger_2 d_2$.   But since the $d_2$ fermion
fills a band, there is precisely one $d_2$ fermion per rung.  At longer wavelengths
down the ladder, the one-dimensional $d_2$ density does not fluctuate at all.
Within the gauge theory the effective constraint is thus $\sum_{k_y,P} d_{P}^{(k_y)\dagger} d_{P}^{(k_y)} = \sum_{k_y} \partial_x \theta^{(k_y)}/\pi \equiv \rho_A =  0$.   Here, $\rho_A$ is the total ``gauge charge" density.  Note we are dropping the subscript $\alpha$ on the partons and the bosonic fields since this theory only has $\alpha=1$.

The imaginary-time bosonized Lagrangian density is given by Eq.~(\ref{eqn:Lmassive}):
\begin{equation}
\label{L30}
{\cal L}=\sum_{k_y=0,+,-}\frac{i}{\pi}\left(\partial_x\theta^{(k_y)}\right)\left(\partial_{\tau}\phi^{(k_y)}\right)+M\theta_c^2+{\cal H}_{\mathrm{kin}}\,,
\end{equation}
where we have adopted the shorthand $\pm\leftrightarrow\pm2\pi/3$, $\theta_c$ is given by Eq.~(\ref{eqn:thetac}), and ${\cal H}_{\mathrm{kin}}$ by Eq.~(\ref{eqn:Hkin}).

To proceed, we make a canonical transformation:
\begin{align}
\theta_\sigma &\equiv \frac{1}{\sqrt{2}}\left(\theta^{(+)} - \theta^{(-)}\right), \\
\theta_\rho &\equiv \frac{1}{\sqrt{6}}\left(\theta^{(+)} + \theta^{(-)} - 2 \theta^{(0)}\right), \\
\theta_c &\equiv \frac{1}{\sqrt{3}}\left(\theta^{(+)} + \theta^{(-)} + \theta^{(0)}\right),
\label{eqn:U1transform}
\end{align}
and identical definitions for the $\phi^{(k_y)}$ fields.
These can be inverted as
\begin{align}
\label{eqn:inversion}
\theta^{(0)} &= - \sqrt{\frac{2}{3}}  \theta_\rho + \sqrt{\frac{1}{3}} \theta_c\, , \\
\theta^{(+)} &= \sqrt{\frac{1}{2}} \theta_\sigma + \sqrt{\frac{1}{6}} \theta_\rho + \sqrt{\frac{1}{3}} \theta_c\, , \\
\theta^{(-)} &= - \sqrt{\frac{1}{2}} \theta_\sigma +\sqrt{\frac{1}{6}} \theta_\rho + \sqrt{\frac{1}{3}} \theta_c\, .
\end{align}

We now integrate out the massive field $\theta_c$ to obtain
\begin{equation}
{\cal H}_{\mathrm{kin}} = \sum_{\mu = \rho,\sigma} \frac{v_{\mu}}{2\pi}
\left[ g_\mu(\partial_x \phi_{\mu})^2
       + \frac{1}{g_\mu} (\partial_x \theta_{\mu})^2 \right] ,
\label{eqn:30Hkin}
 \end{equation}
 with $g_\sigma = 1$,
 \begin{equation}
 \label{eqn:30grho}
 g_\rho = 3 \sqrt{\frac{v_\sigma v_0}{v_\sigma    v_0 + 2(v_0 + v_\sigma)^2}}\,,
 \end{equation}
  \begin{equation}
 v_\rho = \sqrt{\frac{v_\sigma v_0(v_\sigma+2v_0)}{ v_0 + 2 v_\sigma}}\,.
 \end{equation}
 As it stands, $g_\rho \le 1$, but residual short-range forward scattering interactions
 will renormalize both $g_\rho$ and $g_\sigma$ as well as the two velocities.
 Quadratic cross terms that couple the two sectors, such as $\partial_x \theta_\rho \partial_x \theta_\sigma$
 are precluded by $y \rightarrow -y$ symmetry.  Indeed, under this symmetry, $\theta^{(+)} \leftrightarrow \theta^{(-)}$ so that $\theta_\rho \rightarrow \theta_\rho$ is even and $\theta_\sigma \rightarrow - \theta_\sigma
$ is odd.  The cross term $\partial_x \theta_\rho \partial_x \theta_\sigma$ is odd, and not invariant
under this symmetry, thereby being forbidden.
Thus, the fixed point theory of the DBL$[3,0]$ phase is given by the Hamiltonian in Eq.~(\ref{eqn:30Hkin}), or in Lagrangian form by
\begin{equation}
\label{eqn:30Lagrangian}
{\cal L}'=\sum_{\mu=\rho,\sigma}\frac{i}{\pi}(\partial_x\theta_{\mu})(\partial_{\tau}\phi_{\mu})+{\cal H}_\mathrm{kin}\,,
\end{equation}
and consists of two gapless bosonic modes.

There are two allowed four-fermion interactions other than density-density type (already absorbed in the general Luttinger parameters $g_\rho$ and $g_\sigma$); they can be crudely viewed as ``Cooper channel'' interactions:
\begin{eqnarray}
{\cal H}_1 &=& w_1 \left[d_{R}^{(+)\dagger} d_{L}^{(+)\dagger} d_{L}^{(-)} d_{R}^{(-)} + {\rm H.c.} \right] \label{eqn:30Cooper} \\
&=& 2 w_1 \cos(2\phi^{(+)} - 2\phi^{(-)}) \\
&=& 2 w_1 \cos(2\sqrt{2}\phi_\sigma) ~,\\
{\cal H}_2 &=& w_2 \left[d_{R}^{(0)\dagger} d_{L}^{(0)\dagger} \sum_{a=+,-} d_{L}^{(a)} d_{R}^{(a)} + {\rm H.c.} \right] \label{eqn:30Cooper2} \\
&=& 2 w_2 \sum_{a=+,-} \cos(2\phi^{(a)} - 2\phi^{(0)}) \\
&=& 4 w_2 \cos(\sqrt{2}\phi_\sigma) \cos(\sqrt{6}\phi_\rho) ~.
\end{eqnarray}
By re-expressing these operators in terms of $\phi_\rho,\phi_\sigma,\theta_\rho,\theta_\sigma$ one can readily deduce their scaling dimensions: $\Delta_1 = 2/g_\sigma$, $\Delta_2 = 3/(2g_\rho) + 1/(2g_\sigma)$.  Provided these are irrelevant, the DBL$[3,0]$ phase is stable.  This requires $g_\sigma \leq 1$ for $\Delta_1$ and $g_\rho \leq 3/(4 - 1/g_\sigma)$ for $\Delta_2$.  The bare gauge theory that gave $g_\sigma=1$ and $g_\rho$ in Eq.~(\ref{eqn:30grho}) automatically satisfies these requirements.

In the DBL$[3,0]$ phase the boson operator $b=d_1d_2$ will be short-ranged because the $d_2$ fermion has a gap.  However, power law behavior is expected in the density-density correlator.  To examine this we consider various density bilinears that contribute to $b^\dagger b \sim d_1^\dagger d_1$.  We consider first the dominant ``$2k_F$" contributions at $k_y = \pm 2\pi/3$, listed by their momenta:
\begin{align}
(2k_F, 2\pi/3)&:  d_{R}^{(-)\dagger} d_{L}^{(-)} \sim -i e^{-2i\theta^{(-)}}\,;\\
(k_{F0} + k_F, 2\pi/3)&: d_{R}^{(+)\dagger} d_{L}^{(0)} \sim e^{i(\phi^{(0)} - \phi^{(+)})}e^{-i(\theta^{(0)}+\theta^{(+)})}\,,
\end{align}
which have scaling dimensions
\begin{align}
(2k_F, 2\pi/3)&: \frac{g_\rho}{6} + \frac{g_\sigma}{2}\,;\\
(k_{F0} + k_F, 2\pi/3)&: \frac{1}{8}\left(\frac{g_\rho}{3} + \frac{3}{g_\rho} \right) + \frac{1}{8}\left(g_\sigma + \frac{1}{g_\sigma}\right).
\end{align}
[Our writing here and below is somewhat schematic, e.g., we do not spell out all complex pre-factors and Klein factors as they do not affect the power-laws.  For the $(k_{F0} + k_F, 2\pi/3)$, there is another contribution at the same momentum, and a more precise expression would read
$d_{R}^{(+)\dagger} d_{L}^{(0)} + d_{R}^{(0)\dagger} d_{L}^{(+)} \sim -i \eta^{(+)} \eta^{(0)} e^{-i(\theta^{(0)}+\theta^{(+)})} \sin(\phi^{(0)} - \phi^{(+)})$.]

There are also ``$2k_F$" contributions at $k_y = 0$:
\begin{align}
(2k_{F0},0): d_{R}^{(0)\dagger} d_{L}^{(0)}& \sim e^{-2i\theta^{(0)}}=
e^{i2 \sqrt{2/3} \theta_\rho}\,; \\
(2k_{F},0): d_{R}^{(+)\dagger} d_{L}^{(-)}& \sim e^{i(\phi^{(-)}-\phi^{(+)})}e^{-i(\theta^{(-)}+\theta^{(+)})}\\
&\sim e^{-i\sqrt{2}\phi_\sigma}e^{-i\sqrt{2/3}\theta_\rho}\,,
\end{align}
with scaling dimensions $2 g_\rho/3$ and $g_\rho/6 + 1/(2g_\sigma)$, respectively (the second equalities follow from setting $\theta_c=0$ consistent with the pinning condition).  It would appear that this leads to a power law decay of density-density correlations at $k_y=0$, but due to the full band for the $d_2$ fermion we know that this cannot be correct.  However, there exist four-fermion operators that will contribute to the density and that have the same momenta and scaling dimensions as the bilinears above:
\begin{align}
d_{L}^{(+)\dagger} d_{R}^{(+)} d_{L}^{(-)\dagger} d_{R}^{(-)}& \sim e^{2i(\theta^{(+)}+\theta^{(-)})} =
e^{i2 \sqrt{2/3} \theta_\rho}\,,\\
d_{L}^{(0)\dagger} d_{R}^{(0)} d_{L}^{(+)\dagger} d_{R}^{(-)} &\sim e^{i(\phi^{(-)}-\phi^{(+)})}e^{i(2\theta^{(0)}+\theta^{(+)}+\theta^{(-)})} \nonumber\\&\sim e^{-i\sqrt{2}\phi_\sigma}e^{-i\sqrt{2/3}\theta_\rho}.
\end{align}
The presence of these operators follows
due to the existence of a six-fermion umklapp term wherein a fermion in each of the three bands is
backscattered, $\prod_{a=0,+,-} d_{L}^{(a)\dagger} d_{R}^{(a)} \sim e^{2i \sum_{a=0,+,-} \theta^{(a)}} \sim e^{i 2 \sqrt{3} \theta_c} \sim 1$.   In a more microscopic implementation of the gauge constraint
wherein the $d_1$ fermion is strictly enslaved to the $d_2$ fermion, one would anticipate
that these two operators exactly cancel one another when contributing to the microscopic density.

Having considered the potential destabilizing interactions and determined that they are irrelevant when short-ranged density-density interactions are ignored, we can conclude that the DBL$[3,0]$ phase is potentially stable.  In light of the DMRG/VMC correspondence shown in Sec.~\ref{sec:3leg_30}, it is evident that this phase is likely the ground state of our model, Eq.~(\ref{eqn:fullmodel}), over a large region of the phase diagram.

Let us briefly consider what happens when both interactions in Eqs.~(\ref{eqn:30Cooper}) and (\ref{eqn:30Cooper2}) become relevant thus pinning the fields $\phi_\rho$ and $\phi_\sigma$.  For simplicity, let us assume that Eqs.~(\ref{eqn:30Cooper}) and (\ref{eqn:30Cooper2}) represent interactions when they are already $O(1)$ after some initial flows, and we now need to minimize ${\cal H}_1 + {\cal H}_2$ semiclassically.  The resulting states depend on the signs of the couplings $w_1$ and $w_2$, and we consider different cases in turn.

When $w_1 < 0$, we can simultaneously minimize ${\cal H}_1$ and ${\cal H}_2$.  For either sign of $w_2$, there is a unique physical state (analyzed as in Ref.~\onlinecite{Lin98_PRB_58_1794}, Sec.~IV.E.1.): $\phi^{(+)} = \phi^{(-)} = \phi^{(0)} + [1 + {\rm sign}(w_2)] \pi/4$.  We did not find any local physical observable that would obtain an expectation value for such pinning of phases for either sign of $w_2$, and hence these states are good candidates for the featureless rung Mott insulator phase.  At this point, we do not know whether the states obtained for $w_2>0$ or $w_2<0$ are qualitatively distinct and cannot be connected by any path; if they are distinct, then we do not know how to interpret the featureless state other than rung Mott insulator.  It could also be that the two cases are only quantitatively distinct and connect to the same rung Mott insulator picture.

When $w_1 > 0$ and sufficiently large, $w_1 > |w_2|/2$, the energy is minimized at $\phi^{(+)} = \phi^{(-)} \pm {\rm acos}(-w_2/2w_1)$, $\phi^{(0)} = (\phi^{(+)} + \phi^{(-)})/2$.  For any $w_2$, there are two physically distinct solutions, and now we have an ``order parameter'' $\sin(2\phi^{(+)} - 2\phi^{(-)})$ that obtains an expectation value and takes opposite signs for the two physically distinct solutions.  The order parameter respects lattice translation and inversion symmetries, but changes sign under both time reversal and mirror symmetry.  A possible state with similar symmetries is a chiral state with spontaneously generated current circulation, which can be realized, e.g., by introducing uniform flux in the $d_1$ hopping problem.  An operator with similar transformation properties as the current circulation is $\chi({\bf r}) = J_b^{\hy}({\bf r}) - J_b^{\hy}({\bf r} + \hat{x})$, whose correlators can be reduced to current-current correlators,
$\langle \chi({\bf r}) \chi({\bf r}') \rangle = C_b^{\hy,\hy}({\bf r}, {\bf r}') + C_b^{\hy,\hy}({\bf r}+\hat{x}, {\bf r}'+\hat{x}) - C_b^{\hy,\hy}({\bf r}, {\bf r}'+\hat{x}) - C_b^{\hy,\hy}({\bf r}+\hat{x}, {\bf r}')$.  Since we have not observed long-range order in current-current correlations in any phase at $\rho=1/3$, we conclude that this phase is not realized in the ring model.
Note that again we do not know whether the different signs of $w_2$ produce qualitatively distinct phases. 

\section{Three-leg ladder at $\rho\neq1/3$} \label{app:3legIncomm}

\begin{figure}[t]
\centerline{\includegraphics[width=\columnwidth]{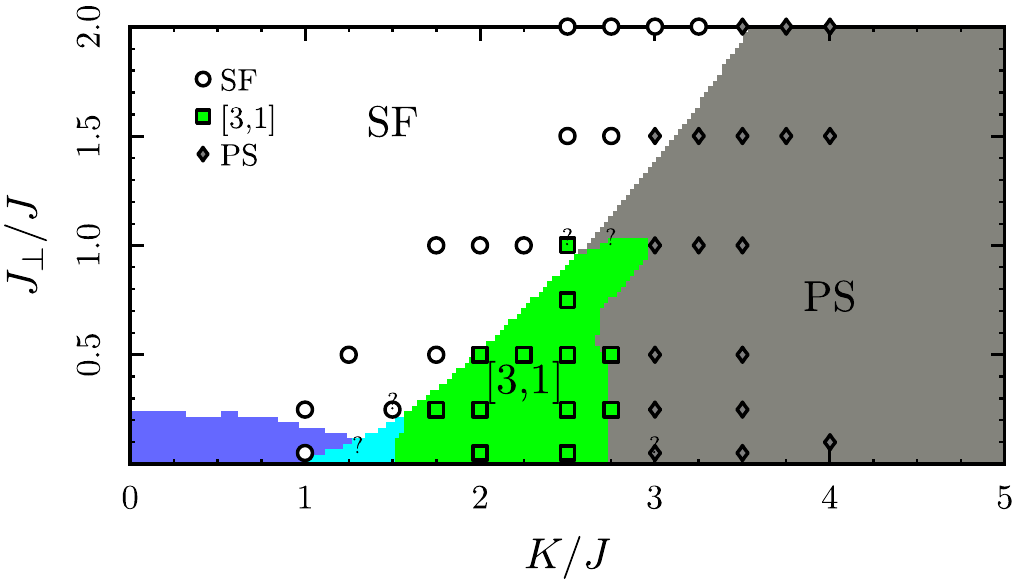}}
\caption{(color online).  Phase diagram of the three-leg system at boson density $\rho=1/4$, using a $24\times 3$ system with $N_b=18$.   The colored regions are delineated using VMC data.  Phase separation is determined with VMC methods by means of a Maxwell construction.  Commensurate densities tend to be more stable for large $K$ and so phase separation for this system generally consists of a region of zero density and a region at $1/3$ density.  DMRG points for the superfluid, DBL$[3,1]$, and phase separation are indicated by white circles, green squares, and gray diamonds, respectively.  Our calculations indicate that the DBL$[3,1]$ phase is never stabilized in the system with isotropic hopping ($J_\perp/J=1$).}
\label{fig:3leg31PD}
\end{figure}

\begin{figure}[t]
\centerline{\subfigure{\includegraphics[width=\corrScale\columnwidth]{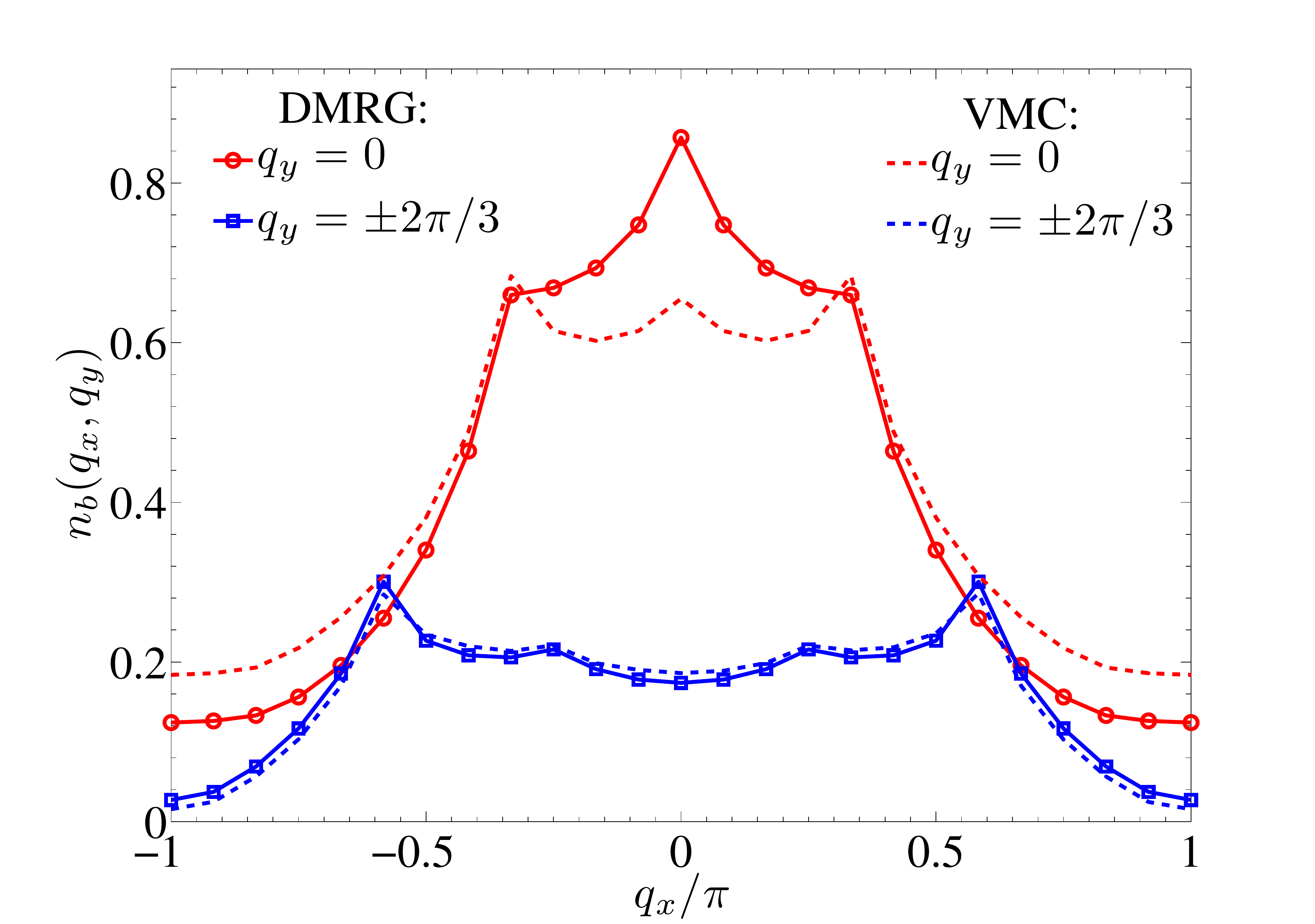}}}
\centerline{\subfigure{\includegraphics[width=\corrScale\columnwidth]{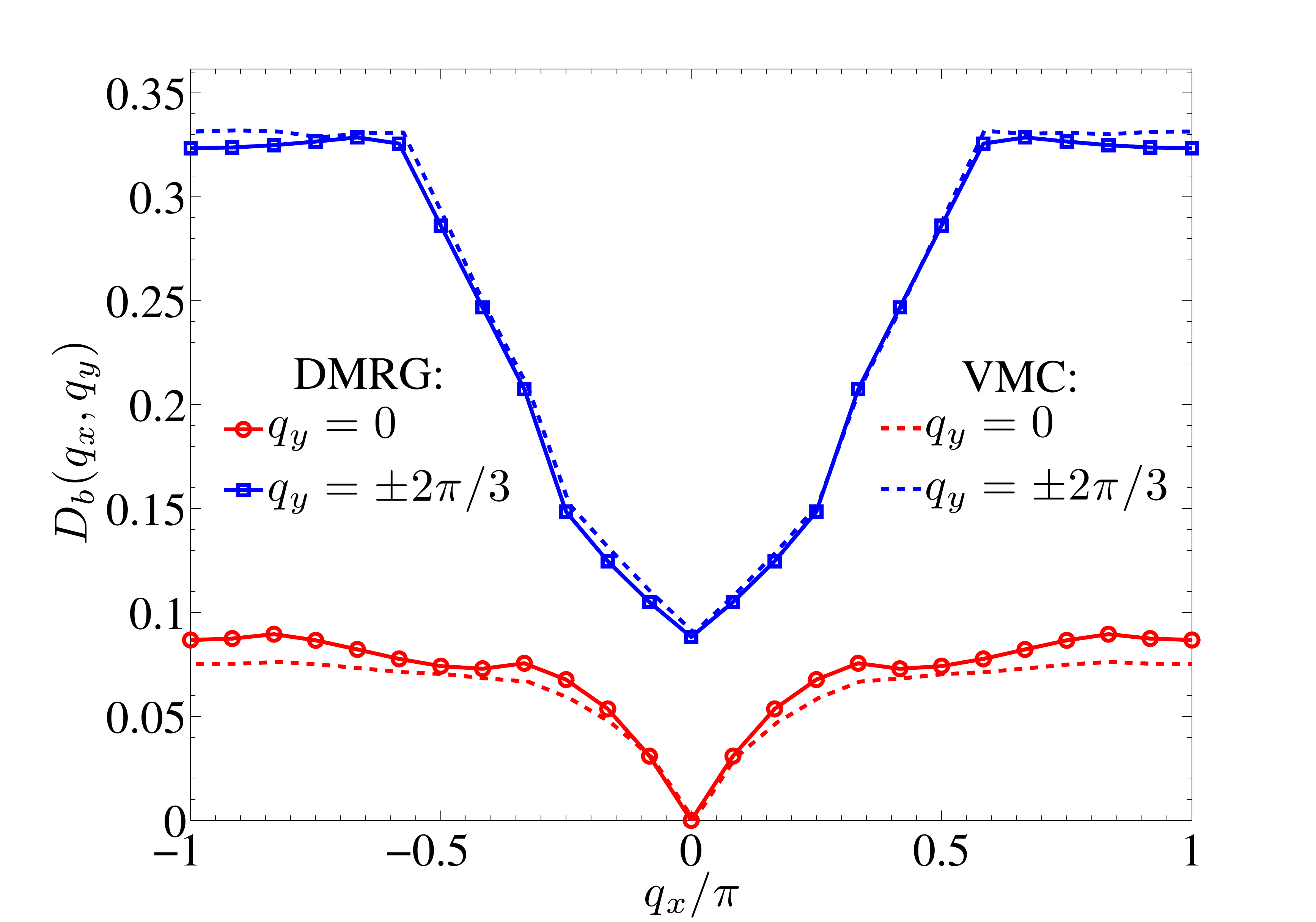}}}
\caption{(color online).  Boson momentum distribution function (top panel) and density-density structure factor (bottom panel) for a typical point in the DBL$[3,1]$ phase.  The parameters are $J_\perp/J=0.75$ and $K/J=2.5$ on a system of size $24\times3$ with $N_b=18$ ($\rho=1/4$).  DMRG data are joined with solid curves while the VMC data are joined with dashed curves.}
\label{fig:3leg31_DMRG_VMCcf}
\end{figure}

In this paper, we have focused on the novel DBL$[3,0]$ phase for the three-leg ladder at commensurate density $\rho=1/3$.  We did explore, however, the non-commensurate densities on the three-leg ladder, and we shall summarize the results of those studies in this appendix.

\subsection{DBL[3,1] at $\rho=1/4<1/3$} \label{app:3leg_31}

For densities smaller than the commensurate value, our data suggest that a stable DBL$[3,1]$ phase exists over a reasonable portion of the phase diagram (see Fig.~\ref{fig:3leg31PD}).  In particular, we studied $\rho=1/4$ where the energy-optimized VMC state shows excellent qualitative agreement with the DMRG results.  In the top panel of Fig.~\ref{fig:3leg31_DMRG_VMCcf}, we show the comparison of the two data sets for the boson momentum distribution function while the bottom panel shows the density-density structure factor; the overall agreement is quite striking.  The results are for a $24\times3$ system with 18 bosons.  The DBL$[3,1]$ state is constructed as follows: the $d_1$ partons partially fill all three bands with ten orbitals occupied for the $k_y=0$ band and four orbitals occupied for each of the $k_y=\pm2\pi/3$ bands; meanwhile, the $d_2$ partons only fill the $k_y=0$ band.  The structure factors show enhanced features at several wave vectors as predicted by the corresponding gauge theory for this phase.  Also, an analysis of the entanglement entropy with DMRG indeed reveals an effective central charge $c\simeq3$, as we expect from the gauge theory which leads to a theory of $c=3=3+1-1$ 1D gapless modes.

We do not give more attention to this phase since it is a rather straightforward extension of the DBL$[2,1]$ phase on the two-leg ladder that only exists for densities less than $1/N$, where $N$ is the number of legs.  The DBL$[3,1]$ on three legs, like the DBL$[2,1]$ on two legs, has exactly the same number of gapless modes as the number of legs and the $d_2$ determinant in the model wave function can be thought of as a Jordan-Wigner string, which is truly 1D in nature.  Furthermore, for $N=4$ at $\rho<1/4$, we find that the system phase separates right out of the superfluid and no such DBL$[4,1]$ state is observed.  Therefore, the DBL$[3,1]$ is not extensible to 2D and is of limited interest.  Nonetheless, its appearance is well supported by the underlying gauge theory giving further credence to our theoretical approach.

\subsection{Lack of DBL$[n,m]$ phases for $\rho>1/3$} \label{app:3leg_noBMs}

Here we summarize what we know about the situation for $1/3<\rho\leq1/2$ on the three-leg ladder where we had hoped to find a DBL$[3,2]$ phase with one more gapless mode than the number of legs.  In this putative phase, the $d_1$ partons partially fill all three bands $k_y=0,\pm 2\pi/3$, whereas the $d_2$ partons fully fill the $k_y=0$ band and partially fill the $k_y=\pm 2\pi/3$ bands.  Such a phase would be extensible to 2D and constitute substantive progress in our DBL search.  However, we performed an extensive study of the three-leg system at boson densities $\rho=1/2$ and $5/12$, and no such phase was found.  In fact, the VMC calculations do not even favor a DBL$[3,2]$ in any region of the phase diagram.  Instead, with the VMC, we find a state in which the $d_1$ partons do partially fill all three bands, while the $d_2$ partons fully fill the $k_y=0$ band but partially fill only one of the remaining $k_y=\pm 2\pi/3$ bands; we denote this state DBL$[3,1]^\dagger$ (with the dagger to distinguish it from the DBL$[3,1]$ discussed above in Appendix~\ref{app:3leg_31}).  Due to the population imbalance of the $k_y=\pm 2\pi/3$ bands, this wave function breaks time reversal symmetry and has a finite boson current in the $\hy$ direction:  $\langle J_b^{\hy}\rangle\neq 0$ [see Eq.~(\ref{eqn:current})] .  This prompted us to scan the phase diagram with DMRG, considering current-current correlations [see Eq.~(\ref{eqn:cccorr})] in addition to the single-boson Green's function and density-density correlation function.

\begin{figure}
\centerline{\subfigure{\includegraphics[width=\corrScale\columnwidth]{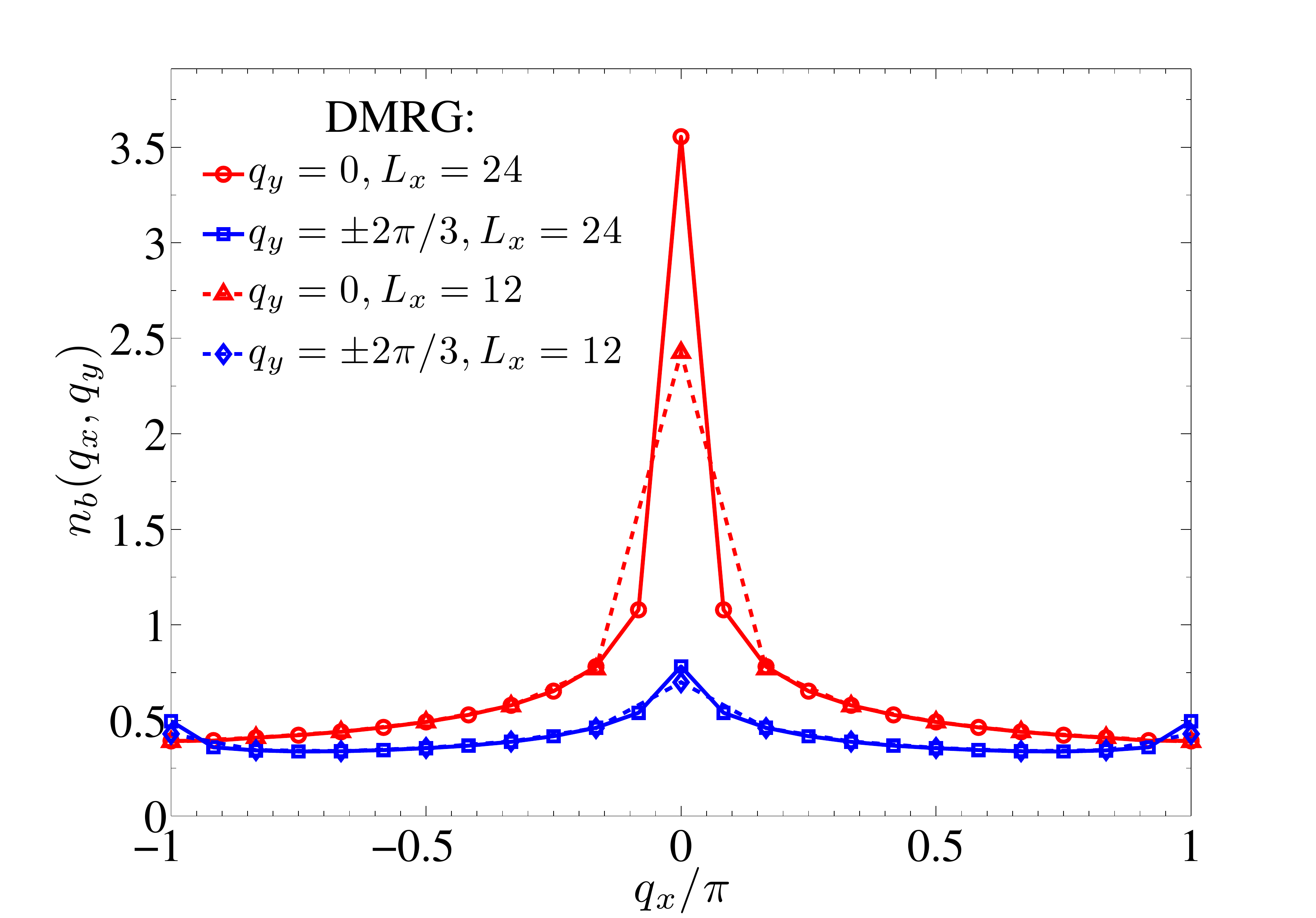}}}
\centerline{\subfigure{\includegraphics[width=\corrScale\columnwidth]{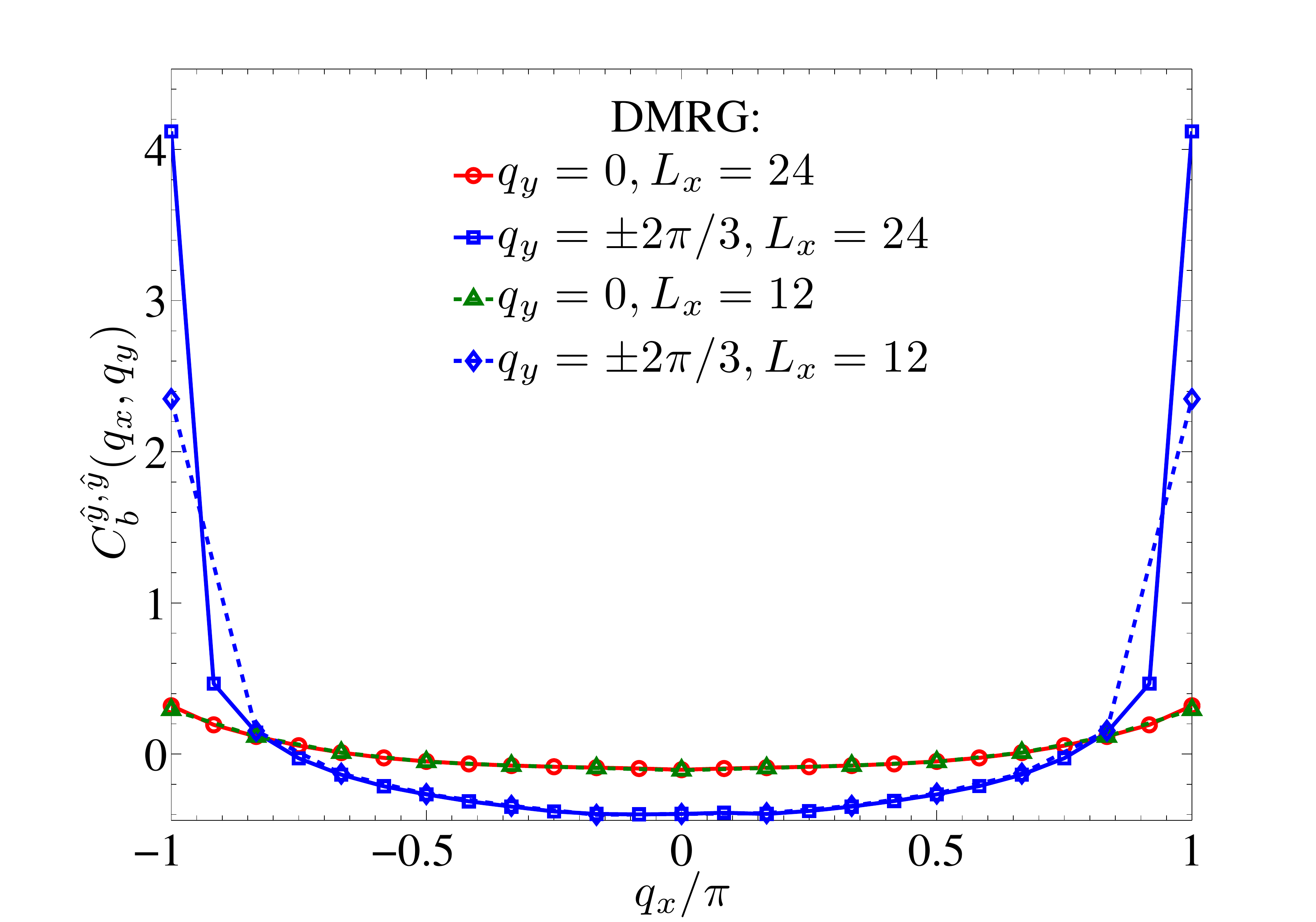}}}
\centerline{\subfigure{\includegraphics[width=\columnwidth]{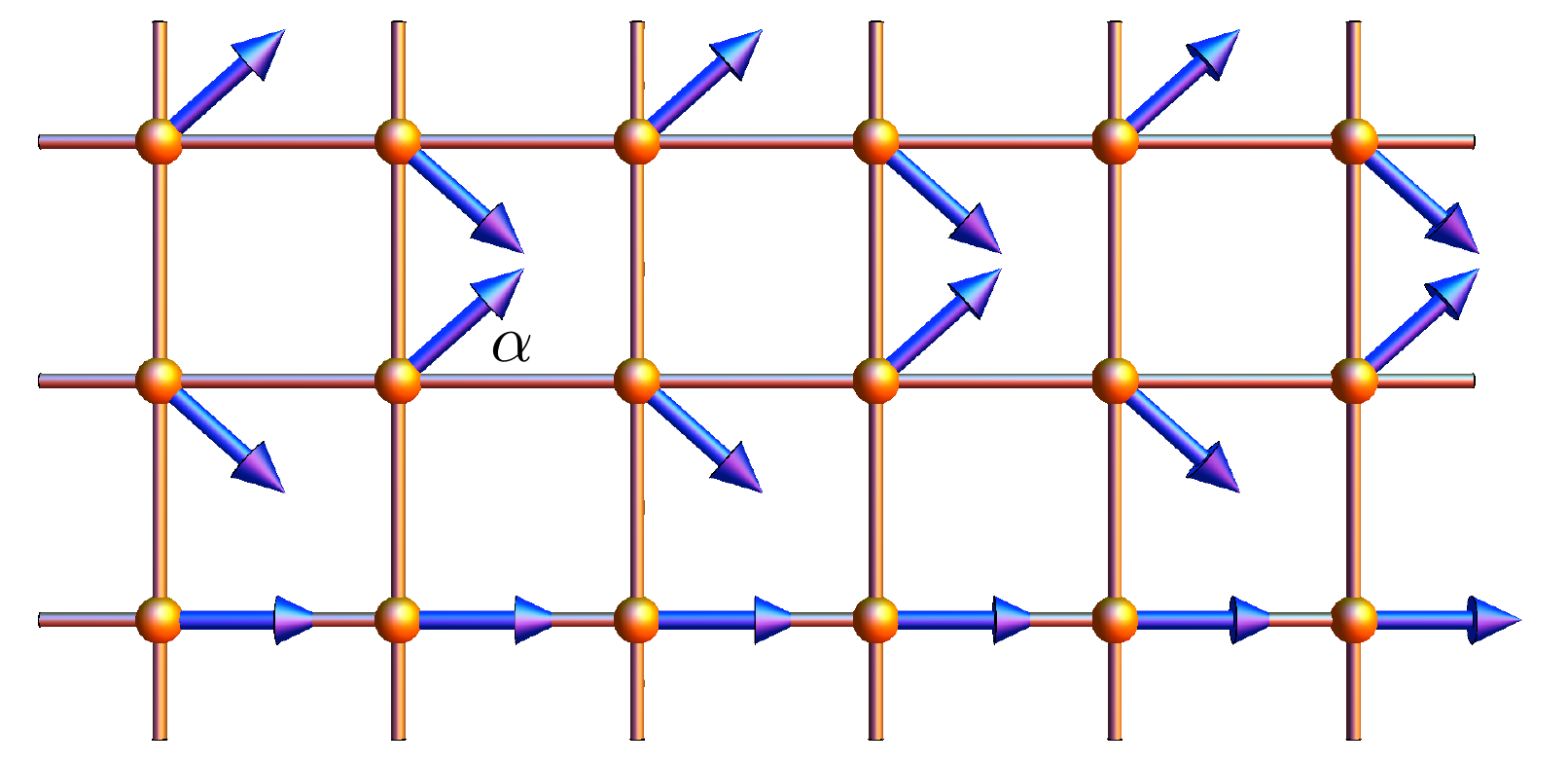}}}
\caption{(color online).  Boson momentum distribution function (top panel) and $\hy$-$\hy$ current-current structure factor (middle panel) for a typical point in the putative ``bond-chiral superfluid'' (BCSF) region at $\rho=1/2$ on the three-leg ladder.  The parameters are $J_\perp/J=1$ and $K/J=3$ on two system sizes:  $L_x=24,12$.  As explained in the text, the features appearing in these correlations can be qualitatively rationalized by a classical solution, $b_{\rvec}=\sqrt{\rho}\exp{(i\phi_\rvec)}$, with phases given by Eq.~(\ref{eqn:classicalsoln}).  In the bottom panel, we display this solution by showing a vector $(\cos\phi_{\rvec},\sin\phi_{\rvec})$ at each site.  Since true long-range order of the boson phase is prohibited for our three-leg ladder, this is a potentially appropriate description only on short length scales.  Note, however, that the current-current correlations can still be truly long ranged.}
\label{fig:3legBCSF}
\end{figure}

We first focus on density $\rho=1/2$, which is not commensurate on the three-leg ladder but, by particle-hole symmetry, is the largest possible boson density in our system.  For simplicity, we also specialize our discussion here to $J_\perp/J=1$.  While the VMC favors a DBL$[3,1]^\dagger$ for intermediate $K$ (between a superfluid at small $K$ and phase separation at large $K$), we were able to rule out this phase with the DMRG.  The state DBL$[3,1]^\dagger$ would have long-range, non-oscillatory $\hy$-$\hy$ current-current correlations, i.e., $C_b^{\hy,\hy}[\rvec=(x,0),\rvec'=(0,0)]\rightarrow \langle J_b^{\hy}(x,0)\rangle\langle J_b^{\hy}(0,0)\rangle\neq 0$ as $x\rightarrow\infty$, and a corresponding Bragg peak in $C_b^{\hy,\hy}(\qvec)$ at $\qvec=0$.  Interestingly, the DMRG measurement of $C_b^{\hy,\hy}(\qvec)$ for $K/J\gtrsim 2$ shows clear Bragg peaks, not at $\qvec=0$, but at wave vectors on the corners of the Brillioun zone of our three-leg ladder: $\qvec=(\pi,\pm 2\pi/3)$.  There is also a weaker feature at $\qvec=(\pi,0)$.  On the other hand, the boson momentum distribution function $n_b(\qvec)$ shows a distinct, superfluid-like singularity at $\qvec=0$, as well as weaker features at $\qvec=(0,\pm 2\pi/3),(\pi, \pm 2\pi/3)$.  A set of DMRG measurements of $n_b(\qvec)$ and $C_b^{\hy,\hy}(\qvec)$ for a characteristic point ($J_\perp/J=1,K/J=3$) exhibiting these features is shown in Fig.~\ref{fig:3legBCSF}.  The density-density structure factor (data not shown) is rather unremarkable:  it shows (1) no sign of charge ordering and (2) $|q_x|$ behavior around $q_x=0$ at $q_y=0$ clearly indicating a compressible phase.  Finally, for very large $K$, e.g., $K/J=10$, we find that the three-leg half-filled system phase separates into $\rho=1/3$ and $\rho=2/3$ insulating regions.

Although we have not attempted to understand this phase as an instability out of either DBL$[3,2]$ or DBL$[3,1]^\dagger$, it does bear some resemblance to the phase, denoted ``bond-chiral superfluid'' (BCSF), found recently \cite{Schaffer09_PRB_80_014503} in a linear spin-wave treatment of the 2D $J$-$K$ model [see Eq.~(\ref{eqn:model})]; this 2D phase is characterized by the coexistence of superfluidity with static order in bond chirality (boson rung currents) at wave vector $\qvec=(\pi,\pi)$.  To make this connection more concrete, we have performed a simple classical analysis of our model, Eq.~(\ref{eqn:fullmodel}), on the three-leg ladder, and we will now describe the results of this study.

We first replace the boson operator with a classical $c$-number, i.e., $b_{\rvec}\rightarrow \sqrt{\rho}\exp(i\phi_{\rvec})$, to obtain the classical energy:
\begin{align}
\label{eqn:classicalmodel}
\mathcal{H} =&~ \mathcal{H}_\mathrm{hop} + \mathcal{H}_K, \\
\mathcal{H}_\mathrm{hop} =& -J\sum_{\rvec} 2\rho \cos{(\phi_{\rvec}-\phi_{\rvec+\hx})} \nonumber \\ 
& -J_\perp \sum_{\rvec} 2\rho \cos{(\phi_{\rvec}-\phi_{\rvec+\hy})}, \\
\mathcal{H}_K =&~ K \sum_{\rvec}2\rho^2\cos{(\phi_{\rvec}-\phi_{\rvec+\hx}+\phi_{\rvec+\hx+\hy}-\phi_{\rvec+\hy})}.
\end{align}
Formally, this corresponds to relaxing the hard-core constraint and computing the energy $\langle H\rangle=\mathcal{H}$ with respect to a product of local coherent states with amplitudes $\sqrt{\rho}\exp(i\phi_{\rvec})$.  Strictly speaking, such a state cannot actually exist in our quasi-1D setting because it represents a superfluid with true long-range order.  However, as we describe next, on short scales a classical state with a particular phase pattern $\phi_\rvec$ can qualitatively describe much of the DMRG data in Fig.~\ref{fig:3legBCSF}, while the actual three-leg quantum phase would correspond to a slowly varying global rotation of the phase pattern on long scales.

Numerical analysis of Eq.~(\ref{eqn:classicalmodel}) on small periodic three-leg systems indicates that for substantial $K$ the classical ground state has an $\rvec$-dependent phase pattern, $\phi_{\rvec=(x,y)}=\phi_y(x)$ (where $y=1,2,3$), of the form
\begin{equation}
\label{eqn:classicalsoln}
\phi_1(x) = 0, ~ \phi_{2,3}(x) = \pm(-1)^x\alpha\,.
\end{equation}
We depict this solution in the bottom panel of Fig.~\ref{fig:3legBCSF} by drawing at each site a vector $(\cos\phi_{\rvec},\sin\phi_{\rvec})$.  This classical state has finite boson currents $J_b^{\hat{\mu}}(\rvec)=-2\rho\sin{(\phi_{\rvec}-\phi_{\rvec+\hat\mu})}$ and thus long-range order in the current-current correlator $C_b^{\hy,\hy}(\rvec,\rvec')$.  The structure factor $C_b^{\hy,\hy}(\qvec)$ has Bragg peaks, i.e., $C_b^{\hy,\hy}(\qvec)/L_x=\gamma$ with $\gamma$ a constant, at momenta $\qvec=(\pi,\pm2\pi/3),(\pi,0)$ with amplitudes $\gamma=\frac{4}{3} \rho^2 (\sin\alpha + \sin2 \alpha )^2,\,\frac{64}{3} \rho^2 \sin^4{\frac{\alpha}{2}} \sin^2\alpha$, respectively.  To get a very rough idea of the numbers involved, Eq.~(\ref{eqn:classicalmodel}) is minimized at $\rho=1/2$ and $J_\perp/J=1$, $K/J=3$ by Eq.~(\ref{eqn:classicalsoln}) with $\alpha=0.73$, which gives Bragg peaks at $\qvec=(\pi,\pm2\pi/3),(\pi,0)$ with respective amplitudes $\gamma=0.92,0.04$.  This is in qualitative agreement with the apparent large Bragg peak in the DMRG data at $(\pi,\pm2\pi/3)$ but only a very weak feature at $(\pi,0)$ (see Fig.~\ref{fig:3legBCSF}, middle panel).  (Note that the ratio $K/J$ is even somewhat arbitrary in this analysis since it is affected by the normalization of the boson field, which we take as $b^\dagger_\rvec b_\rvec=\rho$, cf.~Ref.~\onlinecite{Schaffer09_PRB_80_014503} which formally takes $b^\dagger_\rvec b_\rvec=1/4=\rho/2$.)  The classical solution also has a Bragg peak in the $\hx$-$\hx$ current-current structure factor $C_b^{\hx,\hx}(\qvec)$ at $\qvec=(\pi,\pm2\pi/3)$ with amplitude $\gamma=4\rho^2\sin^2{2\alpha}$.  We have measured $\hx$-$\hx$ current-current correlations with the DMRG (data not shown), and although they appear weaker than the corresponding $\hy$-$\hy$ correlations, we still cannot rule out a Bragg peak at $(\pi,\pm2\pi/3)$.  Finally, one can compute the boson momentum distribution within this naive framework.  This calculation predicts Bose condensation at wave vectors $\qvec=(0,0),(0,\pm2\pi/3),(\pi,\pm2\pi/3)$, which are interestingly the same wave vectors at which we see features in the DMRG.  Specifically, at these wave vectors, $n_b(\qvec)/L_x=\frac{1}{3}\rho(1+2 \cos\alpha)^2,\,\frac{4}{3}\rho\sin^4\frac{\alpha }{2},\,\rho\sin^2\alpha$\,, respectively.  To again get a feel for the numbers, these three amplitudes evaluated at $\rho=1/2$ and $\alpha=0.73$ give $1.03,0.01,0.22$.  Again, this simple classical analysis agrees fairly well qualitatively with the DMRG, which indeed indicates possible quasi-condensation at momenta $\qvec=(0,0),(0,\pm2\pi/3),(\pi,\pm2\pi/3)$ (see Fig.~\ref{fig:3legBCSF}, top panel).

The fact that such an exceedingly crude classical analysis qualitatively reproduces much of the DMRG data is quite remarkable.  However, we caution that the above picture is likely not complete and could still be qualitatively wrong.  Indeed, more work can certainly be done to characterize this unusual quantum phase.  For example, we have been unable to extract a meaningful central charge with the DMRG data, which points to $c\simeq2$-$3$ for a $24\times3$ system, while the quasi-1D BCSF phase described above should lead to a Luttinger liquid with $c=1$ gapless mode.  It may be that the DMRG entanglement entropy is anomalously large due to the large ground state degeneracy in the putative BCSF phase, and $c=1$ scaling only emerges at much longer length scales.

Upon doping the half-filled three-leg system to boson density $\rho=5/12$, the putative BCSF phase discussed above appears to persist, at least for sizable $J_\perp$ and intermediate $K$, e.g., $J_\perp/J=1$ and $2\lesssim K/J \lesssim 3.5$.  For smaller values of $J_\perp$, e.g., $=J_\perp/J=0.1$, we rather remarkably find DMRG evidence for the DBL$[3,1]^\dagger$ state in which the $d_1$ partons equally fill the three bands.  In this case, there is a conserved number of  bosons in each chain and thus zero rung current in the projected wave function:  $\langle J_b^{\hy}\rangle=0$.  Loosely speaking, the small value of $J_\perp$ has a tendency to suppress rung currents and seems to stabilize the decoupled chains limit of so-called DBL$[3,1]^\dagger$, though this is the only incarnation of this phase that we have been able to realize in the DMRG.  Also, DBL$[3,1]^\dagger$ does not appear to exist at $\rho=1/2$, even at small $J_\perp$, which is at least somewhat surprising.  Because the DBL$[3,1]^\dagger$ is obviously very far removed from 2D, we do not pursue understanding it any further.

In conclusion, the identified putative three-leg BCSF phase (see Fig.~\ref{fig:3legBCSF}) is itself rather interesting and worthy of further investigation.  We stress that we found no evidence for such a phase on the four-leg ladder at any density; e.g., we found no evidence of long-range order in current-current correlations.  In that case, our DBL$[4,2]$ is stabilized for densities slightly below $\rho=1/2$ (e.g., $\rho=5/12$), while at $\rho=1/2$, the conventional $\qvec=0$ superfluid was the only compressible phase identified.  Thus, exactly how all of these $\rho>1/3$ three-leg results and our $\rho>1/4$ four-leg results (see Sec.~\ref{sec:4leg}) extrapolate to 2D, if at all, is not presently clear.  In Sec.~\ref{sec:4leg}, we presented strong evidence in support of the existence and stability of a very 2D-like DBL phase on four legs, namely, the so-called DBL$[4,2]$; however, as we have discussed in this appendix, an analogous phase does not appear to exist for the three-leg system.  Generally speaking, we do believe odd-leg systems are somewhat pathological when approaching 2D in this context because they are not bipartite, thus putting the half-filled system on qualitatively different footing than what we would expect in 2D.  Still, to further investigate this seemingly strong even-odd effect, in addition to going to more legs (which is becoming exceedingly difficult at this point), it may be insightful to modify the transverse ($\hy$ direction) boundary condition for the boson model [Eq.~(\ref{eqn:fullmodel})].  In this work, we have only considered periodic transverse boundary conditions in both the three- and four-leg systems.  However, the structure of the DBL$[4,2]$ identified on four legs suggests that an anti-periodic boundary condition may be desirable to stabilize the sought-after DBL$[3,2]$ for $\rho>1/3$ on three legs.  Namely, because the structure of the stable four-leg DBL$[4,2]$ relies on anti-periodic transverse boundary conditions for both the $d_1$ and $d_2$ partons, a more natural three-leg DBL$[3,2]$ configuration than that discussed at the beginning of this appendix might consist of a periodic boundary condition for $d_1$ and an anti-periodic boundary condition for $d_2$.  This configuration is only possible with an anti-periodic boundary condition for the boson system.  We leave such further exploration for future work.

These effects illustrate the degree of difficulty encountered when trying to extrapolate results from quasi-1D ladder studies to make definitive statements about 2D physics, especially when trying to understand a phase as complicated as the DBL in a nontrivial model such as the $J$-$K$ model.  However, we believe our four-leg results of Sec.~\ref{sec:4leg_42} are more representative of the 2D system than the three-leg results summarized above in this appendix:  firstly, in contrast to the three-leg ladder, the four-leg ladder is a bipartite lattice (also, four is greater than three), and secondly, fully periodic boundary conditions are the most natural choice for approaching full 2D.

\section{Two-leg ladder at $\rho=1/2$} \label{app:2legHalfFilling}

We now address the situation on the two-leg ladder at half-filling ($\rho=1/2$).  In this system, it is in principle possible for a gapless Mott insulating phase, i.e., a DBL$[2,0]$ in our naming convention, to exist for large $K$.  A nontrivial solvable limit is when $J=0$, in which the number of bosons in each rung is conserved and is exactly one for $\rho=1/2$. In this regime, we can map the model exactly to the spin-1/2 XY chain with an in-plane magnetic field (identifying a boson on the upper chain as spin up and a boson on the lower chain as spin down):
\begin{equation}
H_{\rm XY} = K \sum_i(\sigma_i^+\sigma_{i+1}^- + \hc) - J_\perp\sum_i \sigma^x_i,
\end{equation}
where $\sigma^{x,y,z}_i$ are the Pauli matrices for the pseudospin on rung $i$. When $J_\perp=0$, the XY model is solved by a Jordan-Wigner transformation and has one 1D gapless mode corresponding to the free fermion. $J_\perp$ is relevant at this XY fixed point and immediately causes the formation of a canted magnet (uniform moment along $x$ and staggered moment along $y$) before entering the trivial pseudospin paramagnet (staggered moment along $y$ goes to zero) at large $J_\perp$. \cite{Dmitriev02_PRB_65_172409} In the two-leg ladder boson language, this corresponds to a staggered rung current state and the featureless rung Mott insulator, respectively; the transition between these two gapped phases is continuous and is of the Ising universality class.  The phase diagram at $J=0$ can be logically extended into the entire $K/J-J_\perp/J$ plane without the introduction of an exotic GMI phase.  Indeed, we have verified this with DMRG and found no evidence for a GMI phase at nonzero $J_\perp$.  In addition to the staggered rung current phase mentioned above, a $(\pi,\pi)$ CDW appears at intermediate $K/J$ and small $J_\perp/J$.  Remarkably, a bosonization analysis of the putative two-leg GMI phase predicts an allowed instability toward either $(\pi,\pi)$ static order in the density or $\pi$ static order in the rung currents, both of which we actually observe in the DMRG.

Finally, we note that two very recent works \cite{Carrasquilla11_PRB_83_245101, Crepin11_PRB_84_054517} have considered in detail hard-core bosons hopping on the two-leg ladder without ring exchange, i.e., our $J$-$J_{\perp}$-$K$ model with $K=0$.   In these works, it was shown that at half-filling the system enters a rung Mott insulating phase for any finite $J_\perp$, although the charge gap grows exponentially slowly with $J_\perp$, i.e., as $\exp(-aJ/J_\perp)$, and is extremely difficult to deal with numerically due to a large value of the constant factor $a$ in the exponential. \cite{Crepin11_PRB_84_054517}  In our DMRG study of the two-leg half-filled system, we observed what looked like a superfluid on finite-size systems for small $K$; however, we did not perform an extensive finite-size analysis of this system as in Ref.~\onlinecite{Crepin11_PRB_84_054517}.  It is therefore possible that the apparent superfluid is a rung Mott phase on long length scales, and there are only direct transitions between the rung Mott phase and the staggered rung current and $(\pi,\pi)$ CDW phases mentioned above.

To conclude, although we were able to successfully identify a GMI phase on the three-leg ladder (see Sec.~\ref{sec:3leg} and Ref.~\onlinecite{Block11_PRL_106_046402}), analogous phases do not exist on the two- and four-leg systems, at least with our model, boundary conditions, etc.  This again suggests a rather strong even-odd effect similar to what we found above for conducting DBL phases at incommensurate densities.


\begin{thebibliography}{10}

\bibitem{Ashcroft76_SSP}
N.~W. Ashcroft and N.~D. Mermin, {\em Solid State Physics} (Thomson, United
  States, 1976).

\bibitem{Baym91_FermiLiquids}
G. Baym and C. Pethick, {\em {L}andau {F}ermi-Liquid Theory: Concepts and
  Applications} (Wiley-VCH, Germany, 1991).

\bibitem{Troyer05_PRL_94_170201}
M. Troyer and U.-J. Wiese, Phys. Rev. Lett. {\bf 94},  170201  (2005).

\bibitem{Wolf06_PRL_96_010404}
M.~M. Wolf, Phys. Rev. Lett. {\bf 96},  010404  (2006).

\bibitem{Gioev06_PRL_96_100503}
D. Gioev and I. Klich, Phys. Rev. Lett. {\bf 96},  100503  (2006).

\bibitem{Barthel06_PRA_74_022329}
T. Barthel, M.-C. Chung, and U. Schollw\"ock, Phys. Rev. A {\bf 74},  022329
  (2006).

\bibitem{Swingle10_PRL_105_050502}
B. Swingle, Phys. Rev. Lett. {\bf 105},  050502  (2010).

\bibitem{Corboz09_PRB_80_165129}
P. Corboz and G. Vidal, Phys. Rev. B {\bf 80},  165129  (2009).

\bibitem{Corboz10_PRA_81_010303}
P. Corboz, G. Evenbly, F. Verstraete, and G. Vidal, Phys. Rev. A {\bf 81},
  010303  (2010).

\bibitem{Corboz10_PRB_81_165104}
P. Corboz, R. Or\'us, B. Bauer, and G. Vidal, Phys. Rev. B {\bf 81},  165104
  (2010).

\bibitem{Lee06_RevModPhys_78_17}
P.~A. Lee, N. Nagaosa, and X.-G. Wen, Rev. Mod. Phys. {\bf 78},  17  (2006).

\bibitem{Motrunich05_PRB_72_045105}
O.~I. Motrunich, Phys. Rev. B {\bf 72},  045105  (2005).

\bibitem{Lee05_PRL_95_036403}
S.-S. Lee and P.~A. Lee, Phys. Rev. Lett. {\bf 95},  036403  (2005).

\bibitem{Sheng09_PRB_79_205112}
D.~N. Sheng, O.~I. Motrunich, and M.~P.~A. Fisher, Phys. Rev. B {\bf 79},
  205112  (2009).

\bibitem{Block11_PRL_106_157202}
M.~S. Block, D.~N. Sheng, O.~I. Motrunich, and M.~P.~A. Fisher, Phys. Rev.
  Lett. {\bf 106},  157202  (2011).

\bibitem{Grover11_PRL_107_067202}
Y. Zhang, T. Grover, and A. Vishwanath, Phys. Rev. Lett. {\bf 107},  067202
  (2011).

\bibitem{Motrunich07_PRB_75_235116}
O.~I. Motrunich and M.~P.~A. Fisher, Phys. Rev. B {\bf 75},  235116  (2007).

\bibitem{Sheng08_PRB_78_054520}
D.~N. Sheng, O.~I. Motrunich, S. Trebst, E. Gull, and M.~P.~A. Fisher, Phys. Rev. B {\bf 78},  054520  (2008).

\bibitem{Block11_PRL_106_046402}
M.~S. Block, R.~V. Mishmash, R.~K. Kaul, D.~N. Sheng, O.~I. Motrunich, and M.~P.~A. Fisher, Phys. Rev. Lett. {\bf 106},  046402  (2011).

\bibitem{Fisher08_condmat0812.2955}
M.~P.~A. Fisher, O.~I. Motrunich, and D.~N. Sheng, arXiv:0812.2955v1
  [cond-mat.str-el]  (2008).

\bibitem{White92_PRL_69_2863}
S.~R. White, Phys. Rev. Lett. {\bf 69},  2863  (1992).

\bibitem{White93_PRB_48_10345}
S.~R. White, Phys. Rev. B {\bf 48},  10345  (1993).

\bibitem{Schollwock05_RevModPhys_77_259}
U. Schollwock, Rev. Mod. Phys. {\bf 77},  259  (2005).

\bibitem{Ceperley77_PRB_16_3081}
D. Ceperley, G.~V. Chester, and M.~H. Kalos, Phys. Rev. B {\bf 16},  3081
  (1977).

\bibitem{Buchler05_PRL_95_040402}
H.~P. B\"uchler, M. Hermele, S. D. Huber, M. P. A. Fisher, and P. Zoller, Phys. Rev. Lett. {\bf 95},  040402  (2005).

\bibitem{Feiguin09_PRL_103_025303}
A.~E. Feiguin and M.~P.~A. Fisher, Phys. Rev. Lett. {\bf 103},  025303  (2009).

\bibitem{Mandel03_Nature_425_937}
O. Mandel {\it et~al.}, Nature {\bf 425},  937  (2003).

\bibitem{Mandel03_PRL_91_010407}
O. Mandel {\it et~al.}, Phys. Rev. Lett. {\bf 91},  010407  (2003).

\bibitem{Feiguin11_PRB_83_115104}
A.~E. Feiguin and M.~P.~A. Fisher, Phys. Rev. B {\bf 83},  115104  (2011).

\bibitem{Chubukov92_PRB_45_7889}
A. Chubukov, E. Gagliano, and C. Balseiro, Phys. Rev. B {\bf 45},  7889
  (1992).

\bibitem{Lauchli05_PRL_95_137206}
A. L\"auchli, J.~C. Domenge, C. Lhuillier, P. Sindzingre, and M. Troyer, Phys. Rev. Lett. {\bf 95},  137206  (2005).

\bibitem{Lauchli03_PRB_67_100409}
A. L\"auchli, G. Schmid, and M. Troyer, Phys. Rev. B {\bf 67},  100409  (2003).

\bibitem{Schaffer09_PRB_80_014503}
R. Schaffer, A.~A. Burkov, and R.~G. Melko, Phys. Rev. B {\bf 80},  014503
  (2009).

\bibitem{Tay11_PRB_83_205107}
T. Tay and O.~I. Motrunich, Phys. Rev. B {\bf 83},  205107  (2011).

\bibitem{Tay11_PRB_83_235122}
T. Tay and O.~I. Motrunich, Phys. Rev. B {\bf 83},  235122  (2011).

\bibitem{Hellberg91_PRL_67_2080}
C.~S. Hellberg and E.~J. Mele, Phys. Rev. Lett. {\bf 67},  2080  (1991).

\bibitem{Hellberg92_PRL_68_3111}
C.~S. Hellberg and E.~J. Mele, Phys. Rev. Lett. {\bf 68},  3111  (1992).

\bibitem{Hellberg93_PRB_48_646}
C.~S. Hellberg and E.~J. Mele, Phys. Rev. B {\bf 48},  646  (1993).

\bibitem{Chen93_PRB_47_11548}
Y.~C. Chen and T.~K. Lee, Phys. Rev. B {\bf 47},  11548  (1993).

\bibitem{Chen96_PRB_54_9062}
Y.~C. Chen and T.~K. Lee, Phys. Rev. B {\bf 54},  9062  (1996).

\bibitem{Yang07_CM_19_186218}
H.-Y. Yang and T. Li, Journal of Physics: Condensed Matter {\bf 19},  186218
  (2007).

\bibitem{Paramekanti02_PRB_66_054526}
A. Paramekanti, L. Balents, and M.~P.~A. Fisher, Phys. Rev. B {\bf 66},  054526
   (2002).

\bibitem{Sandvik06_AnnPhys_321_1651}
A. Sandvik and R. Melko, Ann. Phys. {\bf 321},  1651   (2006), {J}uly 2006
  {S}pecial {I}ssue.

\bibitem{Sandvik02_PRL_89_247201}
A.~W. Sandvik, S. Daul, R.~R.~P. Singh, and D.~J. Scalapino, Phys. Rev. Lett.
  {\bf 89},  247201  (2002).

\bibitem{Melko04_PRB_69_100408}
R.~G. Melko, A.~W. Sandvik, and D.~J. Scalapino, Phys. Rev. B {\bf 69},  100408
   (2004).

\bibitem{Rousseau04_PRL_93_110404}
V. Rousseau, G.~G. Batrouni, and R.~T. Scalettar, Phys. Rev. Lett. {\bf 93},
  110404  (2004).

\bibitem{Rousseau05_PRB_72_054524}
V.~G. Rousseau, R.~T. Scalettar, and G.~G. Batrouni, Phys. Rev. B {\bf 72},
  054524  (2005).

\bibitem{Tay10_PRL_105_187202}
T. Tay and O.~I. Motrunich, Phys. Rev. Lett. {\bf 105},  187202  (2010).

\bibitem{Cardy04_JStatMech_P06002}
P. Calabrese and J. Cardy, J. Stat. Mech. {\bf 2004},  P06002  (2004).

\bibitem{Pollman09_PRL_102_255701}
F. Pollmann, S. Mukerjee, A.~M. Turner, and J.~E. Moore, Phys. Rev. Lett. {\bf
  102},  255701  (2009).

\bibitem{Hastings10_PRL_104_157201}
M.~B. Hastings, I. Gonz\'alez, A.~B. Kallin, and R.~G. Melko, Phys. Rev. Lett.
  {\bf 104},  157201  (2010).

\bibitem{Calabrese10_PRL_104_095701}
P. Calabrese, M. Campostrini, F. Essler, and B. Nienhuis, Phys. Rev. Lett. {\bf
  104},  095701  (2010).

\bibitem{Vidal07_PRL_99_220405}
G. Vidal, Phys. Rev. Lett. {\bf 99},  220405  (2007).

\bibitem{Vidal08_PRL_101_110501}
G. Vidal, Phys. Rev. Lett. {\bf 101},  110501  (2008).

\bibitem{Carrasquilla11_PRB_83_245101}
J. Carrasquilla, F. Becca, and M. Fabrizio, Phys. Rev. B {\bf 83},  245101
  (2011).

\bibitem{Crepin11_PRB_84_054517}
F. Cr\'epin, N. Laflorencie, G. Roux, and P. Simon, Phys. Rev. B {\bf 84},
  054517  (2011).

\bibitem{Song10_PRB_82_012405}
H.~F. Song, S. Rachel, and K. Le~Hur, Phys. Rev. B {\bf 82},  012405  (2010).

\bibitem{Lauchli06_PRB_74_144426}
A. L\"auchli, G. Schmid, and S. Trebst, Phys. Rev. B {\bf 74},  144426  (2006).

\bibitem{Lauchli10_PRL_105_267204}
H.-Y. Yang, A.~M. L\"auchli, F. Mila, and K.~P. Schmidt, Phys. Rev. Lett. {\bf
  105},  267204  (2010).

\bibitem{Lai74_JMathPhys_15_1675}
C.~K. Lai, J. Math. Phys. {\bf 15},  1675  (1974).

\bibitem{Sutherland75_PRB_12_3795}
B. Sutherland, Phys. Rev. B {\bf 12},  3795  (1975).

\bibitem{Corboz07_PRB_76_220404}
P. Corboz, A.~M. L\"auchli, K. Totsuka, and H. Tsunetsugu, Phys. Rev. B {\bf
  76},  220404  (2007).

\bibitem{Rachel08_AnnPhys_17_922}
M. F\"{u}hringer, S. Rachel, R. Thomale, M. Greiter, and P. Schmitteckert, Ann. Phys. (Berlin) {\bf 17},  922  (2008).

\bibitem{Fath91_PRB_44_11836}
G. F\'ath and J. S\'olyom, Phys. Rev. B {\bf 44},  11836  (1991).

\bibitem{Itoi97_PRB_55_8295}
C. Itoi and M.-H. Kato, Phys. Rev. B {\bf 55},  8295  (1997).

\bibitem{Schmitt96_JPhysA_29_3951}
A. Schmitt, K.-H. M\"{u}tter, and M. Karbach, Journal of Physics A: Mathematical
  and General {\bf 29},  3951  (1996).

\bibitem{White11_Science_332_1173}
S. Yan, D. A. Huse, and S. R. White, Science {\bf 332}, 1173 (2011).

\bibitem{Stoudenmire11_condmat1105.1374}
E. M. Stoudenmire and S. R. White, arXiv:1105.1374v2 [cond-mat.str-el] (2011).

\bibitem{Lin97_PRB_56_6569}
H.-H. Lin, L. Balents, and M.~P.~A. Fisher, Phys. Rev. B {\bf 56},  6569
  (1997).

\bibitem{Lin98_PRB_58_1794}
H.-H. Lin, L. Balents, and M.~P.~A. Fisher, Phys. Rev. B {\bf 58},  1794
  (1998).

\bibitem{Dmitriev02_PRB_65_172409}
D.~V. Dmitriev, V.~Y. Krivnov, and A.~A. Ovchinnikov, Phys. Rev. B {\bf 65},
  172409  (2002).

\end{thebibliography}

\end{document}